\documentclass[3p]{elsarticle}
\usepackage{lineno,hyperref}
\usepackage{amsmath}
\usepackage{color}
\usepackage[normalem]{ulem}

\journal{arXiv}
\begin{document}

\begin{frontmatter}

\title{Distributions of Historic Market Data -- Implied and Realized Volatility}

 \author[mymainaddress]{M. Dashti Moghaddam}
 \author[mymainaddress]{Zhiyuan Liu}
 \author[mymainaddress]{R. A. Serota\fnref{myfootnote}}
 \fntext[myfootnote]{serota@ucmail.uc.edu}
 
 \address[mymainaddress]{Department of Physics, University of Cincinnati, 
 Cincinnati, Ohio 45221-0011}

\begin{abstract}
We undertake a systematic comparison between implied volatility, as represented by VIX (new methodology) and VXO (old methodology), and realized volatility. We compare visually and statistically distributions of realized and implied variance (volatility squared) and study the distribution of their ratio. We find that the ratio is best fitted by heavy-tailed -- lognormal and fat-tailed (power-law) -- distributions, depending on whether preceding or concurrent month of realized variance is used. We do not find substantial difference in accuracy between VIX and VXO. Additionally, we study the variance of theoretical realized variance for Heston and multiplicative models of stochastic volatility and compare those with realized variance obtained from historic market data.
\end{abstract}


\end{frontmatter}

\section{Introduction\label{Introduction}}
The implied volatility index VIX was created in order to estimate, looking forward, the expected realized volatility. CBOE introduced the original VIX (now VXO) in 1986. It was based on an inverted Black-Scholes formula, where S\&P 100 near-term, at-the-money options were used to calculate a weighted average of volatilities. However, the Black-Scholes formula assumes that the volatility in the stock returns equation is either a constant, or at least does not have a stochastic component, while in reality it was already understood that volatility itself is stochastic in nature. A number of well-studied models of stochastic volatility have emerged, such as Heston (HM) \cite{heston1993closed, dragulescu2002probability} and multiplicative (MM) \cite{nelson1990arch, ma2014model}. Consequently, a need arose for an implied volatility index, which would not only be based on stochastic volatility but would also be agnostic to a particular model of the latter \cite{zhou2003vix, bollerslev2004dynamic}. 

CBOE introduced its current  VIX methodology on September 22, 2003 \cite{whitepaper2003cboe} to fulfill the above requirements and was based on \cite{demeterfi1999more, demeterfi1999guide}, where a closed-form formula for the expected value of realized volatility \cite{barndorff2002econometric} was derived using call and put prices. Notably, it utilized the S\&P 500 index, which is far more representative of the total market, both near-term and next-term options and a broader range of strike prices. CBOE publishes historic data using both methodologies, VIX (new) and VXO (old) dating back to 1990 \cite{cboe2018vixhistoric} (historic stock prices used in calculation of realized volatility can be found at \cite{google2018sp500historic}). Here we call 1990 through September 19, 2003 VIX Archive and VXO Archive and from September 22, 2003 through December 30, 2016 VIX Current and VXO Current.

Naturally, the question arises of whether VIX, designed to be a superior methodology, has a better track record than VXO. The short answer is that it is unclear. All-in-all, VIX/VXO is still too young to have accumulated sufficient amount of data and only time will tell how reliable it is in predicting realized volatility. Still, one of our notable observations discussed below is that the ratio of realized to implied variance (squared volatility) is best fitted with a fat-tailed (power-law) distribution, which clearly signals occasional large discrepancies between prediction and realization. This is not surprising, given that we are trying to predict the future (by pricing options) based on what we know today and thus are unaware of unexpected future events that can spike the volatility. 

On the other hand, we also find that the distribution of the ratio of realized variance of the \emph{preceding} month to the implied volatility, as well as of its inverse, is distributed with lognormal distribution. While the latter is heavy-tailed, this nonetheless shows that VIX is better attuned to the known volatility. We note a recent surmise that VIX can be manipulated \cite{griffin2017manipulation,eisen2017traders} and that Nasdaq is working on its own volatility index \cite{banerji2017nasdaq}. Hopefully, this work will establish a proper framework for testing implied volatility indices.

This paper is the second in a series devoted to analysis of historic market data, the other two discussing, respectively, stock returns \cite{liu2017distributions} and relaxation and correlations \cite{dashti2018correlations}. It is organized as follows. In Section \ref{RV2vVIX2} we give a detailed visual and statistical comparison between realized volatility ($RV$) and implied volatility represented by VIX and VXO. More precisely, we compare distributions of realized variance $RV^2$ with $VIX^2$ and $VXO^2$ and, in particular, we analyze KS statistics of fits of $RV^2/VIX^2$ and $RV^2/VXO^2$ by various distributions, from normal to fat-tailed. In Section \ref{RV2var} we compare the variance of the $RV^2$ distribution against the analytical results obtained using Heston and multiplicative models respectively. We conclude with the discussion of open questions and future work.

\section{Comparing distributions of $RV^2$ and $VIX^2$}\label{RV2vVIX2}
\subsection{Definitions, rescaling and normality}\label{rescaling}

Realized variance (index) is defined as follows
\begin{equation}
RV^2=100^2\times\frac{252}{n}\sum_{i=1}^nr_i^2
\label{RV2}
\end{equation}
where
\begin{equation}
r_i=\ln\frac{S_{i}}{S_{i-1}}
\label{ri}
\end{equation}
are daily returns and $S_{i}$ is the reference (closing) price on day $i$. Time-averaged realized variance can be calculated from stochastic volatility $\sigma_t$ \cite{barndorff2002econometric}, \cite{liu2017distributions} as 
\begin{equation}
\frac{1}{\tau}\int_{0}^{\tau}\sigma_t^2\mathrm{d}t 
\label{V}
\end{equation}
Evaluation of the implied volatility is based on the evaluation of the expectation value of (\ref{V}) \cite{demeterfi1999more, demeterfi1999guide}. VIX uses options prices to estimate this expectation value via the generalized formula \cite{whitepaper2003cboe}
\begin{equation}
VIX^2=(100)^2\times\left(\frac{2}{T} \sum_{i} \frac{\Delta K_i}{K_{i}^2} e^{RT} Q(K_i) - \frac{1}{T} [\frac{F}{K_0}-1]^2\right)
\label{VIX2} 
\end{equation}
where $T$ is the time to expiration; $F$ is the forward index level desired from index option price; $K_0$ is the first strike below the forward index level, $F$; $K_i$ is the strike price of $i$th out-of-money option: a call if $K_i > K_0$, a put if $K_i < K_0$ and both a put and a call if $K_i=K_0$; $\Delta K_i$ is the interval between strike prices, that is half the difference between the strike on either side of $K_i$, $\Delta K_i=(K_{i+1}-K_{i-1})/2$; $R$ is the risk-free interest rate to expiration and $Q(K_i)$ is the midpoint of bid-ask spread for each option with strike $K_i$. This formula is then used for near- and next-term options \cite{whitepaper2003cboe} and the final expression for VIX is effectively an average between the two so the latter and the sum in (\ref{VIX2}) are intended to approximate the time average in (\ref{V}).

VIX and VXO were designed to measure a 30-day expected volatility. However, in their final form $VIX^2$ and $VXO^2$ are annualized by the ratio of $365/30\approx12$ \cite{whitepaper2003cboe}. As is clear from (\ref{RV2}), $RV^2$ is also annualized and for comparison with VIX/VXO, we should take $n=21$, so that $252/21=12$; unlike VIX/VXO, RV is calculated based on the number of trading days. Accordingly, to compare the distributions of $VIX^2$ and $VXO^2$ with $RV^2$, we must rescale one of them with the ratio of their mean values. Table \ref{365252} lists ratios of the mean of $VIX^2$ and $VXO^2$ over the mean of $RV^2$. In what follows, the distributions of $RV^2$ are rescaled with the respective ratios from Table \ref{365252}. We also analyze data for VIX Current and VXO Current both in aggregate form and split nearly evenly for a period covering the financial crisis and after (see Appendix).

\begin{figure}[!htbp]
\centering
\begin{tabular}{cc}
\includegraphics[width = 0.49 \textwidth]{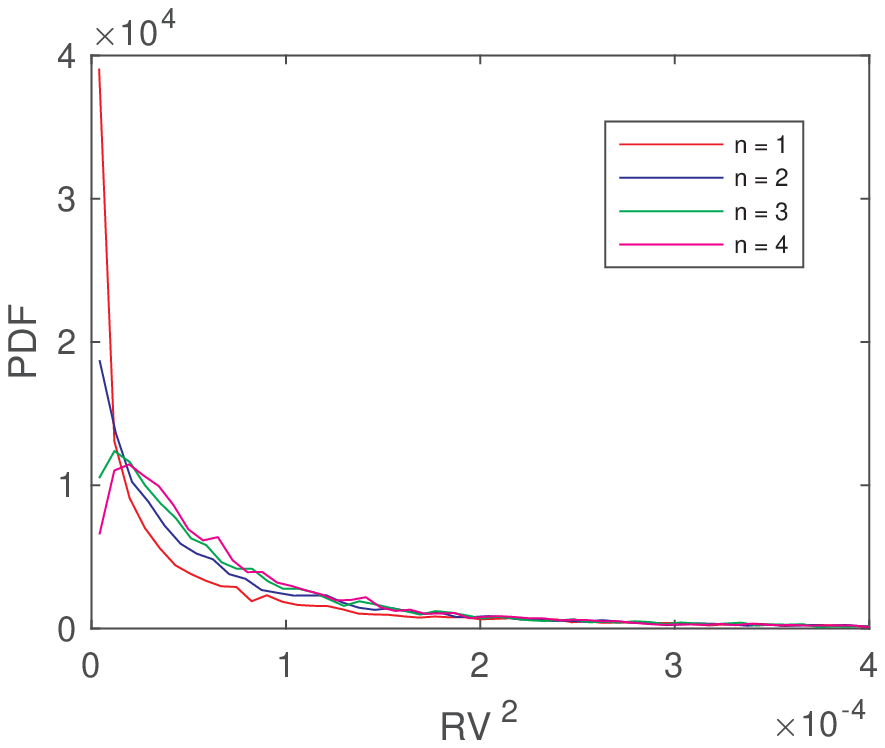} \hspace{0.1cm}
\includegraphics[width = 0.49 \textwidth]{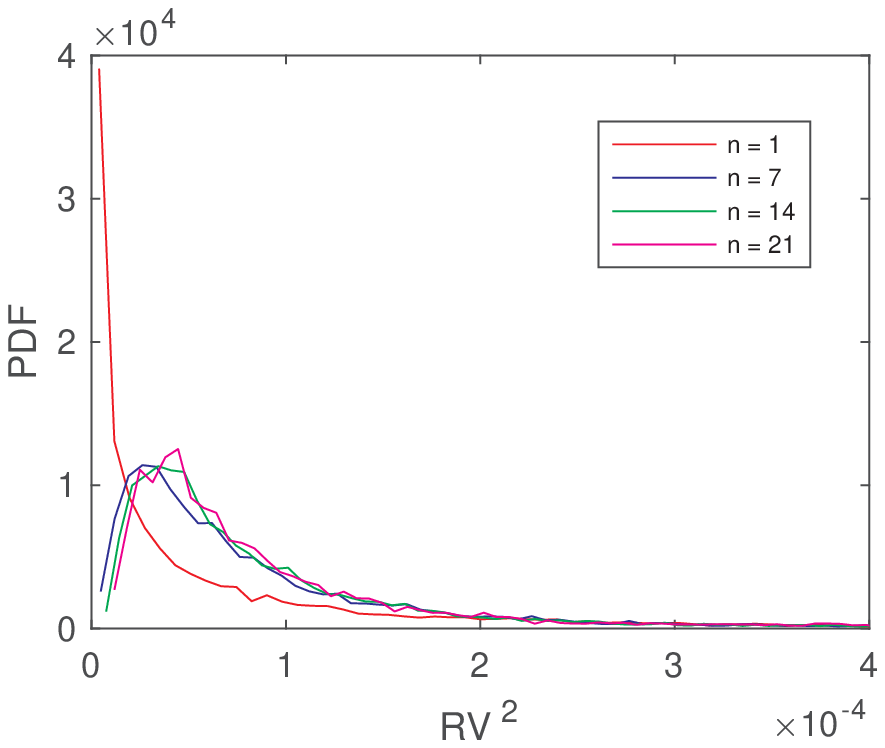}
\end{tabular}
\caption{PDFs of $\frac{1}{n}\sum_{i=1}^nr_i^2$ for $n=$1,2,3,4 (left) and $n=$1,7,14,21 (right).}
\label{sumri2}
\end{figure}

\begin{figure}[!htbp]
\centering
\begin{tabular}{cc}
\includegraphics[width = 0.56 \textwidth]{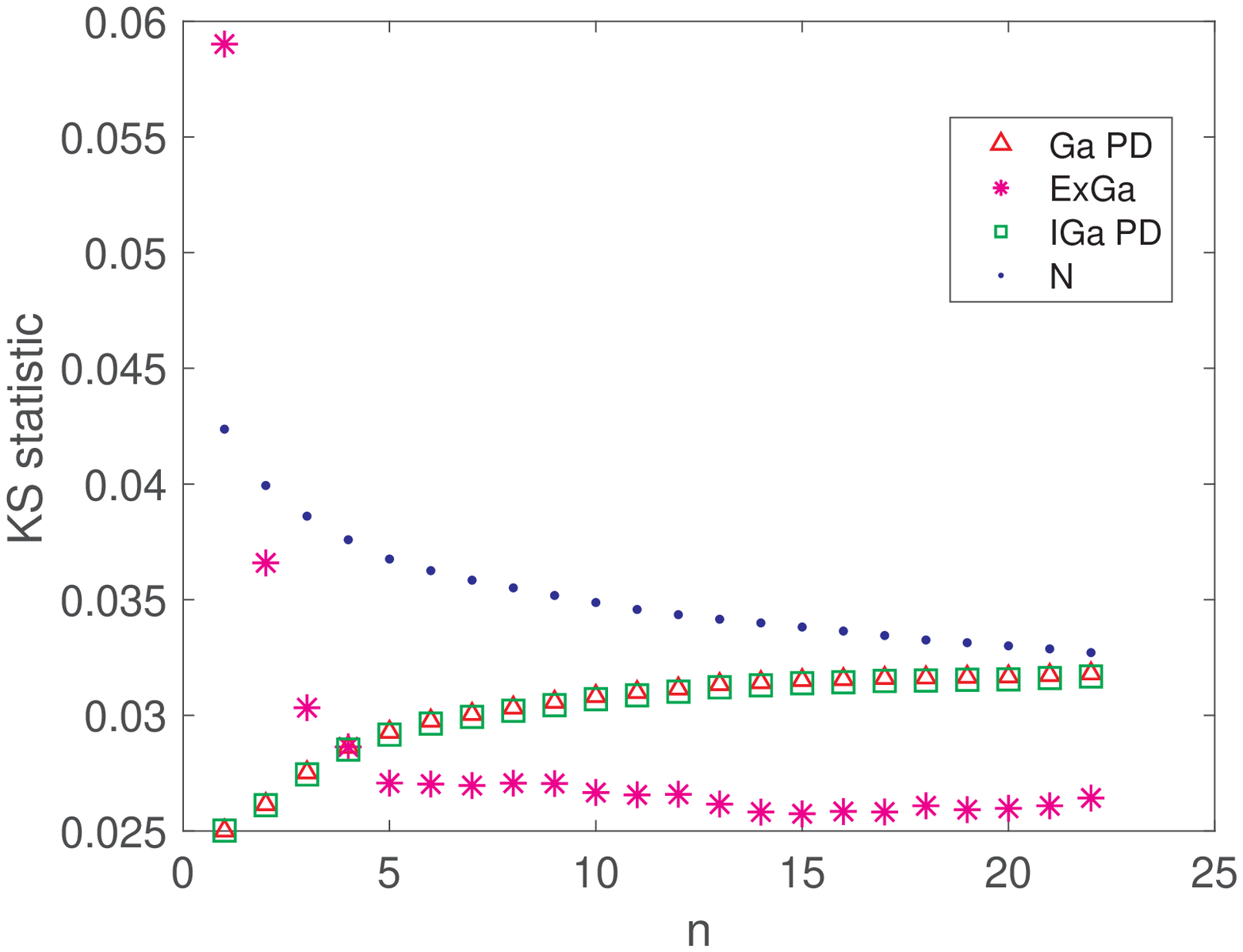}
\includegraphics[width = 0.56 \textwidth]{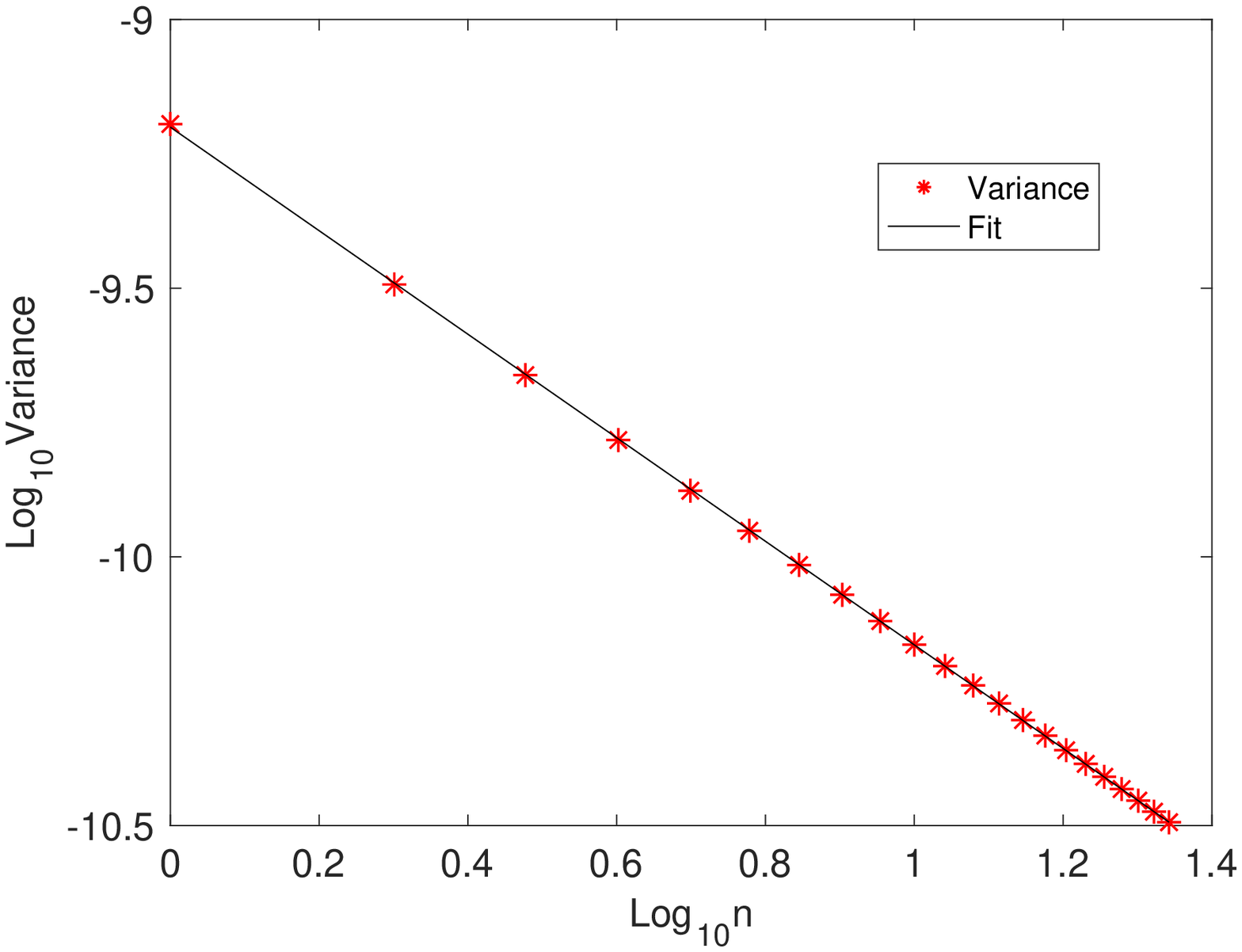}
\end{tabular}
\caption{\textbf{Left:} Kolmogorov-Smirnov statistics for fitting $\frac{1}{n}\sum_{i=1}^nr_i^2$; lower numbers indicate a better fit. N and ExGa are normal and exGaussian distributions respectively. Ga PD and IGa PD are product distributions of gamma and inverse gamma distributions respectively and normal distribution -- which describe the distributions of stock returns in the Heston and multiplicative models \cite{liu2017distributions} -- modified by a change of  variables to squared returns. \textbf{Right:} Log-log plot of the variance of the $RV^2$ distribution versus $n$. The slope of the straight-line fit is -0.9635.}
\label{KSsumri2}
\end{figure}

It should be emphasized that for $n=21$ in ({\ref{RV2}}) the distribution of $RV^2$ should be approaching normal. Fig. \ref{sumri2} hints at that but with an extended tail. The tail may be exponential or power-law, depending on how single-day returns are distributed. While the longer-time returns are better described by the Heston model and exponential tails \cite{liu2017distributions}, single-day returns is still an open question. As always, the tail behavior is hard to pinpoint, especially with smaller data sets. Fig. \ref{KSsumri2} confirms the $RV^2$ distribution approach to modified normality.

\subsection{Visual comparison of realized volatility and VIX/VXO}\label{visual}

\begin{figure}[!htbp]
\centering
\includegraphics[width = 0.8 \textwidth]{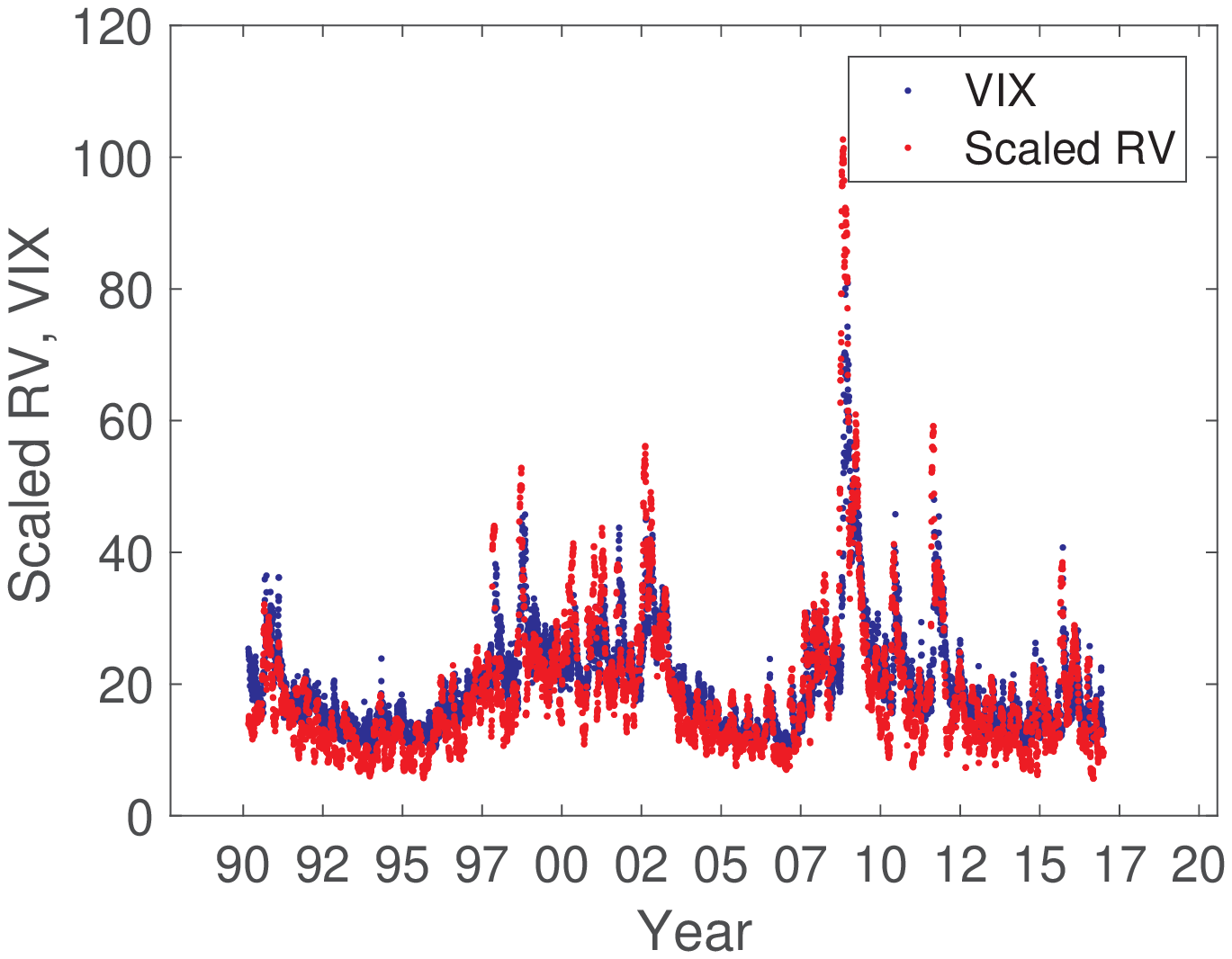}
\includegraphics[width = 0.8 \textwidth]{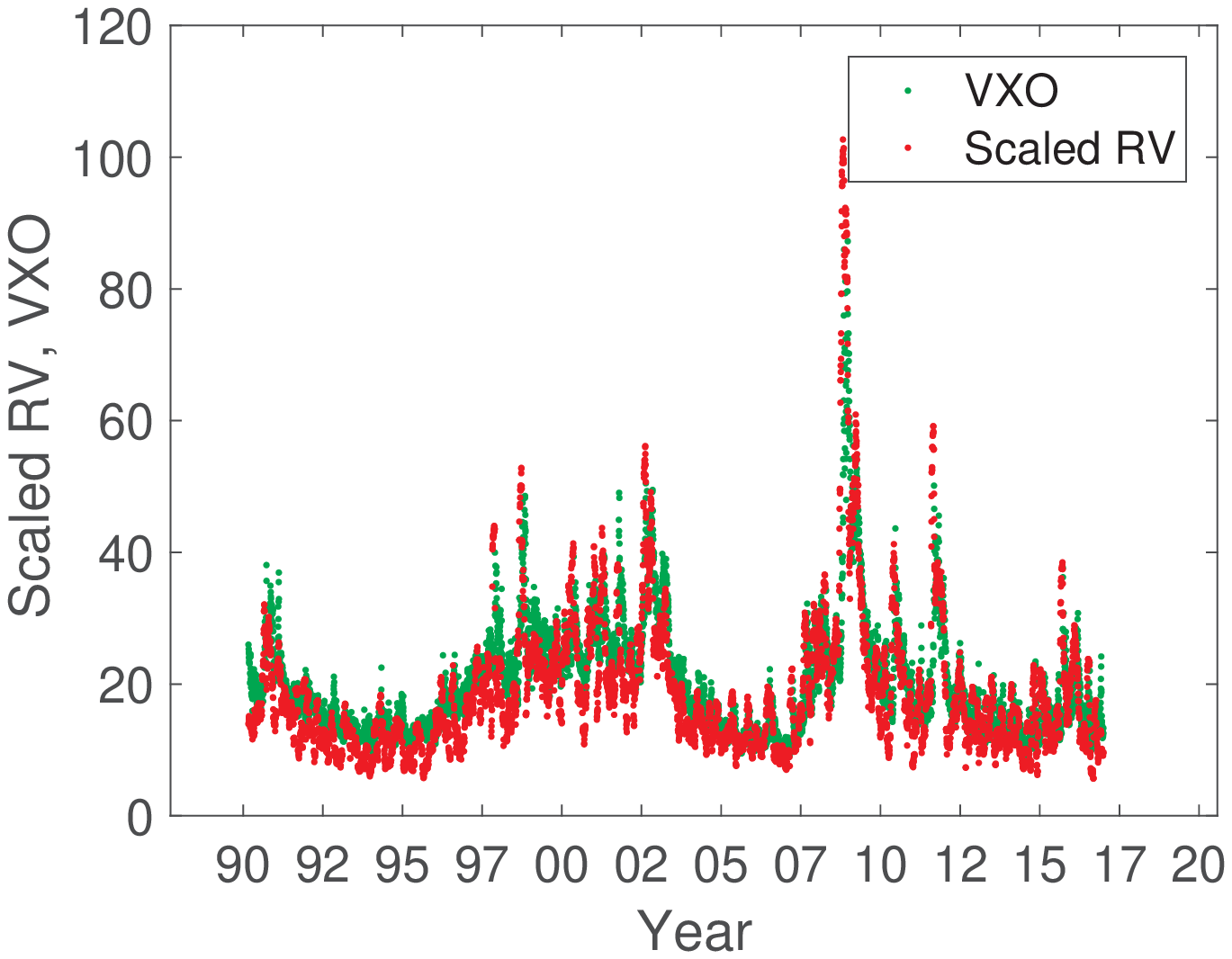}
\caption{ VIX (top) and VXO (bottom) with scaled RV, from Jan 2nd, 1990 to Dec 30th, 2016.}
\label{VIXVXORVPlot}
\end{figure}

\begin{figure}[!htbp]
\centering
\includegraphics[width = 0.8 \textwidth]{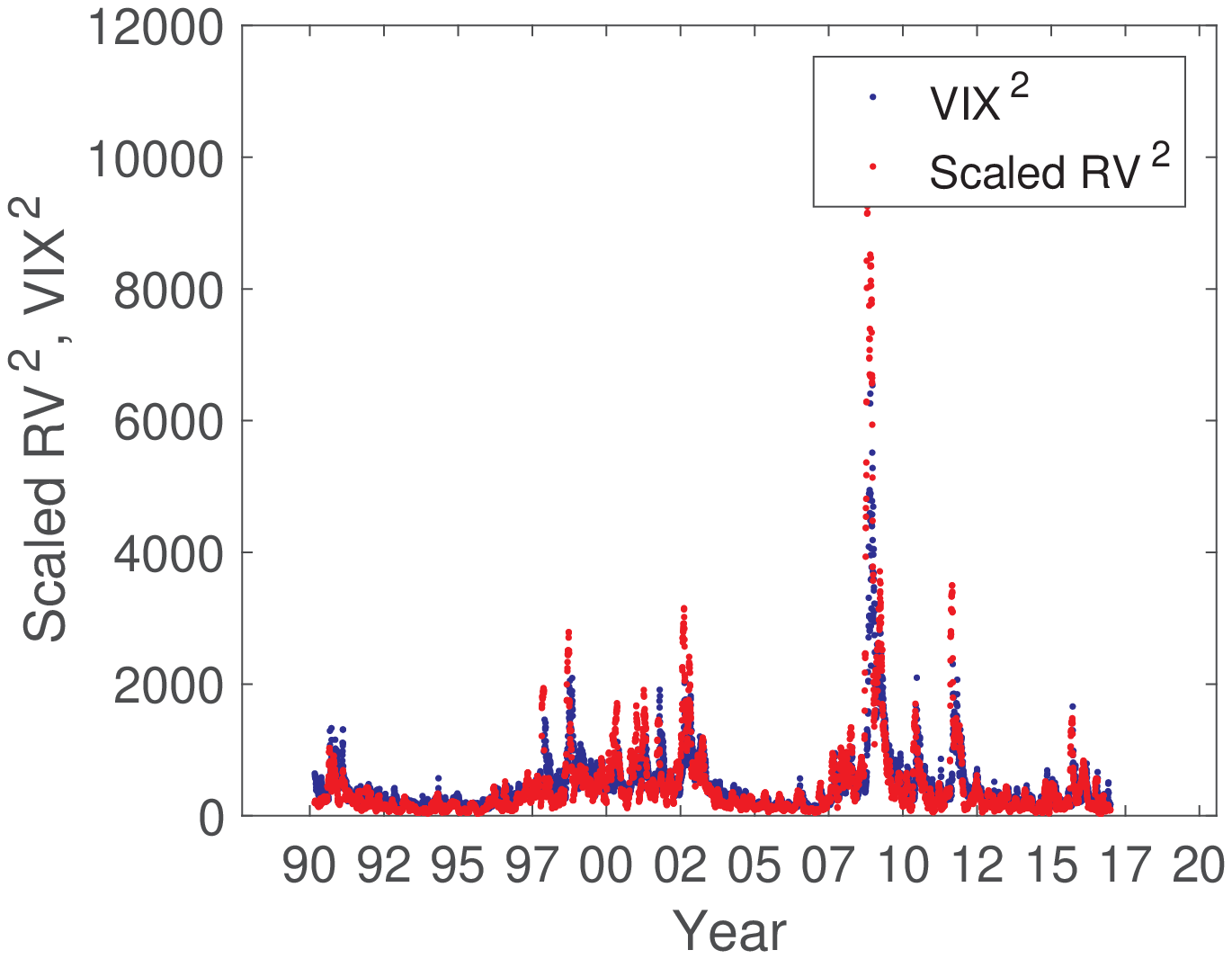}
\includegraphics[width = 0.8 \textwidth]{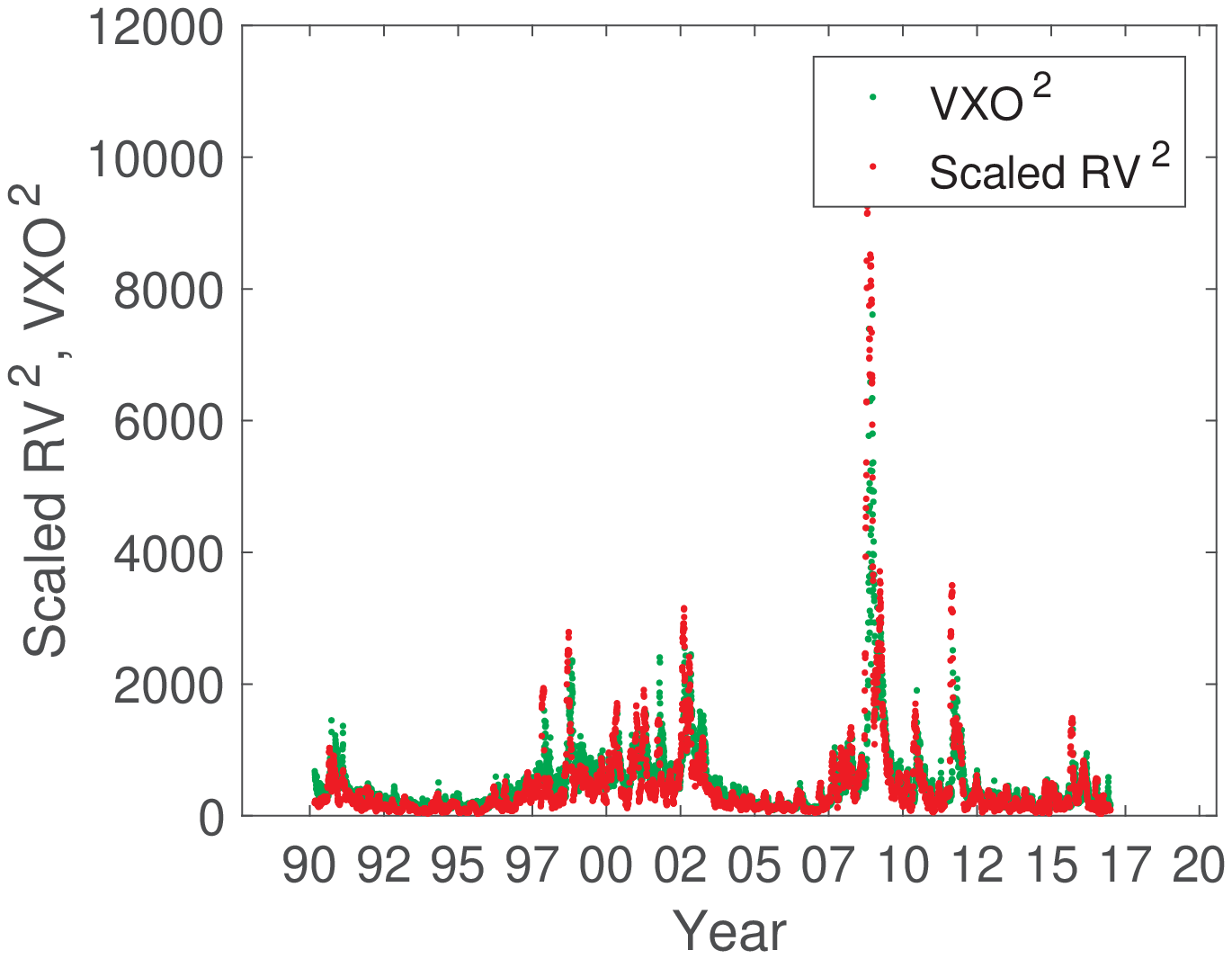}
\caption{${VIX^2}$ (top) and ${VXO^2}$ (bottom) with Scaled ${RV^2} $, from Jan 2nd, 1990 to Dec 30th, 2016.}
\label{VIX2VXO2RV2Plot}
\end{figure}

\begin{figure}[!htbp]
\centering
\begin{tabular}{cc}
\includegraphics[width = 0.49 \textwidth]{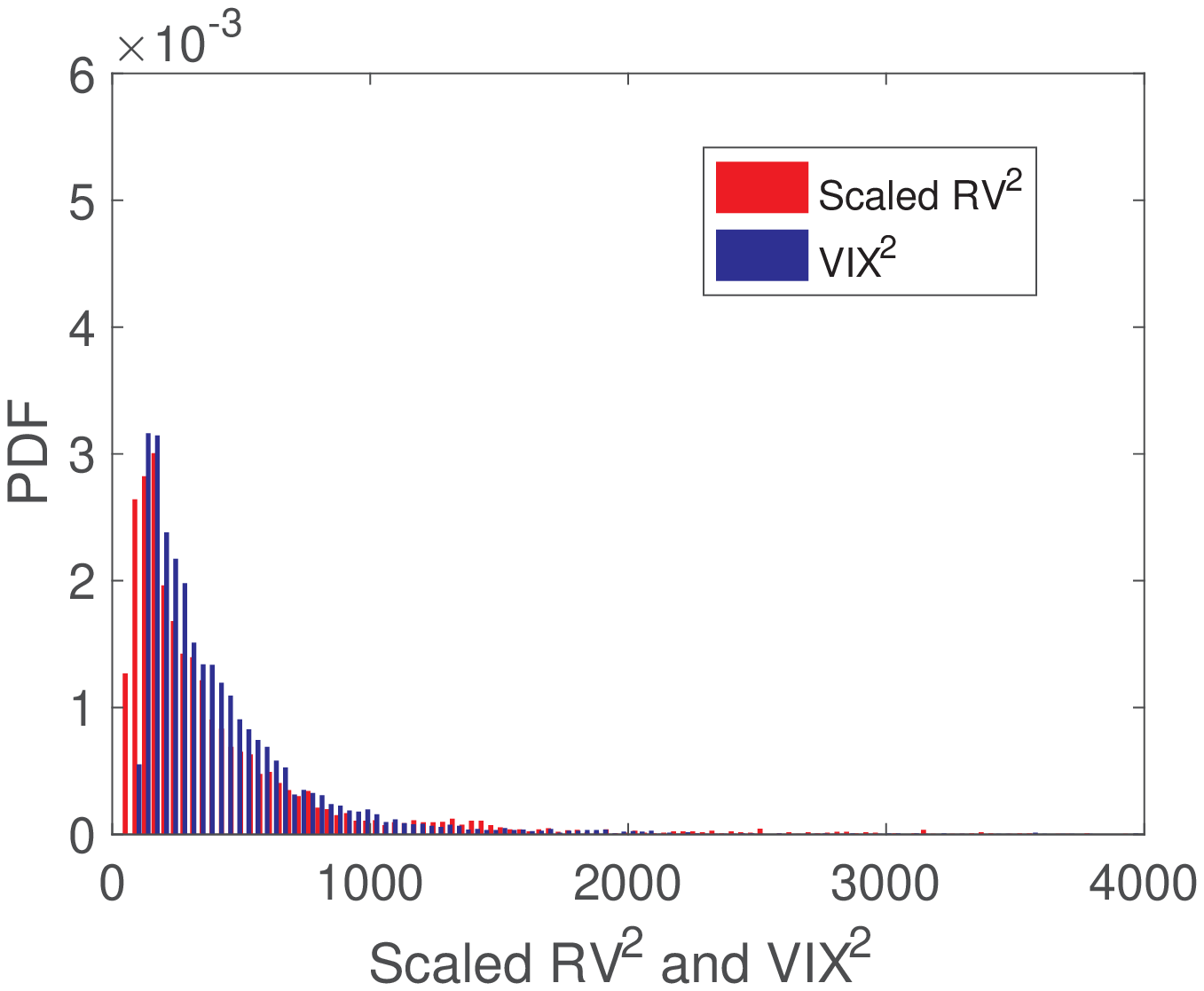} \hspace{0.2cm}
\includegraphics[width = 0.49 \textwidth]{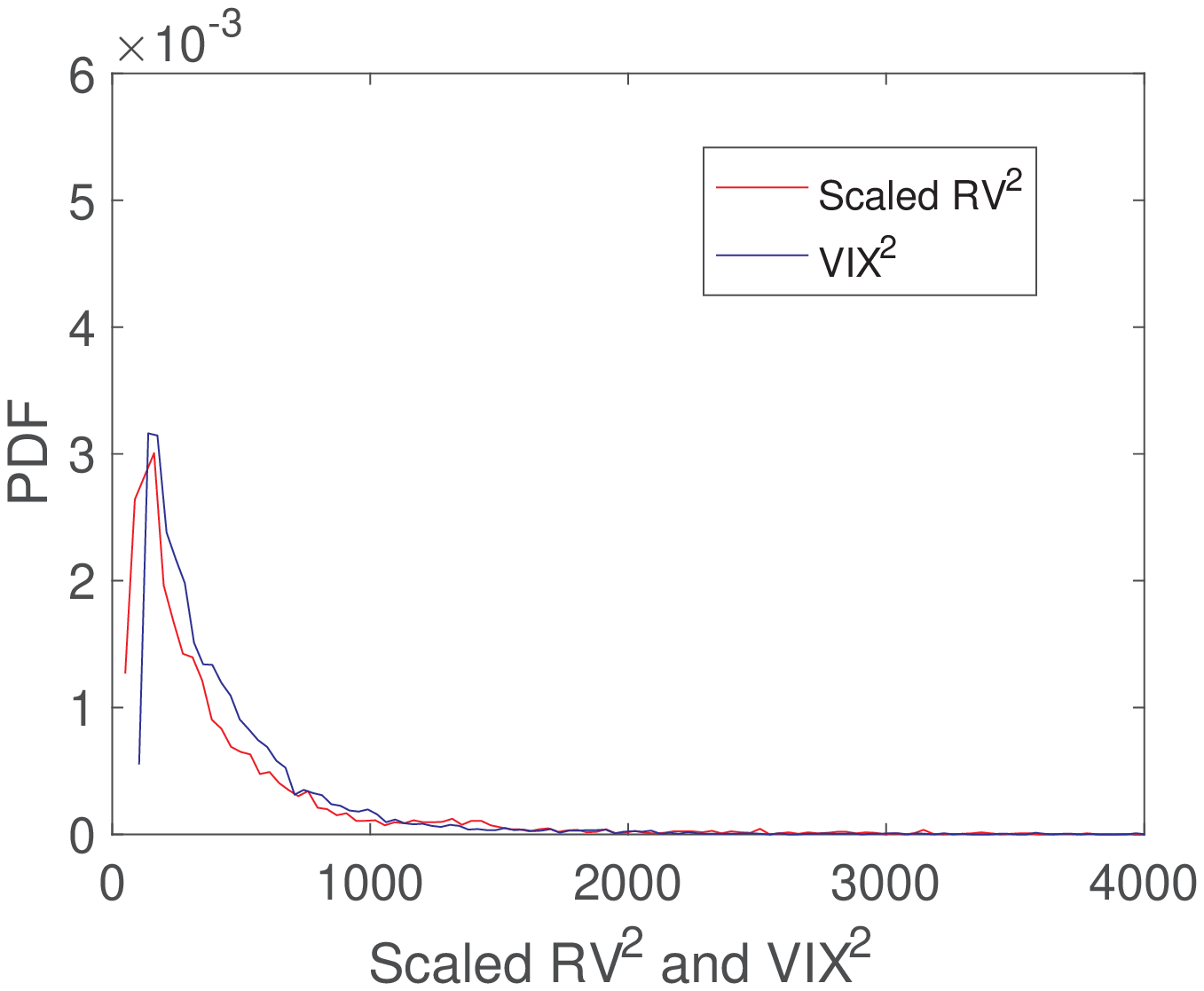}
\end{tabular}
\caption{PDFs of scaled $RV^2$ and $VIX^2$ from Jan 2nd, 1990 to Dec 30th, 2016.}
\label{histogramScaleRVVIX19902016ScaleRVVIXHistogram}
\end{figure}

\begin{figure}[!htbp]
\centering
\begin{tabular}{cc}
\includegraphics[width = 0.49 \textwidth]{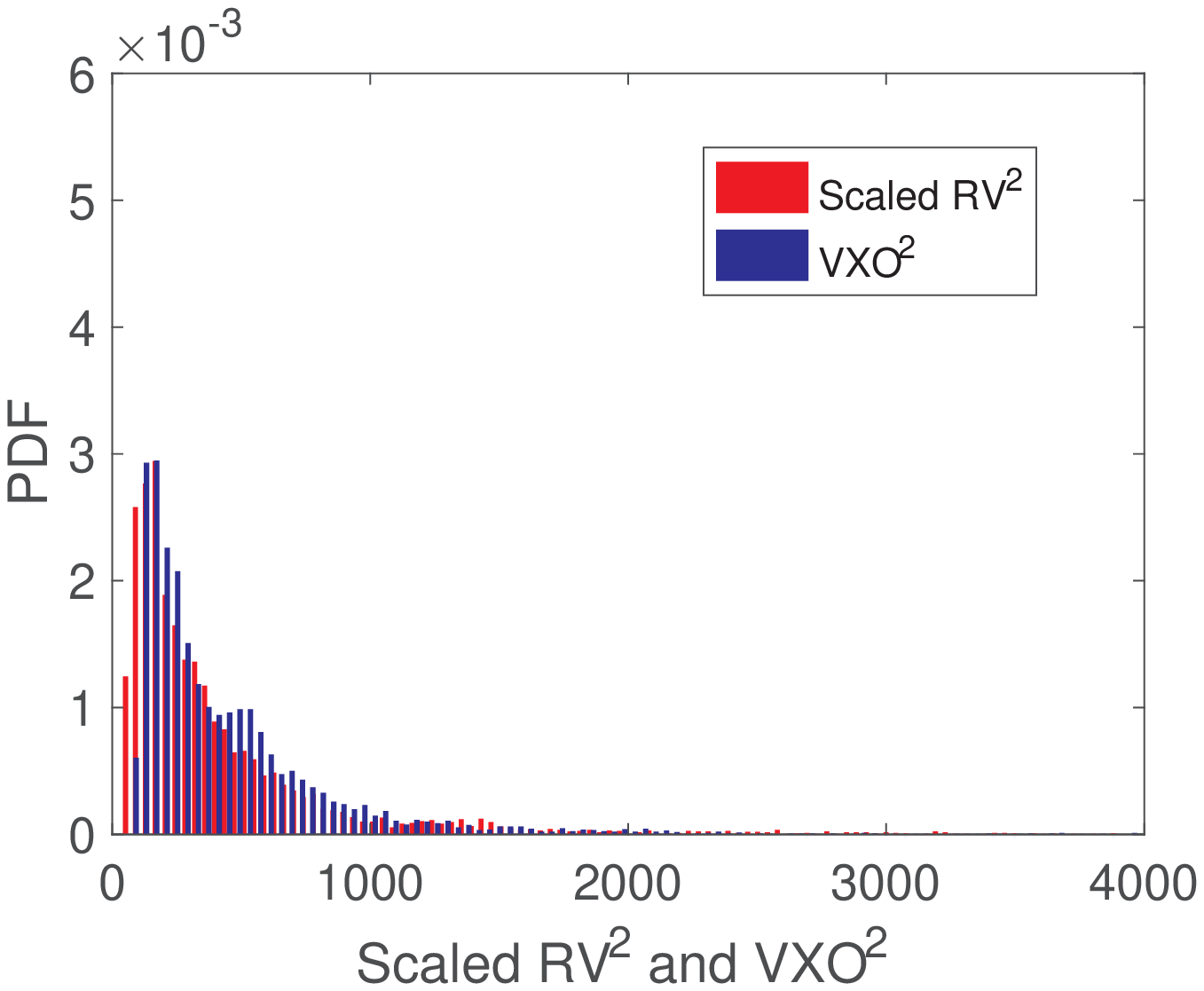} \hspace{0.2cm}
\includegraphics[width = 0.49 \textwidth]{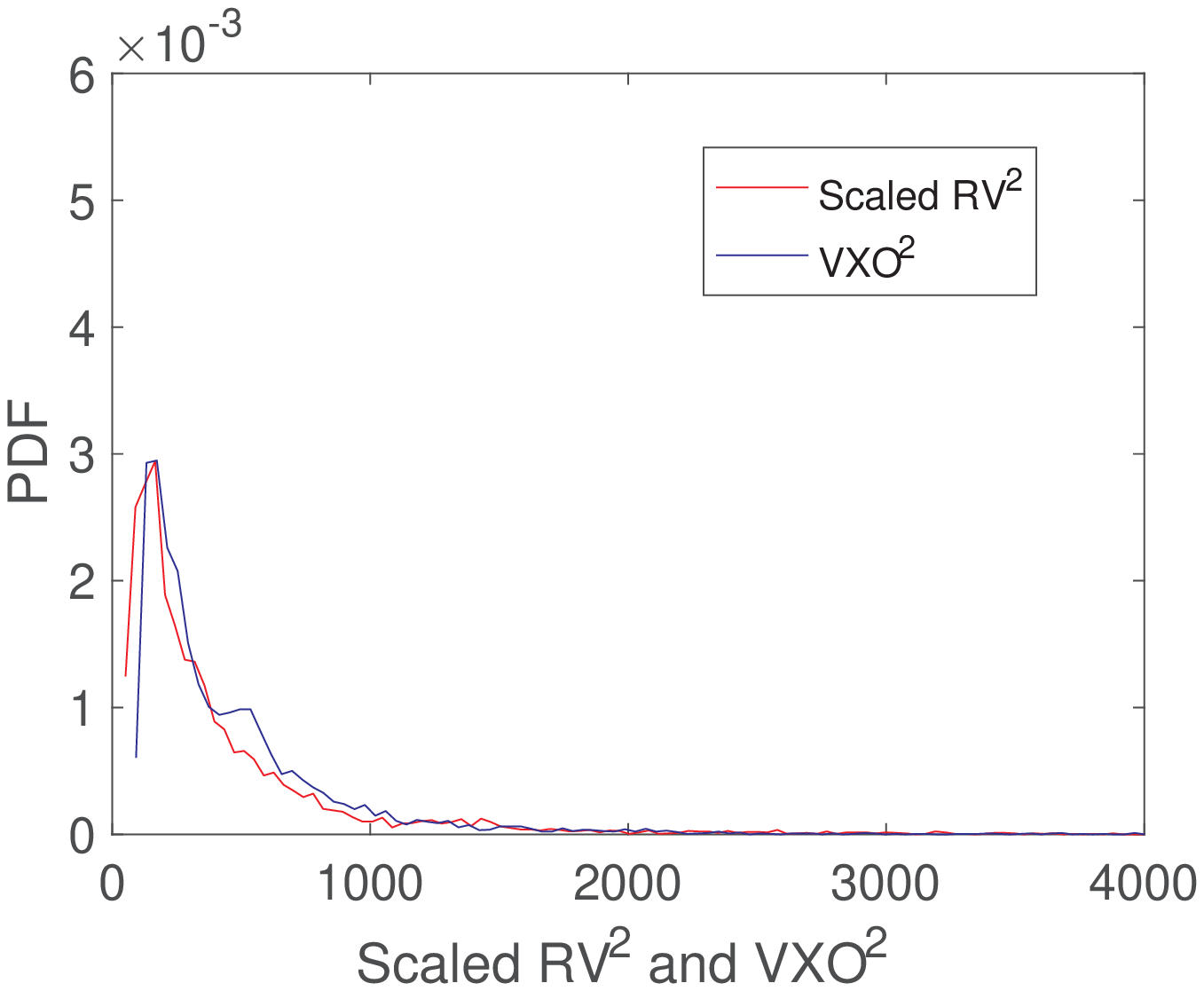}
\end{tabular}
\caption{PDFs of scaled $RV^2$ and  $VXO^2$ from Jan 2nd, 1990 to Dec 30th, 2016.}
\label{histogramScaleRVVXO19902016ScaleRVVXOHistogram}
\end{figure}

\begin{figure}[!htbp]
\centering
\begin{tabular}{cc}
\includegraphics[width = 0.49 \textwidth]{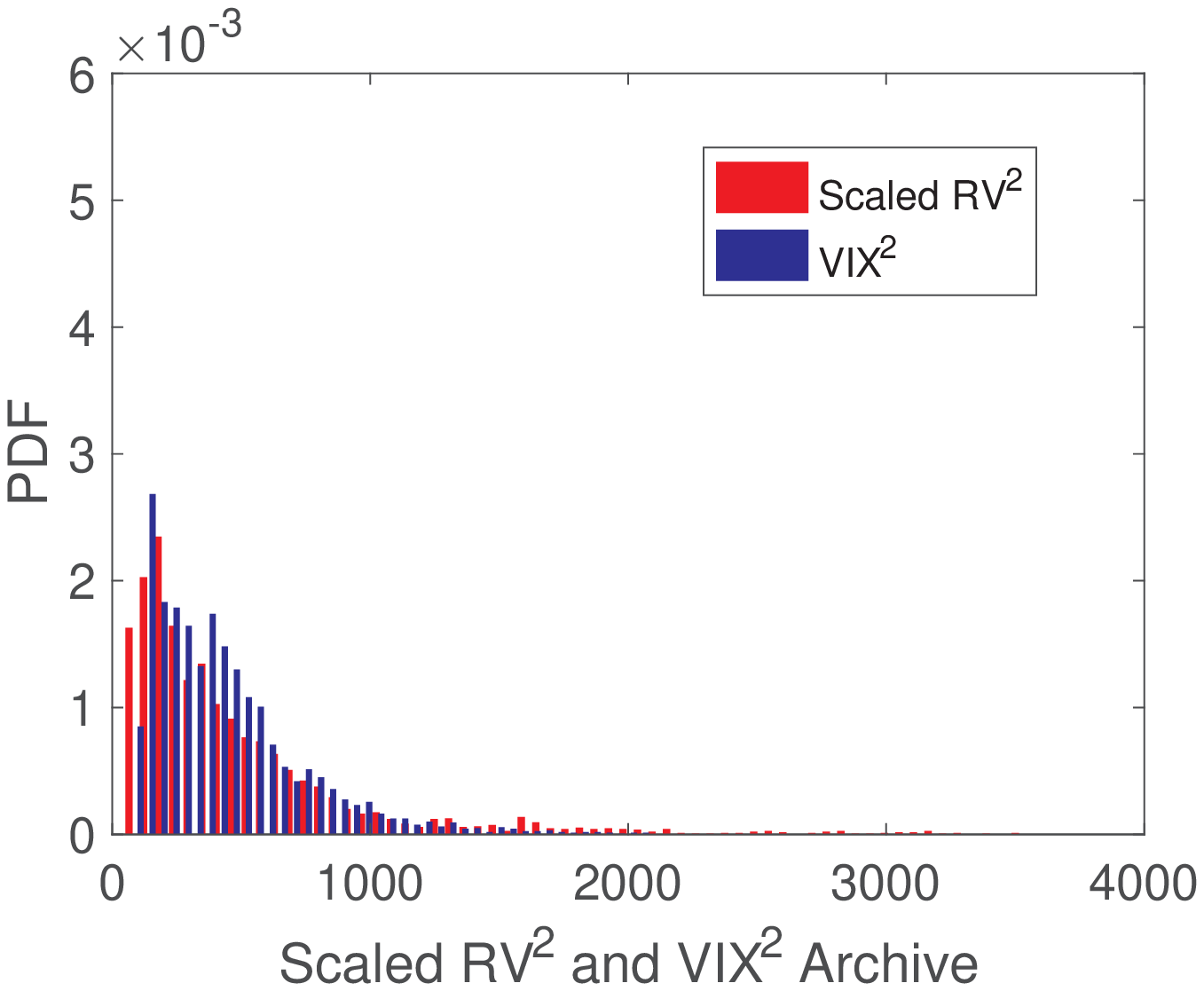} \hspace{0.2cm}
\includegraphics[width = 0.49 \textwidth]{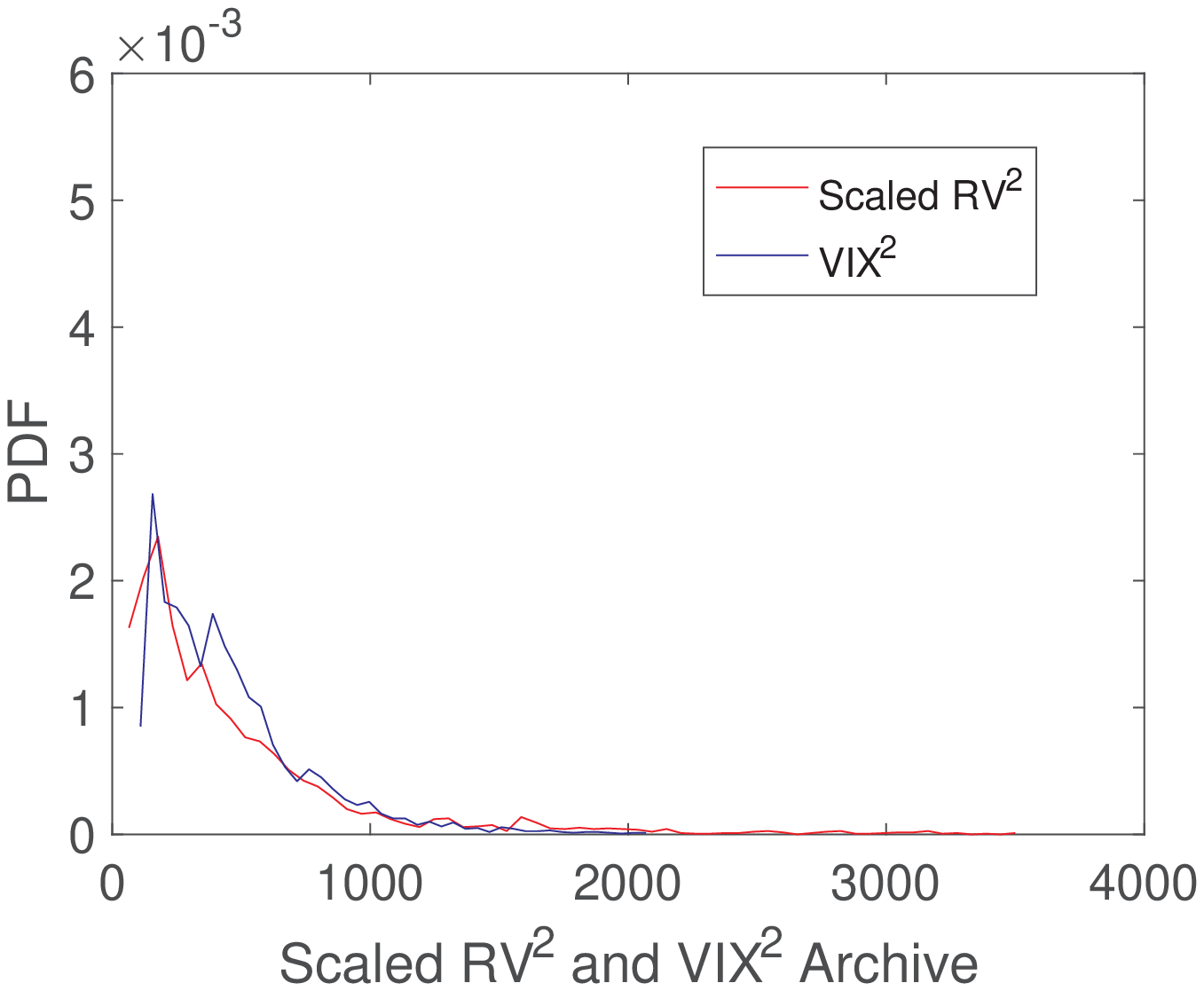}
\end{tabular}
\caption{PDFs of scaled $RV^2$ and $VIX^2$ from Jan 2nd, 1990 to Sep 19th, 2003.}
\label{histogramScaleRVVIX19902003ScaleRVVIXHistogram}
\end{figure}

\begin{figure}[!htbp]
\centering
\begin{tabular}{cc}
\includegraphics[width = 0.49 \textwidth]{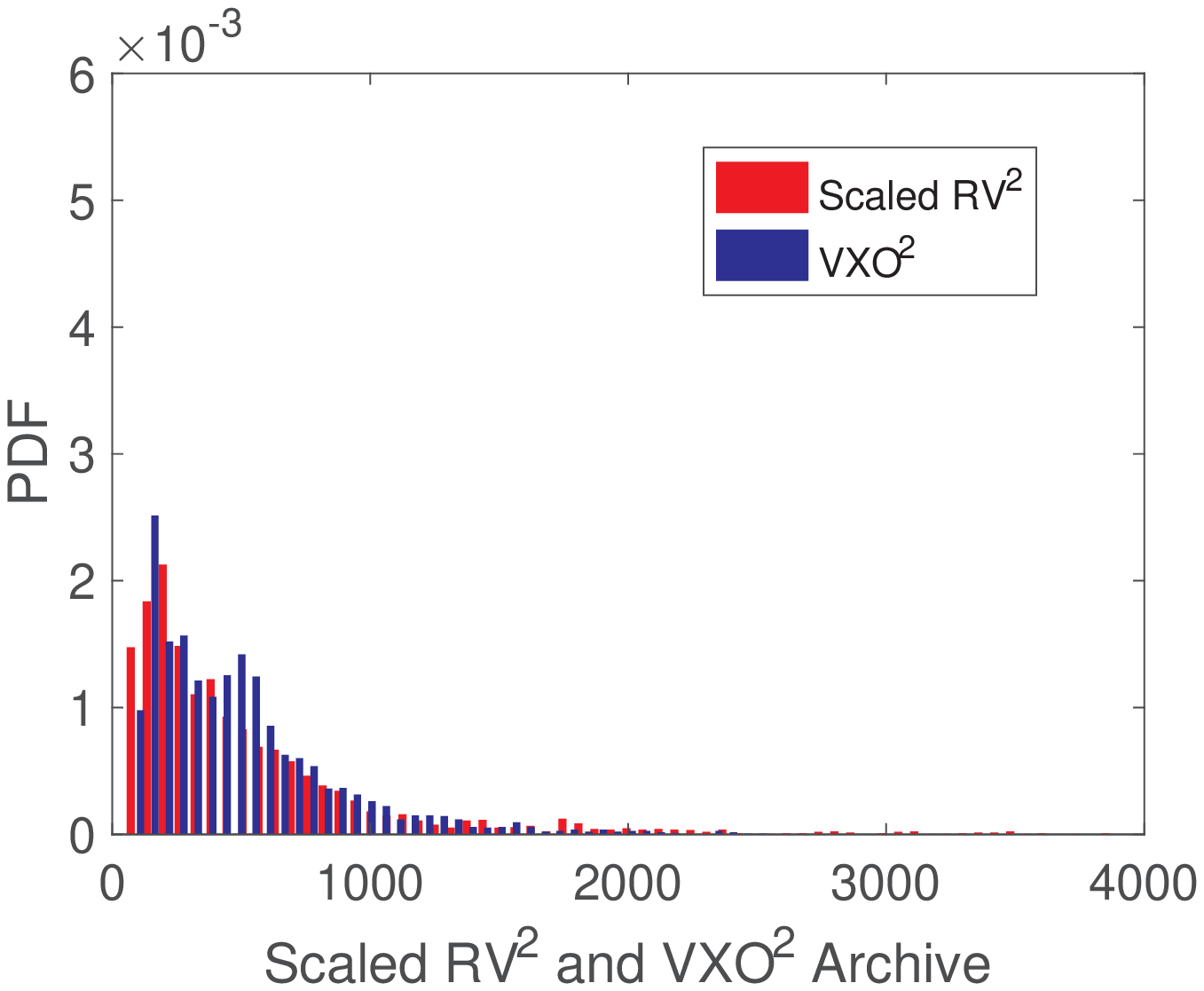} \hspace{0.2cm}
\includegraphics[width = 0.49 \textwidth]{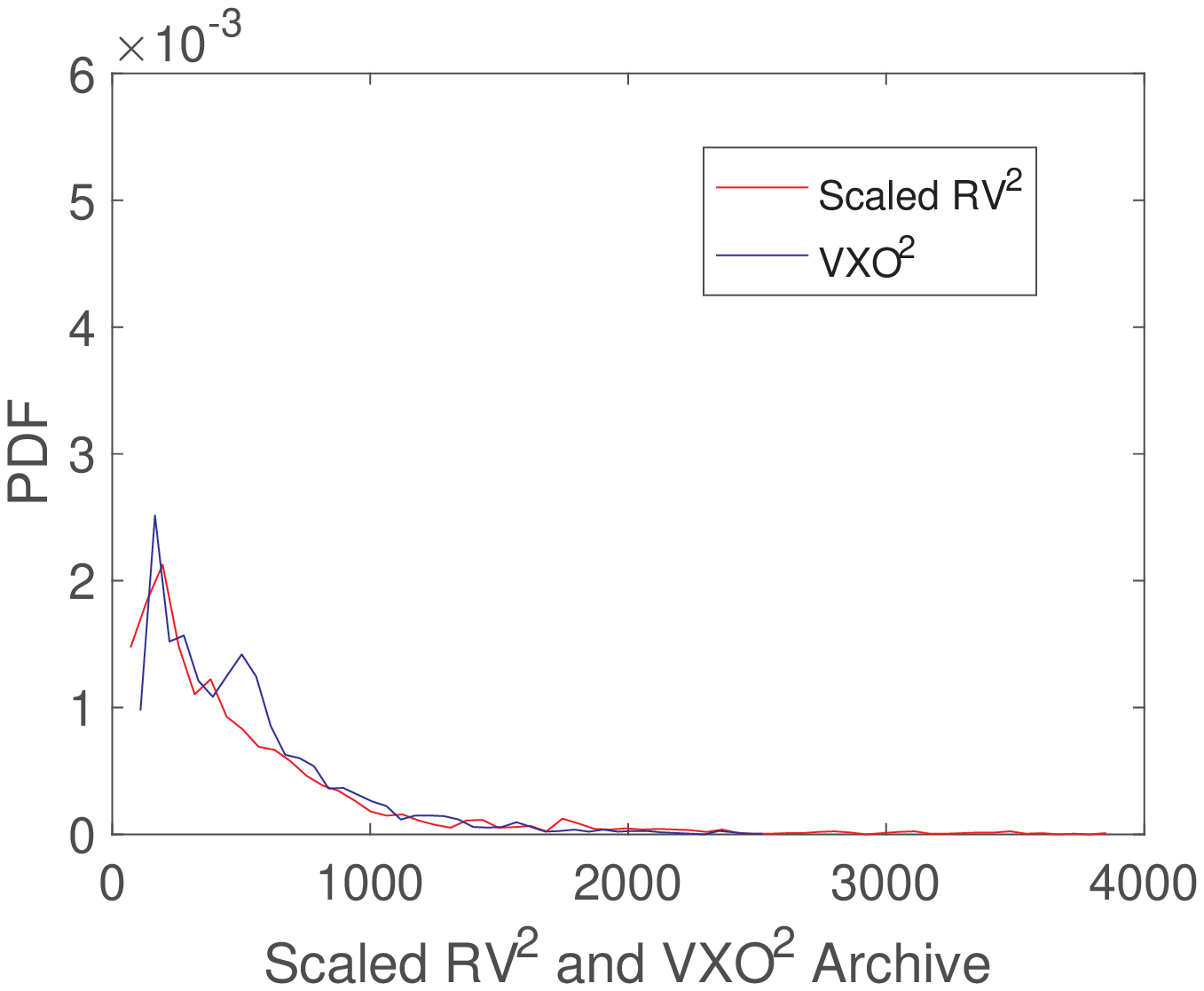}
\end{tabular}
\caption{PDFs of scaled $RV^2$ and $VXO^2$ from Jan 2nd, 1990 to Sep 19th, 2003.}
\label{histogramScaleRVVXO19902003ScaleRVVXOHistogram}
\end{figure}

\begin{figure}[!htbp]
\centering
\begin{tabular}{cc}
\includegraphics[width = 0.49 \textwidth]{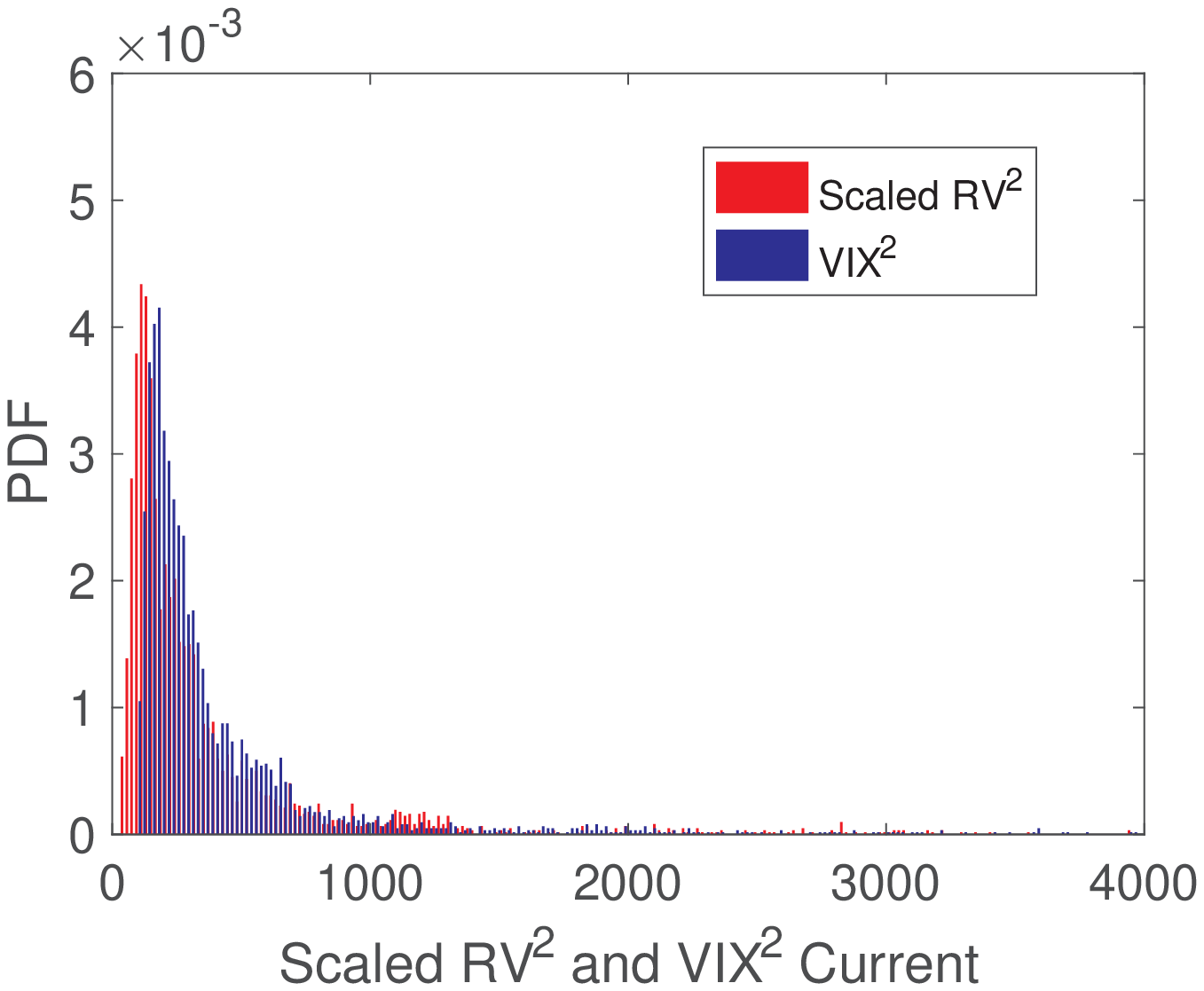} \hspace{0.2cm}
\includegraphics[width = 0.49 \textwidth]{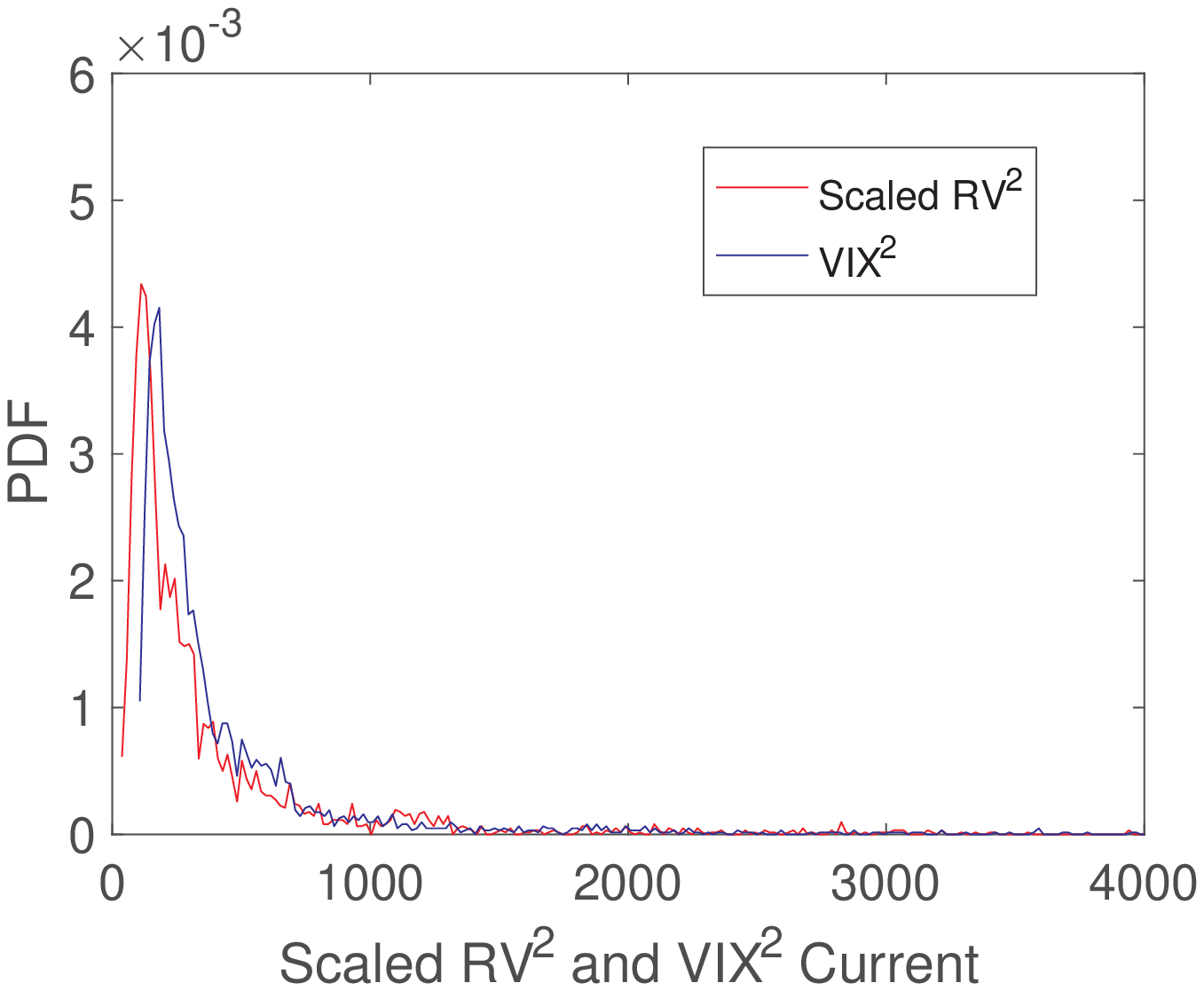}
\end{tabular}
\caption{PDFs of scaled $RV^2$ and  $VIX^2$ from Sep 22nd, 2003 to Dec 30th, 2016.}
\label{histogramScaleRVVIX20032016ScaleRVVIXHistogram}
\end{figure}

\begin{figure}[!htbp]
\centering
\begin{tabular}{cc}
\includegraphics[width = 0.49 \textwidth]{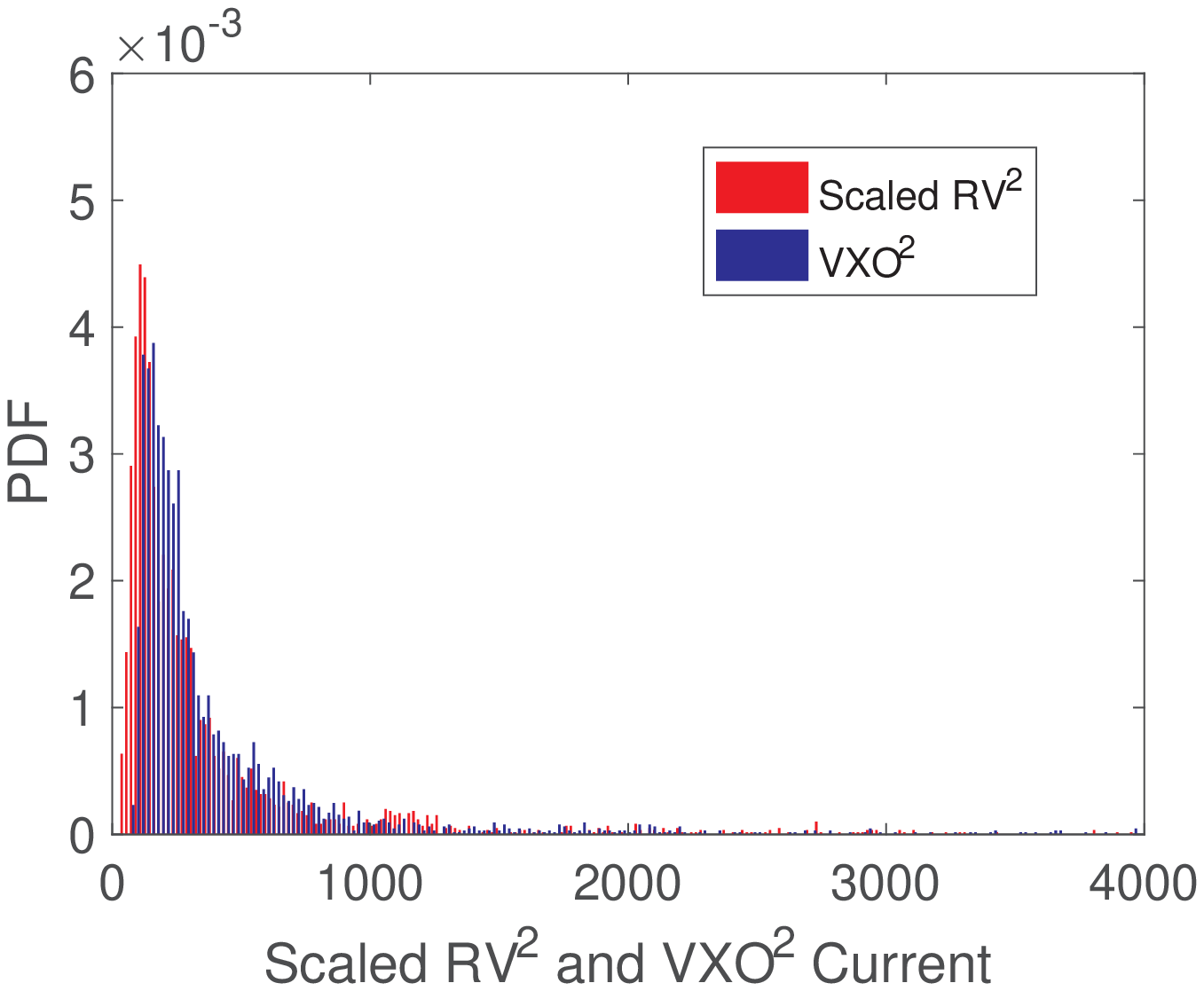} \hspace{0.2cm}
\includegraphics[width = 0.49 \textwidth]{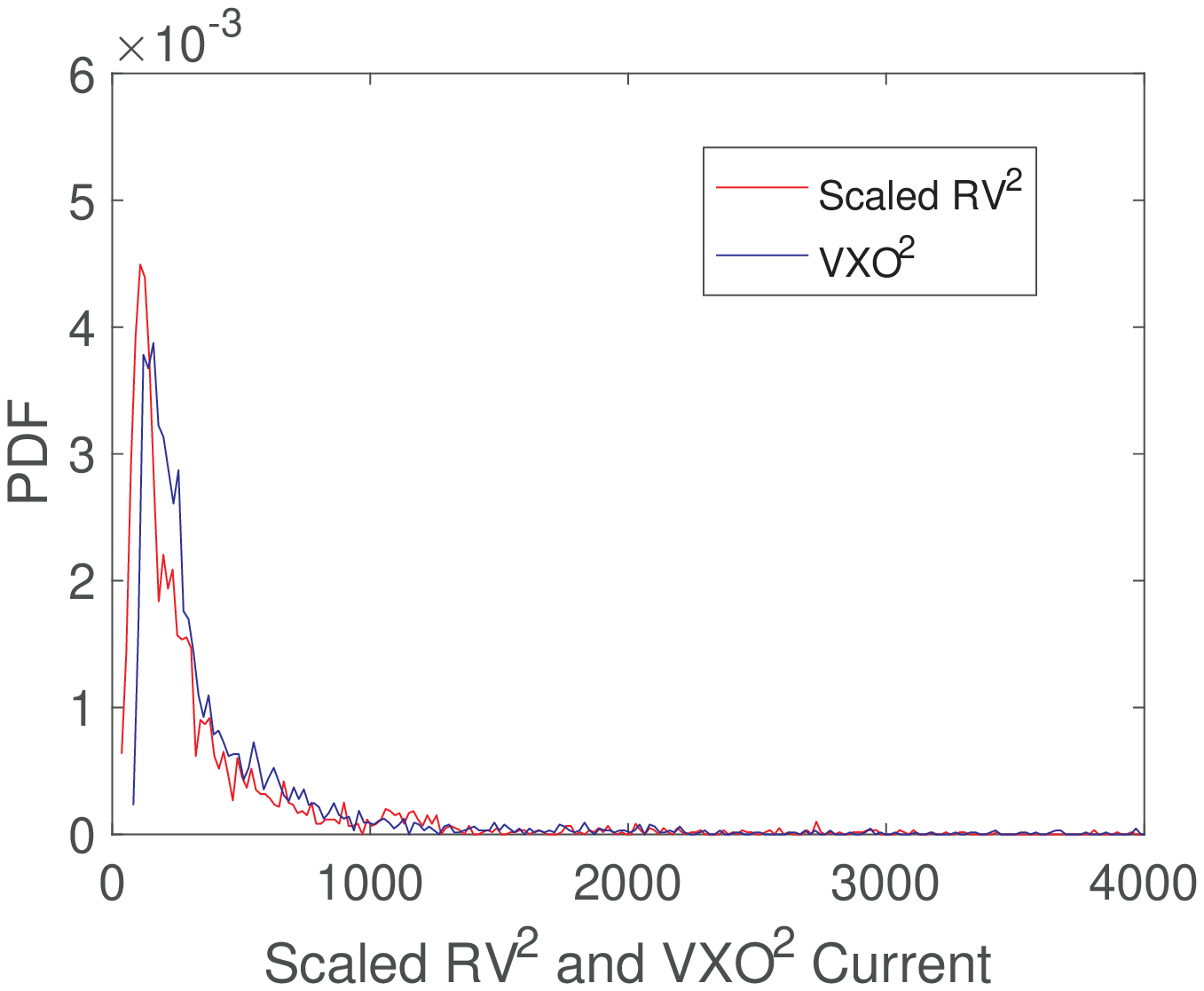}
\end{tabular}
\caption{PDFs of scaled $RV^2$ and $VXO^2$ from Sep 22nd, 2003 to Dec 30th, 2016.}
\label{histogramScaleRVVXO20032016ScaleRVVXOHistogram}
\end{figure}

As previously mentioned, realized variance $RV^2$ is scaled by entries in Table \ref{365252}. In Figs. \ref{VIXVXORVPlot} and \ref{VIX2VXO2RV2Plot} (which is just exaggerated, squared version of \ref{VIXVXORVPlot}) we show scaled RV and scaled $RV^2$ vis-a-vis their volatility indices counterparts. In Figs. \ref{histogramScaleRVVIX19902016ScaleRVVIXHistogram}-\ref{histogramScaleRVVXO20032016ScaleRVVXOHistogram} we show histograms and their contour plots for $RV^2$ vis-a-vis $VIX^2$ and $VXO^2$. KS statistics for comparing the latter two with the scaled $RV^2$ is collected in Table \ref{KStest} (lower numbers correspond to a better match). Further split of the 2003-2016 data is summarized in the Appendix.

\begin{table}[!htb]
\caption{Ratio of mean}
\label{365252}
\begin{minipage}{0.5\textwidth}
\begin{center}
\begin{tabular}{ c c c} 
\multicolumn{2}{c}{$VIX^2$} \\
\hline
   Theory & Ratio  \\ 
    \hline
   $365/252$ & 1.4484  \\ 
    \hline
   $30/21$ & 1.4286  \\ 
    \hline
   Date & Ratio  \\ 
   \hline
    1990-2016 &  1.4911\\  
  \hline
   1990-2003 & 1.6691   \\ 
   \hline
   2003-2016 &  1.3446   \\ 
   \hline
   2003-2010 &   1.2861\\ 
   \hline
   2010-2016 &  1.4104\\  
    \hline
 \hline
 \end{tabular}
\end{center}
\end{minipage}
\begin{minipage}{.5\textwidth}
\begin{center}
\begin{tabular}{ c c c} 
\multicolumn{2}{c}{$VXO^2$} \\
\hline
   Theory & Ratio  \\ 
    \hline
   $365/252$ & 1.4484  \\ 
    \hline
   $30/21$ & 1.4286  \\ 
    \hline
   Date & Ratio  \\ 
   \hline
    1990-2016 &  1.5257\\  
   \hline
   1990-2003 & 1.8372   \\ 
   \hline
   2003-2016 &  1.2985   \\ 
   \hline
   2003-2010 &   1.2850\\ 
   \hline
   2010-2016 &  1.3097\\  
   \hline
 \hline
\end{tabular}
\end{center}
  \end{minipage}
\end{table}

\begin{table}[!htb]
\caption{KS test results}
\label{KStest}
\begin{minipage}{0.5\textwidth}
\begin{center}
\begin{tabular}{ c c} 
\multicolumn{2}{c}{$VIX^2$} \\
\hline
   Date & KS statistic  \\ 
    \hline
    1990-2016 &  0.1723\\  
   \hline
   1990-2003 & 0.1478   \\ 
   \hline
   2003-2016 &  0.2394   \\ 
   \hline
   2003-2010 &   0.2215\\ 
   \hline
   2010-2016 &  0.2734\\  
    \hline
 \hline
 \end{tabular}
\end{center}
\end{minipage}
\begin{minipage}{.5\textwidth}
\begin{center}
\begin{tabular}{ c c } 
\multicolumn{2}{c}{$VXO^2$} \\
\hline
   Date & KS statistic  \\ 
   \hline
   1990-2016 &  0.1589\\
   \hline
   1990-2003 & 0.1632   \\ 
   \hline
   2003-2016 &  0.2157   \\ 
   \hline
   2003-2010 &   0.2034\\ 
   \hline
   2010-2016 &  0.2376\\  
   \hline
 \hline
\end{tabular}
\end{center}
  \end{minipage}
\end{table}

\newpage

\newpage

\newpage

\newpage

\clearpage

\subsection {Ratio distribution}\label{RV2/VIX2}

To further compare the volatilities, we examined the ratios $RV^2/VIX^2$ and $RV^2/VXO^2$. In plots below we show their time series and distribution functions. The latter are fitted using maximum likelihood estimation (MLE) and the parameters of the fits and KS statistics are collected in the tables. Six functions -- normal, lognormal, inverse gamma, gamma, Weibull and inverse Gaussian were used but only three best fits are shown with data PDFs. Clearly, barring VIX Archive, the fat-tailed IGa was the best fit. The hypothesis is that fat tails are due to sudden spikes in RV. The inverse distributions $VIX^2/RV^2$ and $VXO^2/RV^2$ are given to illustrate that there were no unexpected surges in VIX and illustrate consistency of MLE.

\begin{figure}[!htbp]
\centering
\begin{tabular}{cc}
\includegraphics[width = 0.4 \textwidth]{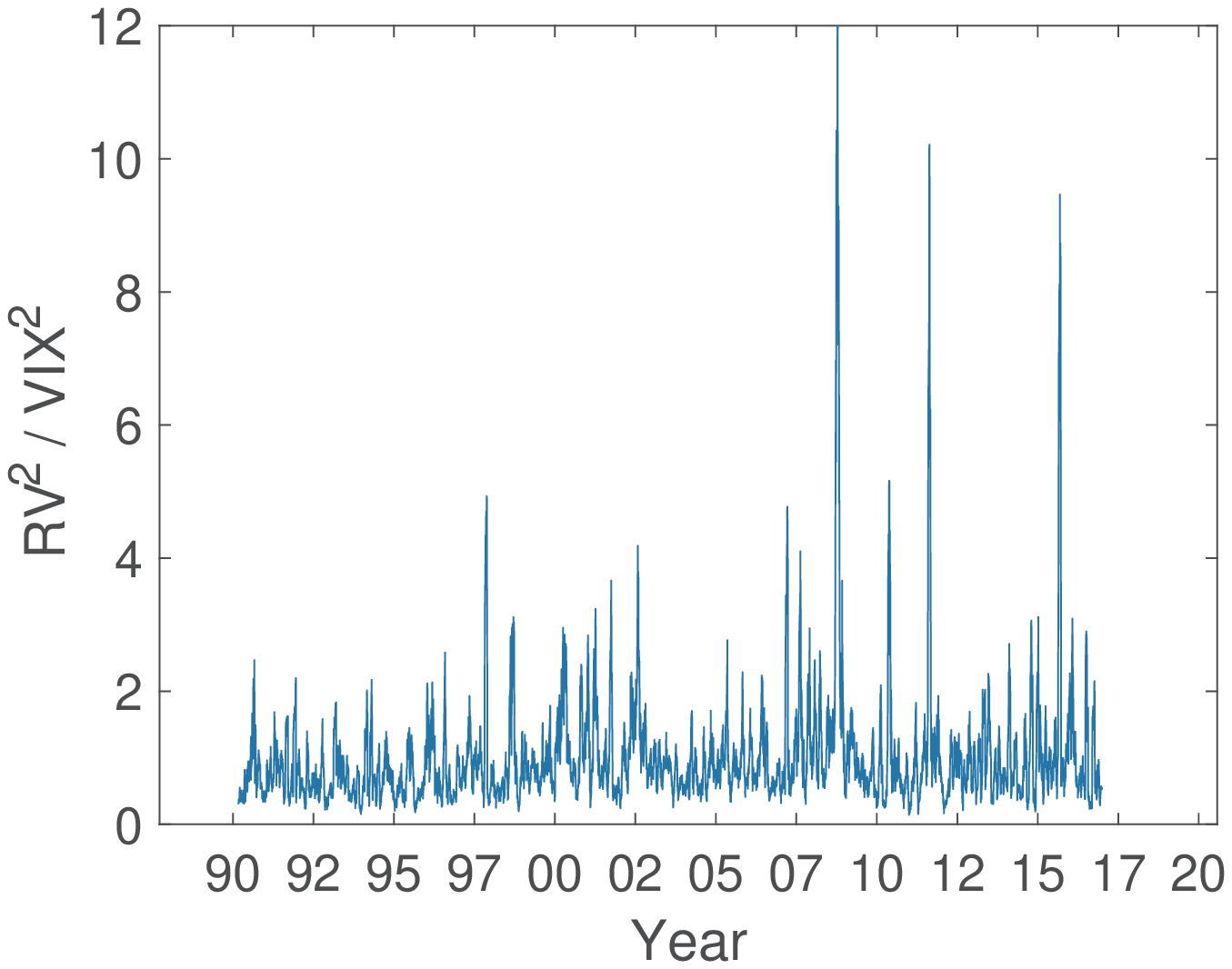} \hspace{0.2cm}
\includegraphics[width = 0.4 \textwidth]{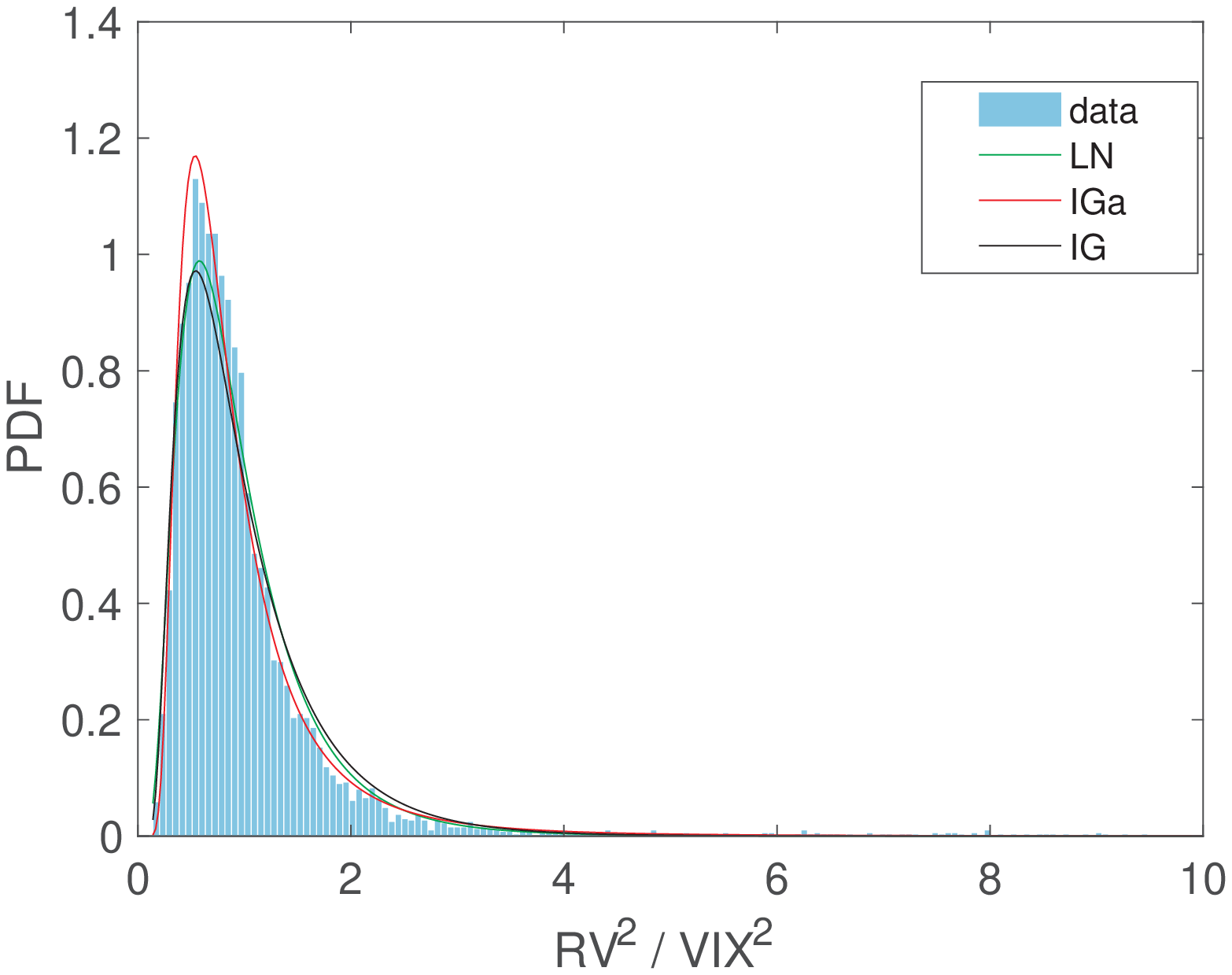}
\end{tabular}
\caption{$\mathrm{RV}^2 / \mathrm{VIX}^2$, from Jan 2nd, 1990 to Dec 30th, 2016.}
\label{RVOverVIXListSRV2OverVIX21990}
\end{figure}

\begin{figure}[!htbp]
\centering
\begin{tabular}{cc}
\includegraphics[width = 0.4 \textwidth]{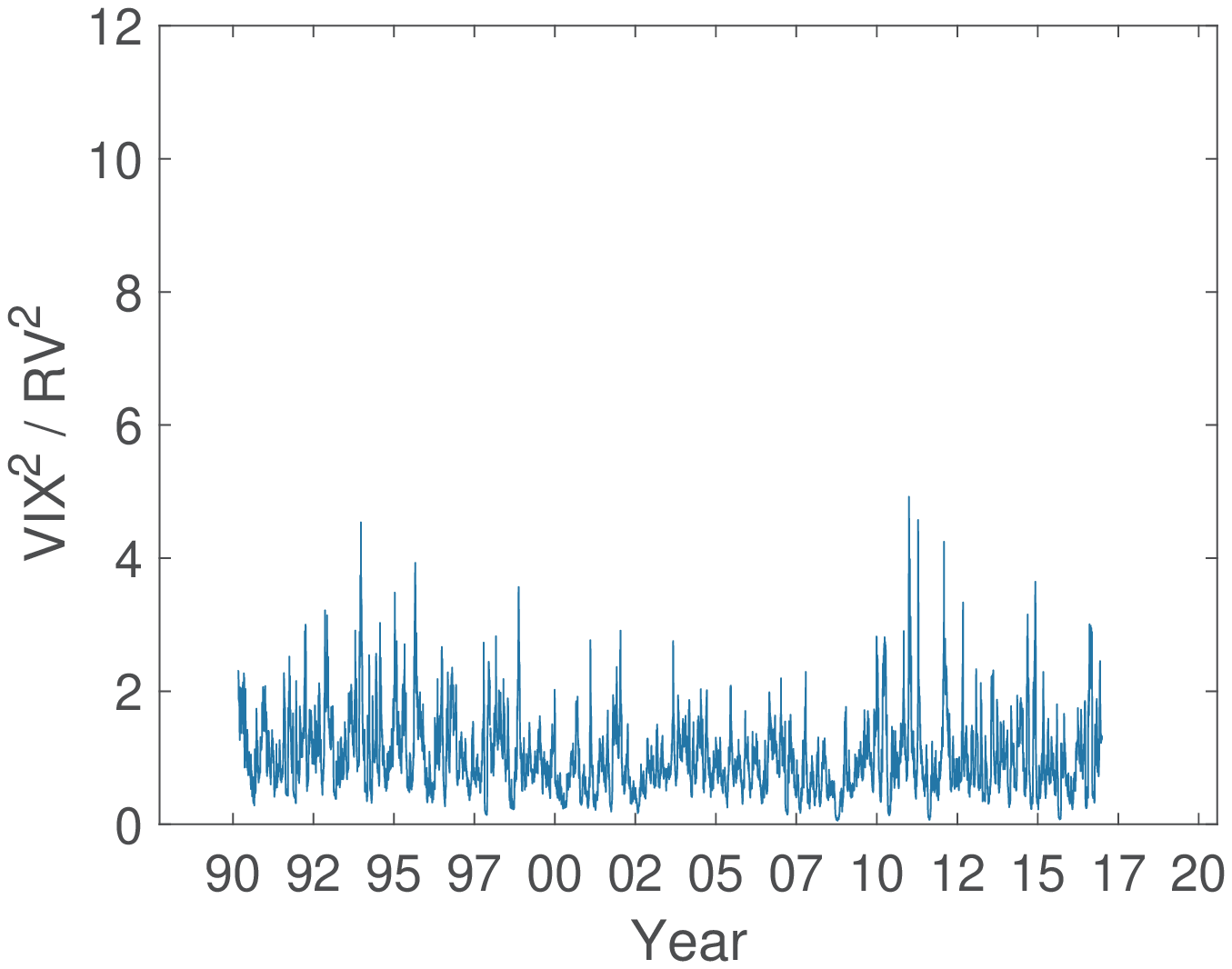} \hspace{0.2cm}
\includegraphics[width = 0.4 \textwidth]{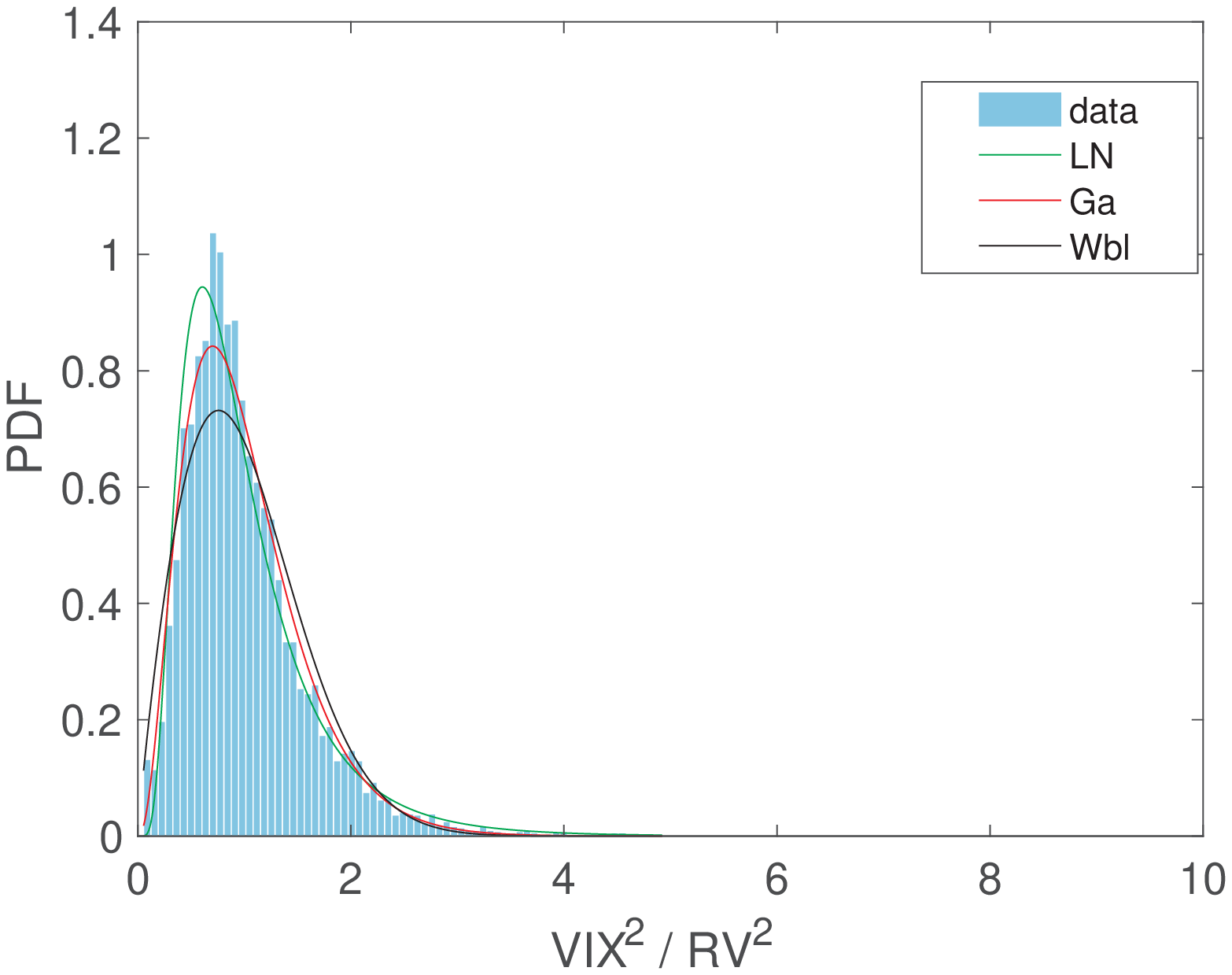}
\end{tabular}
\caption{$ \mathrm{VIX}^2 / \mathrm{RV}^2$, from Jan 2nd, 1990 to Dec 30th, 2016.}
\label{VIXOverRVListSVIX2OverRV21990n}
\end{figure}

\begin{table}[!htb]
\caption{MLE results for ``$\mathrm{RV}^2 / \mathrm{VIX}^2$" and ``$\mathrm{VIX}^2 / \mathrm{RV}^2$"}
\label{MLESRV2OverVIX21990n}
\begin{minipage}{0.5\textwidth}
\begin{center}
\begin{tabular}{ c c c} 
\multicolumn{2}{c}{} \\
 \hline
            type &       parameters &          KS Statistic  \\
\hline
Normal & N(          1.0000,           0.9067) &           0.1940 \\
\hline
LogNormal & LN(         -0.2027,           0.5867) &           0.0446 \\
\hline
IGa & IGa(          3.3595,           2.3466) &           0.0246 \\
\hline
Gamma & Gamma(          2.6219,           0.3814) &           0.0978 \\
\hline
Weibull & Weibull(          1.1124,           1.4009) &           0.1224 \\
\hline
IG & IG(          1.0000,           2.3168) &           0.0607 \\
\hline
\end{tabular}
\end{center}
\end{minipage}
\begin{minipage}{.5\textwidth}
\begin{center}
\begin{tabular}{ c c c} 
\multicolumn{2}{c}{} \\
\hline
            type &       parameters &          KS Statistic  \\
\hline
Normal & N(          1.0000,           0.5626) &           0.0972 \\
\hline
LogNormal & LN(         -0.1562,           0.5867) &           0.0446 \\
\hline
IGa & IGa(          2.6219,           1.8314) &           0.0978 \\
\hline
Gamma & Gamma(          3.3595,           0.2977) &           0.0246 \\
\hline
Weibull & Weibull(          1.1306,           1.8882) &           0.0500 \\
\hline
IG & IG(          1.0000,           2.3168) &           0.0734 \\
\hline
\end{tabular}
\end{center}
  \end{minipage}

\end{table}

\newpage

\begin{figure}[!htbp]
\centering
\begin{tabular}{cc}
\includegraphics[width = 0.4 \textwidth]{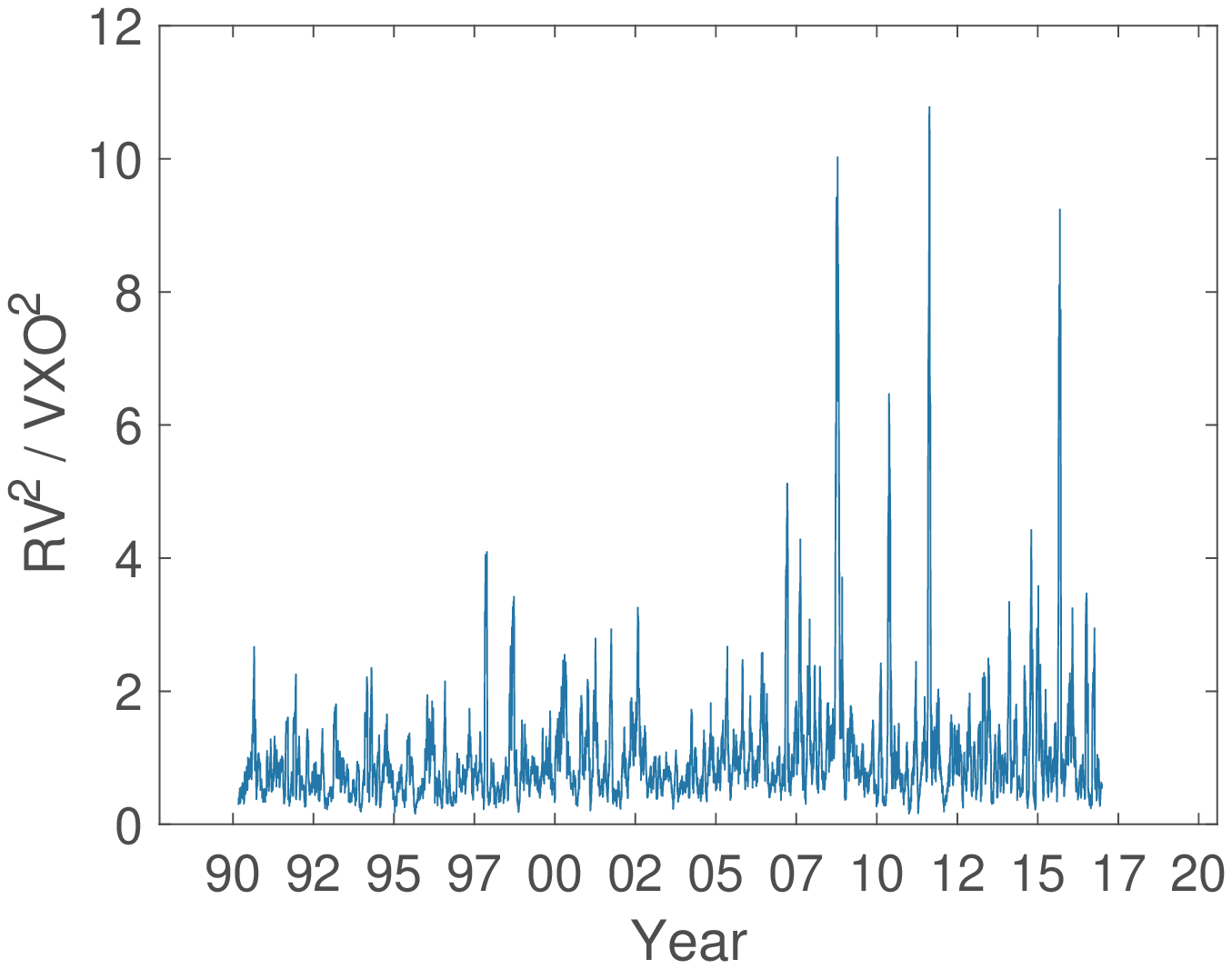} \hspace{0.2cm}
\includegraphics[width = 0.4 \textwidth]{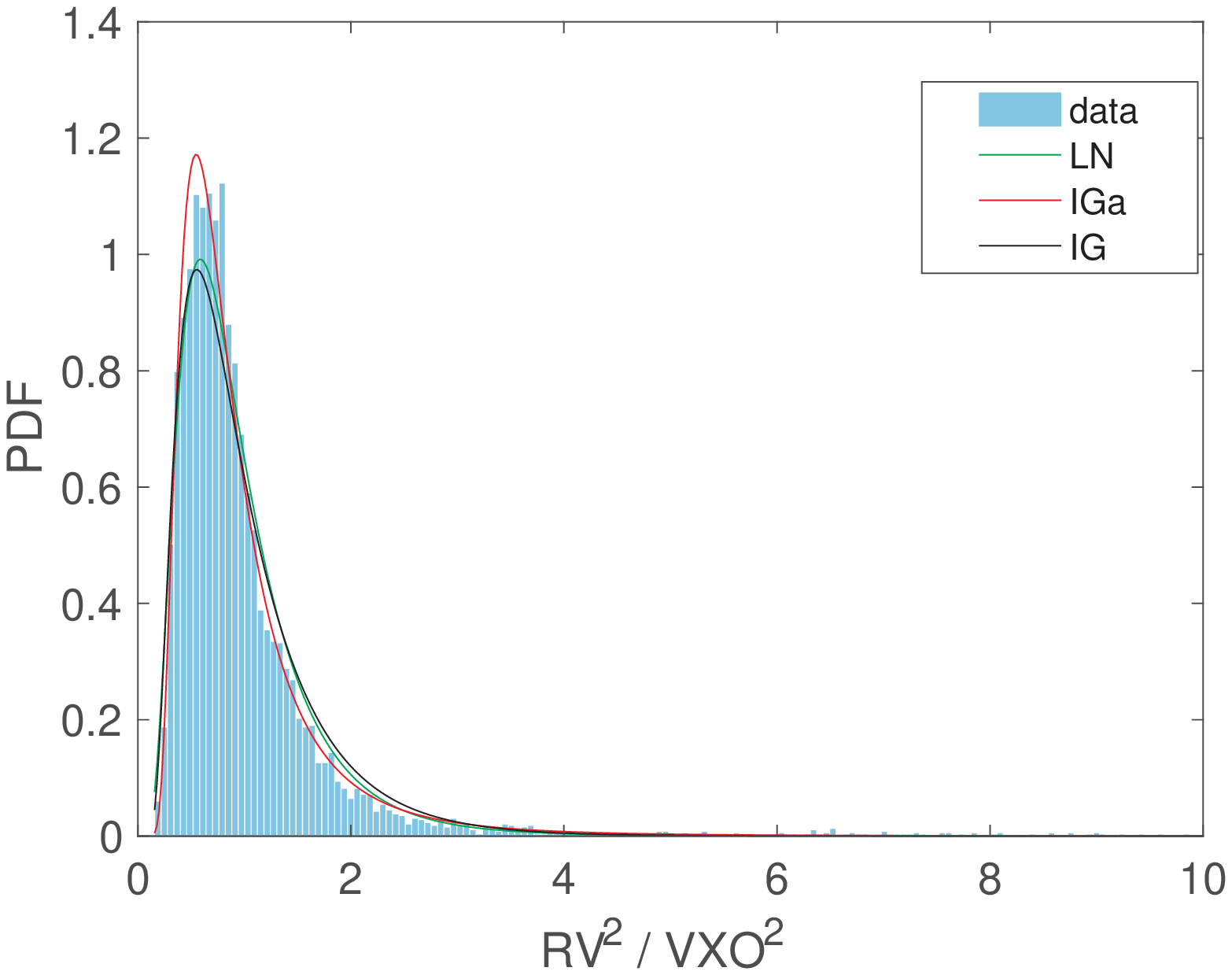}
\end{tabular}
\caption{$\mathrm{RV}^2 / \mathrm{VXO}^2$, Jan 2nd, 1990 to Dec 30th, 2016.}
\label{RVOverVXOListSRV2OverVXO22016}
\end{figure}

\begin{figure}[!htbp]
\centering
\begin{tabular}{cc}
\includegraphics[width = 0.4 \textwidth]{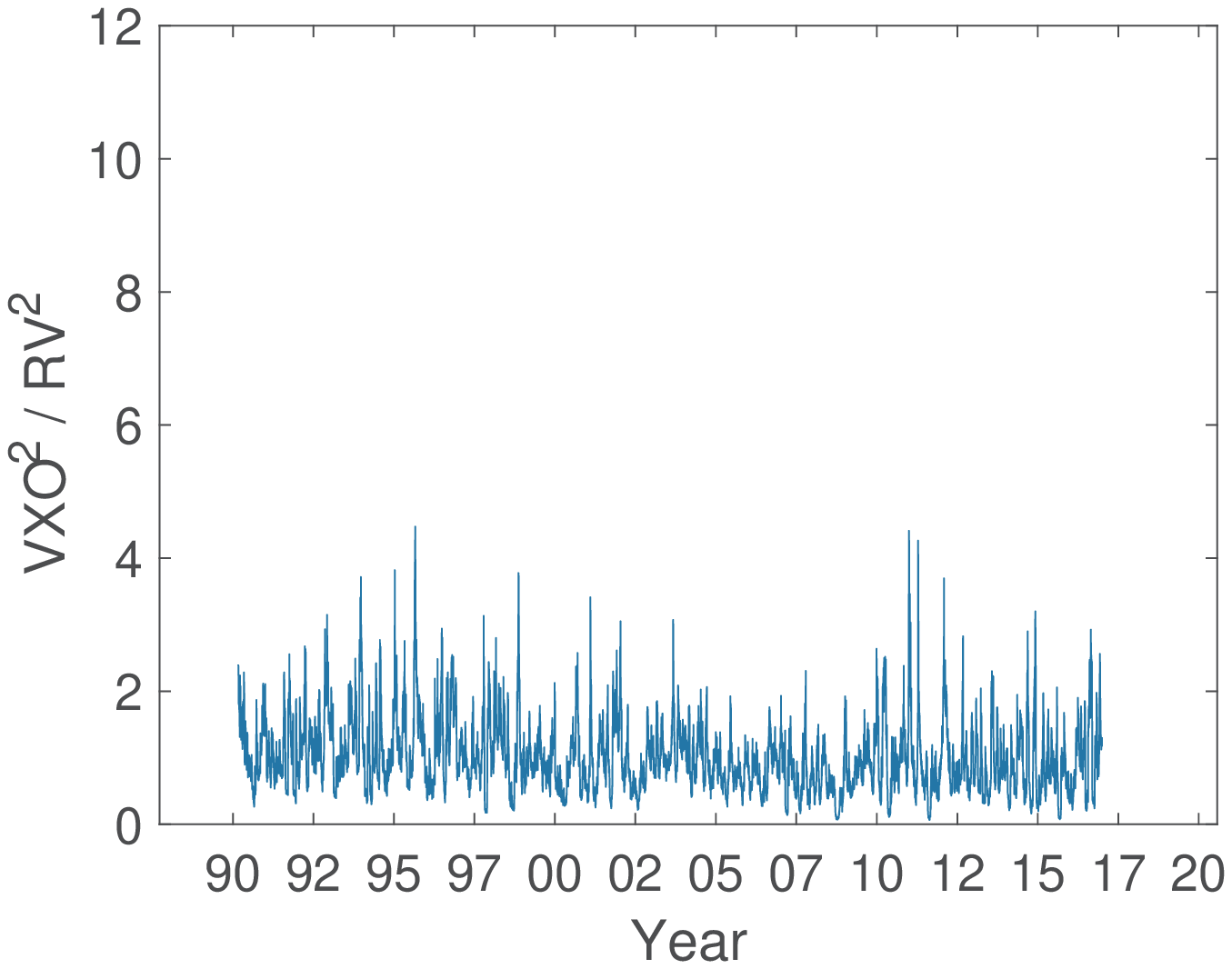} \hspace{0.2cm}
\includegraphics[width = 0.4 \textwidth]{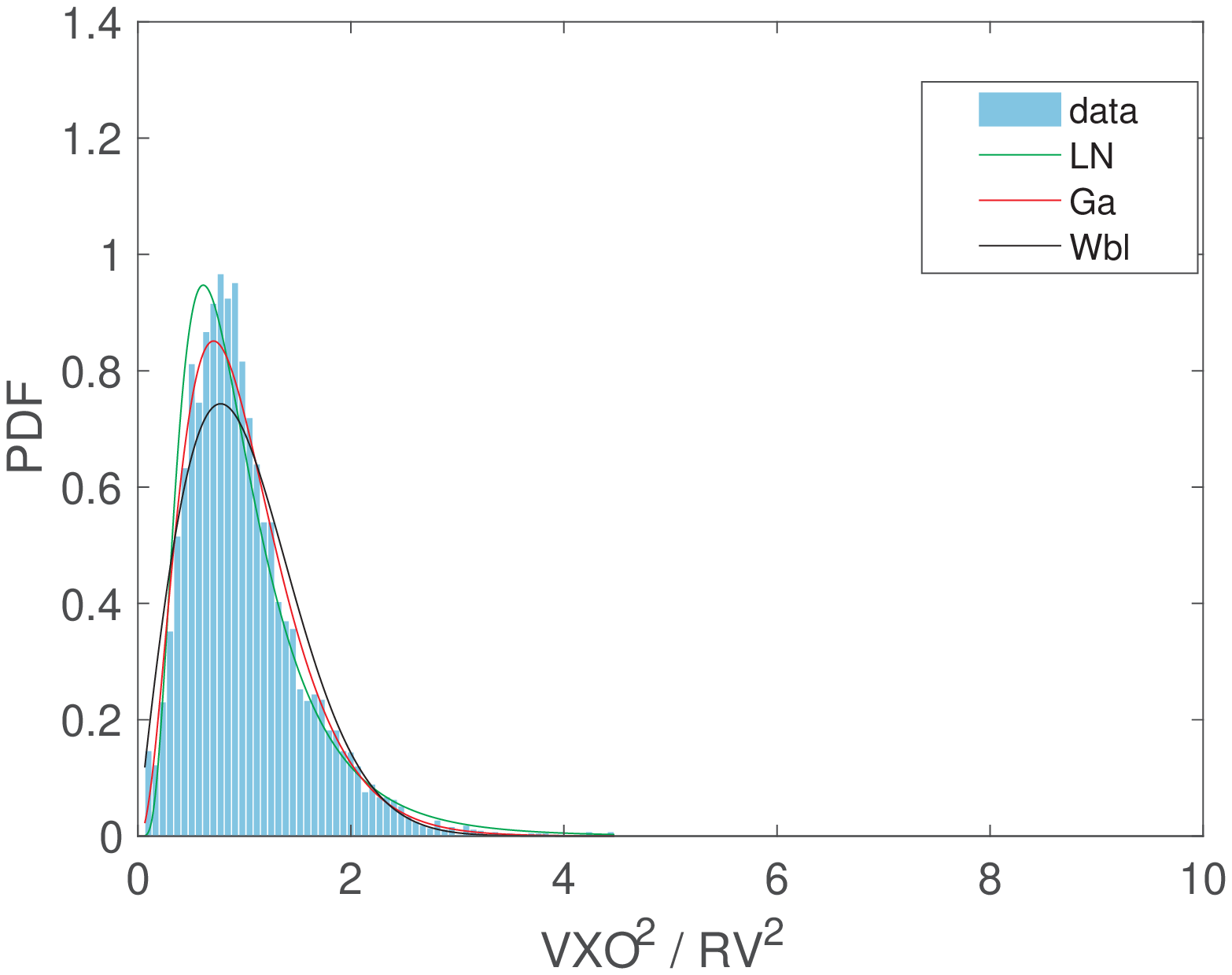}
\end{tabular}
\caption{$ \mathrm{VXO}^2 / \mathrm{RV}^2$, from Jan 2nd, 1990 to Dec 30th, 2016.}
\label{VXOOverRVListSVXO2OverRV22016n}
\end{figure}

\begin{table}[!htb]
\caption{MLE results for ``$\mathrm{RV}^2 / \mathrm{VXO}^2$" and ``$\mathrm{VXO}^2 / \mathrm{RV}^2$"}
\label{MLESRV2OverVXO22016n}
\begin{minipage}{0.5\textwidth}
\begin{center}
\begin{tabular}{ c c c} 
\multicolumn{2}{c}{} \\
\hline
            type &       parameters &          KS Statistic  \\
\hline
Normal & N(          1.0000,           0.8747) &           0.1910 \\
\hline
LogNormal & LN(         -0.1973,           0.5795) &           0.0449 \\
\hline
IGa & IGa(          3.4629,           2.4438) &           0.0224 \\
\hline
Gamma & Gamma(          2.6897,           0.3718) &           0.0971 \\
\hline
Weibull & Weibull(          1.1150,           1.4256) &           0.1230 \\
\hline
IG & IG(          1.0000,           2.3981) &           0.0611 \\
\hline
\end{tabular}
\end{center}
\end{minipage}
\begin{minipage}{.5\textwidth}
\begin{center}
\begin{tabular}{ c c c} 
\multicolumn{2}{c}{} \\
\hline
            type &       parameters &          KS Statistic  \\
\hline
Normal & N(          1.0000,           0.5467) &           0.0925 \\
\hline
LogNormal & LN(         -0.1513,           0.5795) &           0.0449 \\
\hline
IGa & IGa(          2.6897,           1.8982) &           0.0971 \\
\hline
Gamma & Gamma(          3.4629,           0.2888) &           0.0224 \\
\hline
Weibull & Weibull(          1.1308,           1.9374) &           0.0499 \\
\hline
IG & IG(          1.0000,           2.3981) &           0.0729 \\
\hline
\end{tabular}
\end{center}
  \end{minipage}

\end{table}

\newpage

\begin{figure}[!htbp]
\centering
\begin{tabular}{cc}
\includegraphics[width = 0.4 \textwidth]{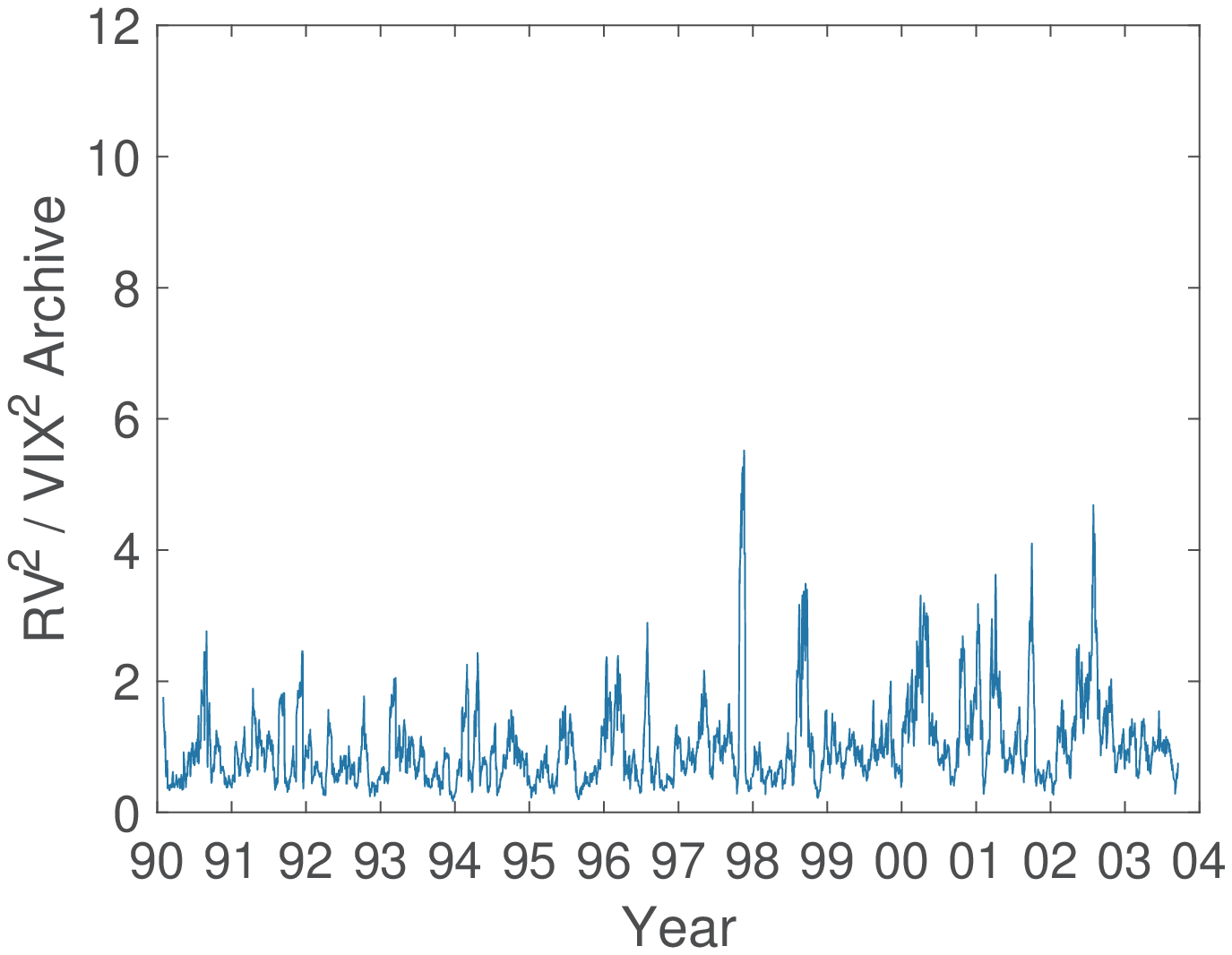} \hspace{0.2cm}
\includegraphics[width = 0.4 \textwidth]{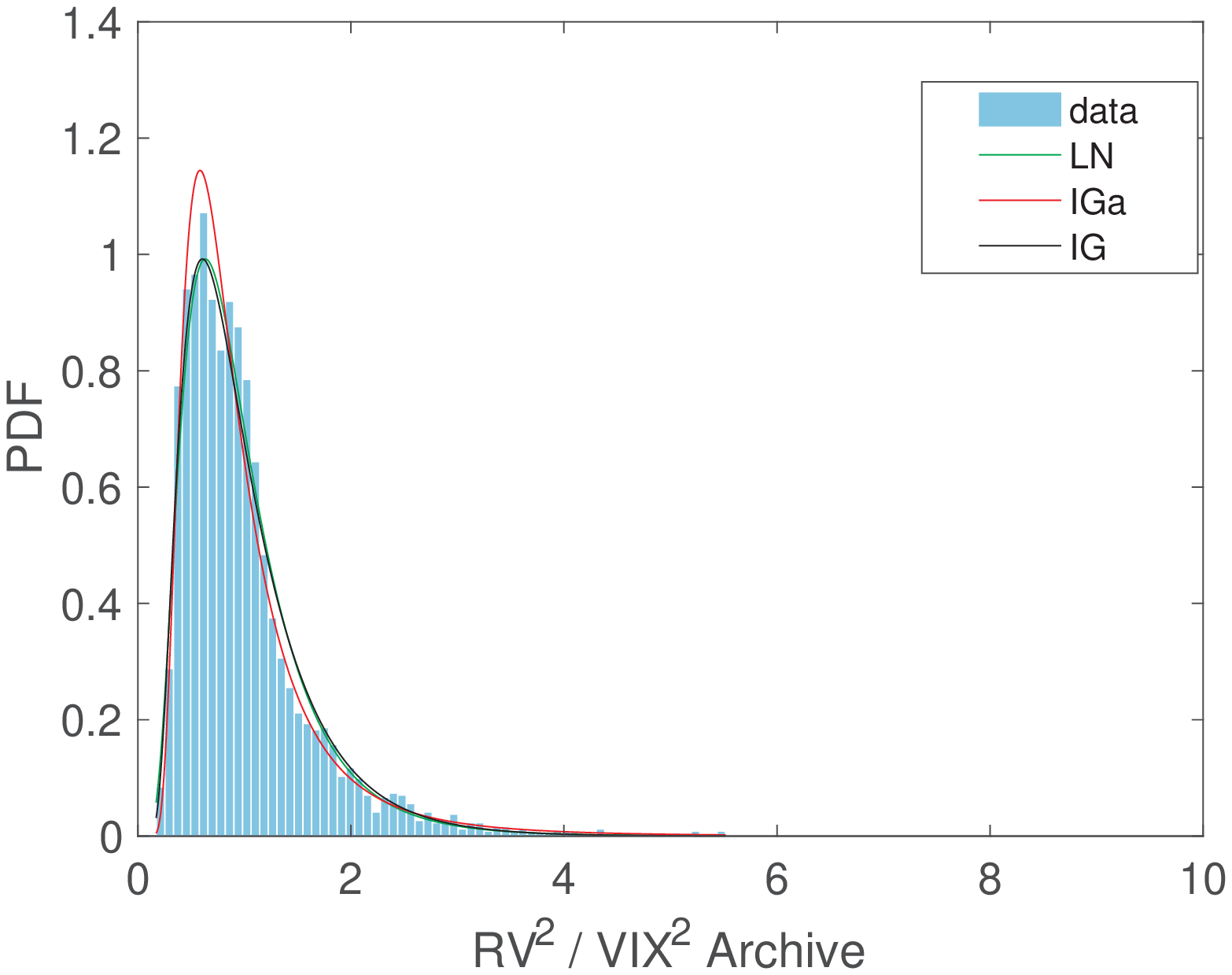}
\end{tabular}
\caption{$\mathrm{RV}^2 / \mathrm{VIX}^2$, from Jan 2nd, 1990 to Sep 19th, 2003.}
\label{RVOverVIXListSRV2OverVIX21990}
\end{figure}

\begin{figure}[!htbp]
\centering
\begin{tabular}{cc}
\includegraphics[width = 0.4 \textwidth]{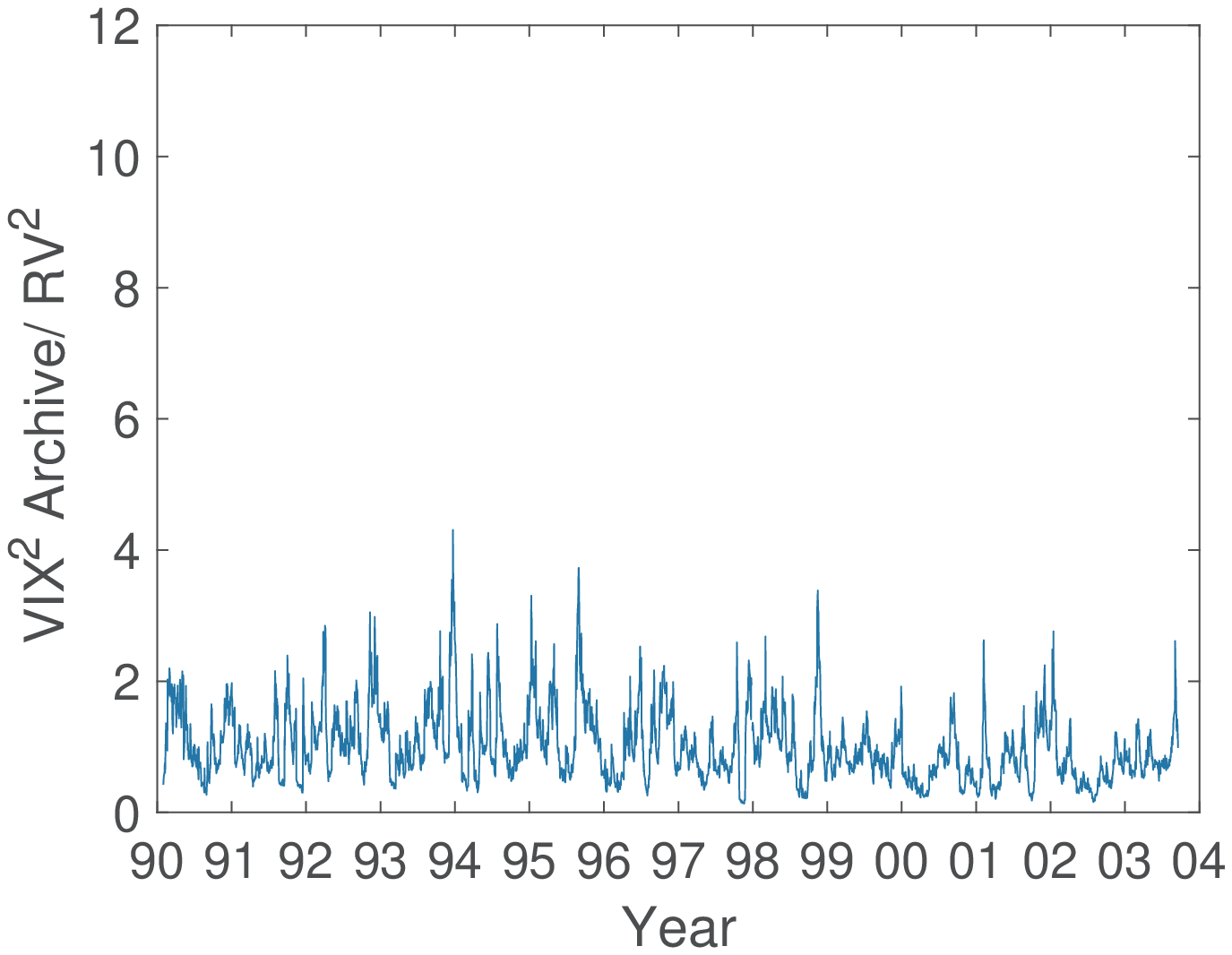} \hspace{0.2cm}
\includegraphics[width = 0.4 \textwidth]{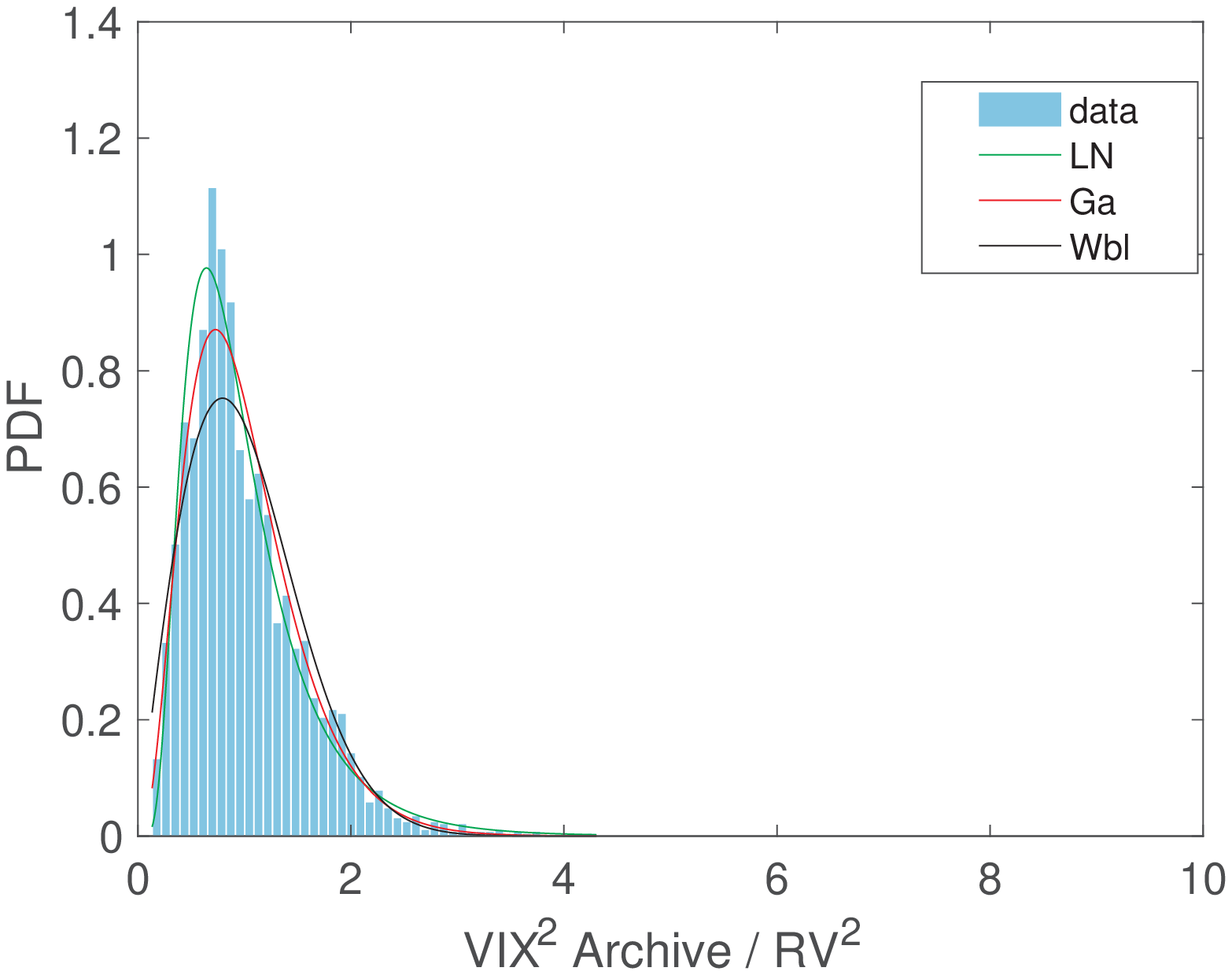}
\end{tabular}
\caption{$ \mathrm{VIX}^2 / \mathrm{RV}^2$, from Jan 2nd, 1990 to Sep 19th, 2003.}
\label{VIXOverRVListSVIX2OverRV21990n}
\end{figure}

\begin{table}[!htb]
\caption{MLE results for ``$\mathrm{RV}^2 / \mathrm{VIX}^2$" and ``$\mathrm{VIX}^2 / \mathrm{RV}^2$"}
\label{MLESRV2OverVIX21990n}
\begin{minipage}{0.5\textwidth}
\begin{center}
\begin{tabular}{ c c c} 
\multicolumn{2}{c}{} \\
 \hline
            type &       parameters &          KS Statistic  \\
\hline
Normal & N(          1.0000,           0.6380) &           0.1421 \\
\hline
LogNormal & LN(         -0.1570,           0.5460) &           0.0275 \\
\hline
IGa & IGa(          3.6963,           2.7429) &           0.0358 \\
\hline
Gamma & Gamma(          3.3420,           0.2992) &           0.0640 \\
\hline
Weibull & Weibull(          1.1310,           1.7263) &           0.0913 \\
\hline
IG & IG(          1.0000,           2.8769) &           0.0344 \\
\hline
\end{tabular}
\end{center}
\end{minipage}
\begin{minipage}{.5\textwidth}
\begin{center}
\begin{tabular}{ c c c} 
\multicolumn{2}{c}{} \\
\hline
            type &       parameters &          KS Statistic  \\
\hline
Normal & N(          1.0000,           0.5372) &           0.1011 \\
\hline
LogNormal & LN(         -0.1413,           0.5460) &           0.0275 \\
\hline
IGa & IGa(          3.3420,           2.4800) &           0.0640 \\
\hline
Gamma & Gamma(          3.6963,           0.2705) &           0.0358 \\
\hline
Weibull & Weibull(          1.1328,           1.9825) &           0.0578 \\
\hline
IG & IG(          1.0000,           2.8768) &           0.0419 \\
\hline
\end{tabular}
\end{center}
  \end{minipage}

\end{table}

\newpage

\begin{figure}[!htbp]
\centering
\begin{tabular}{cc}
\includegraphics[width = 0.4 \textwidth]{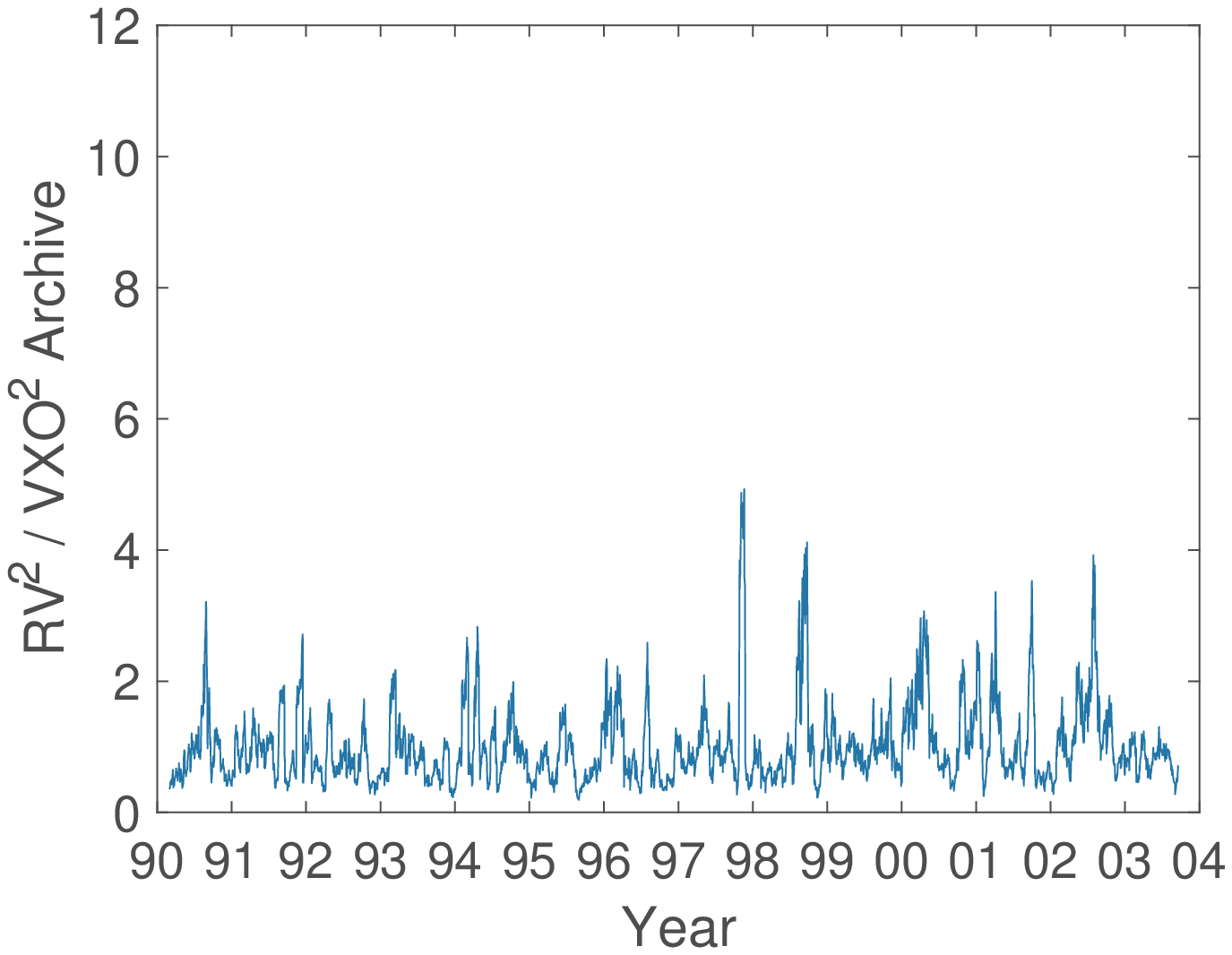} \hspace{0.2cm}
\includegraphics[width = 0.4 \textwidth]{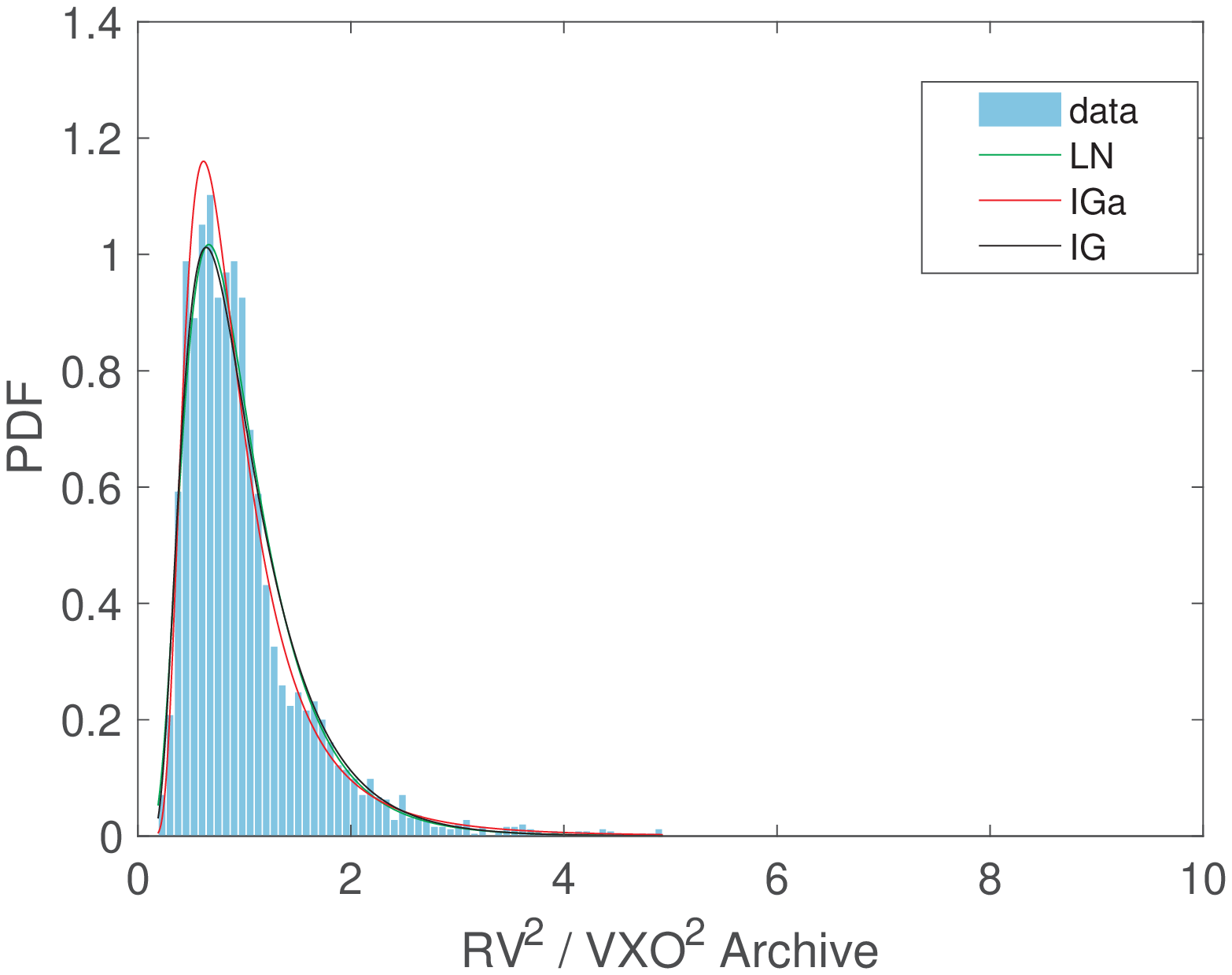}
\end{tabular}
\caption{$\mathrm{RV}^2 / \mathrm{VXO}^2$, from Jan 2nd, 1990 to Sep 19th, 2003.}
\label{RVOverVXOListSRV2OverVXO21990}
\end{figure}

\begin{figure}[!htbp]
\centering
\begin{tabular}{cc}
\includegraphics[width = 0.4 \textwidth]{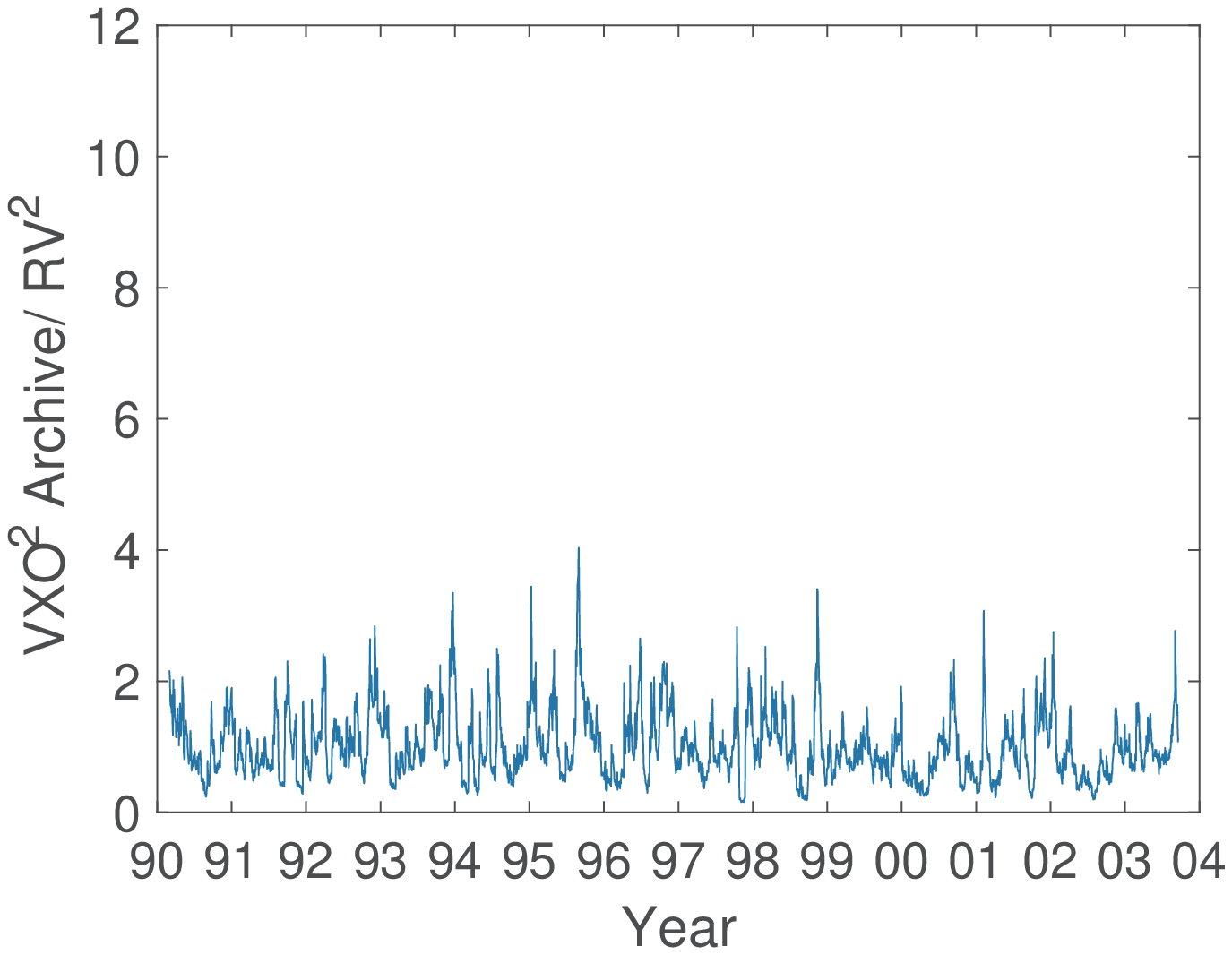} \hspace{0.2cm}
\includegraphics[width = 0.4 \textwidth]{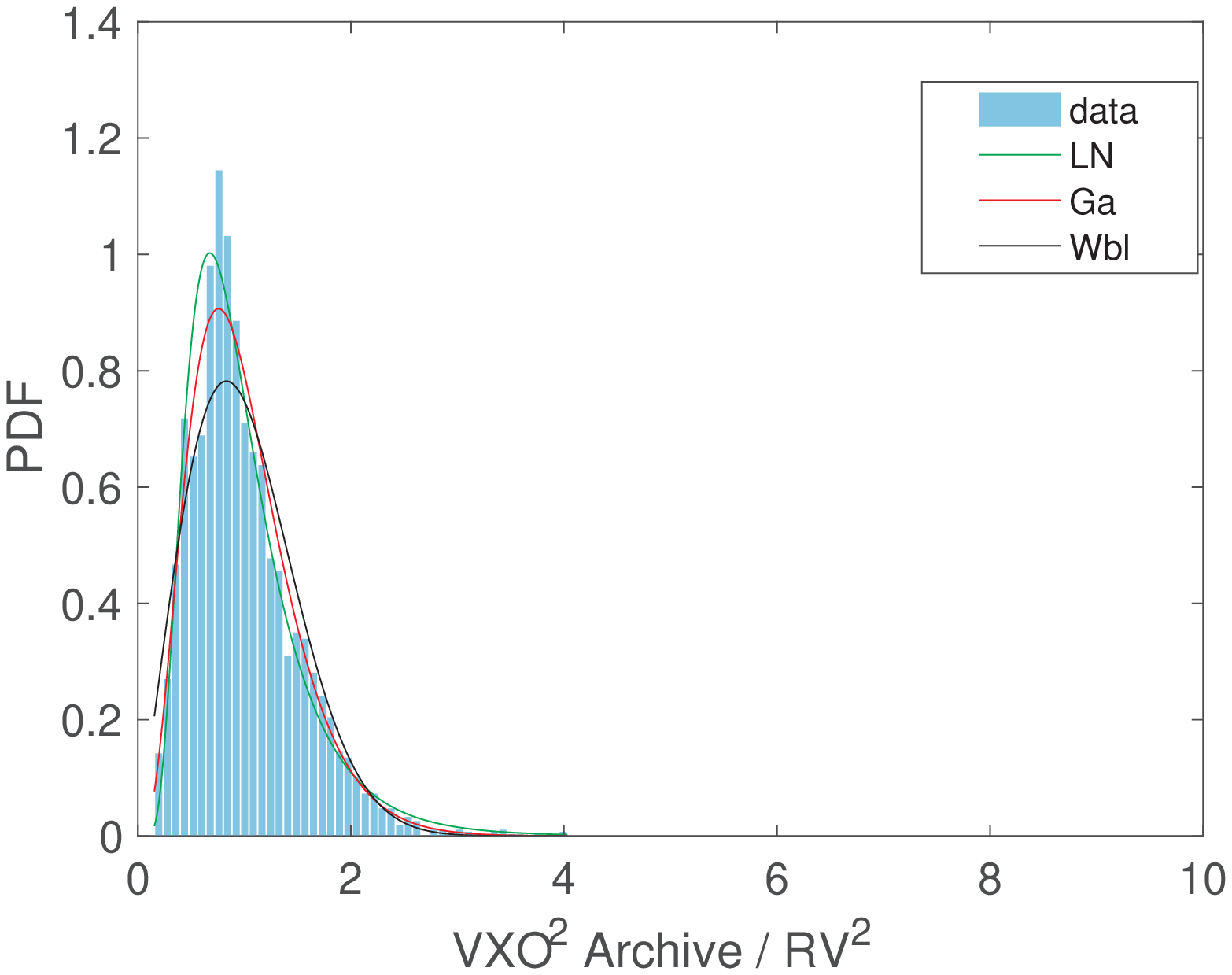}
\end{tabular}
\caption{$ \mathrm{VXO}^2 / \mathrm{RV}^2$, from Jan 2nd, 1990 to Sep 19th, 2003.}
\label{VXOOverRVListSVXO2OverRV21990n}
\end{figure}

\begin{table}[!htb]
\caption{MLE results for ``$\mathrm{RV}^2 / \mathrm{VXO}^2$" and ``$\mathrm{VXO}^2 / \mathrm{RV}^2$"}
\label{MLESRV2OverVXO21990n}
\begin{minipage}{0.5\textwidth}
\begin{center}
\begin{tabular}{ c c c} 
\multicolumn{2}{c}{} \\
 \hline
            type &       parameters &          KS Statistic  \\
\hline
Normal & N(          1.0000,           0.6013) &           0.1451 \\
\hline
LogNormal & LN(         -0.1404,           0.5155) &           0.0349 \\
\hline
IGa & IGa(          4.1290,           3.1633) &           0.0269 \\
\hline
Gamma & Gamma(          3.7185,           0.2689) &           0.0723 \\
\hline
Weibull & Weibull(          1.1328,           1.8135) &           0.0979 \\
\hline
IG & IG(          1.0000,           3.2759) &           0.0405 \\
\hline
\end{tabular}
\end{center}
\end{minipage}
\begin{minipage}{.5\textwidth}
\begin{center}
\begin{tabular}{ c c c} 
\multicolumn{2}{c}{} \\
\hline
            type &       parameters &          KS Statistic  \\
\hline
Normal & N(          1.0000,           0.5058) &           0.0911 \\
\hline
LogNormal & LN(         -0.1260,           0.5155) &           0.0349 \\
\hline
IGa & IGa(          3.7185,           2.8489) &           0.0723 \\
\hline
Gamma & Gamma(          4.1290,           0.2422) &           0.0269 \\
\hline
Weibull & Weibull(          1.1326,           2.0978) &           0.0536 \\
\hline
IG & IG(          1.0000,           3.2759) &           0.0481 \\
\hline
\end{tabular}
\end{center}
  \end{minipage}

\end{table}

\newpage

\begin{figure}[!htbp]
\centering
\begin{tabular}{cc}
\includegraphics[width = 0.4 \textwidth]{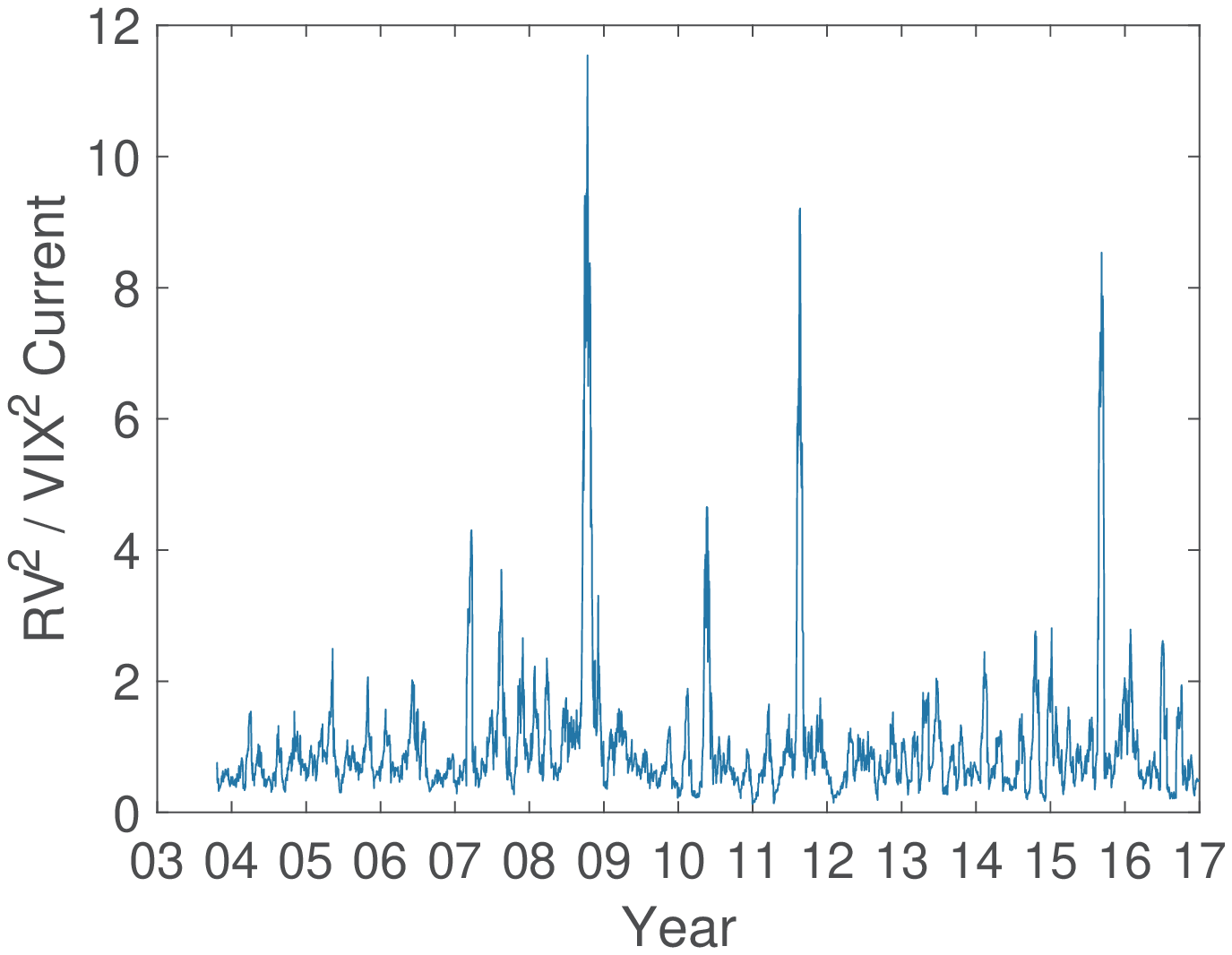} \hspace{0.2cm}
\includegraphics[width = 0.4 \textwidth]{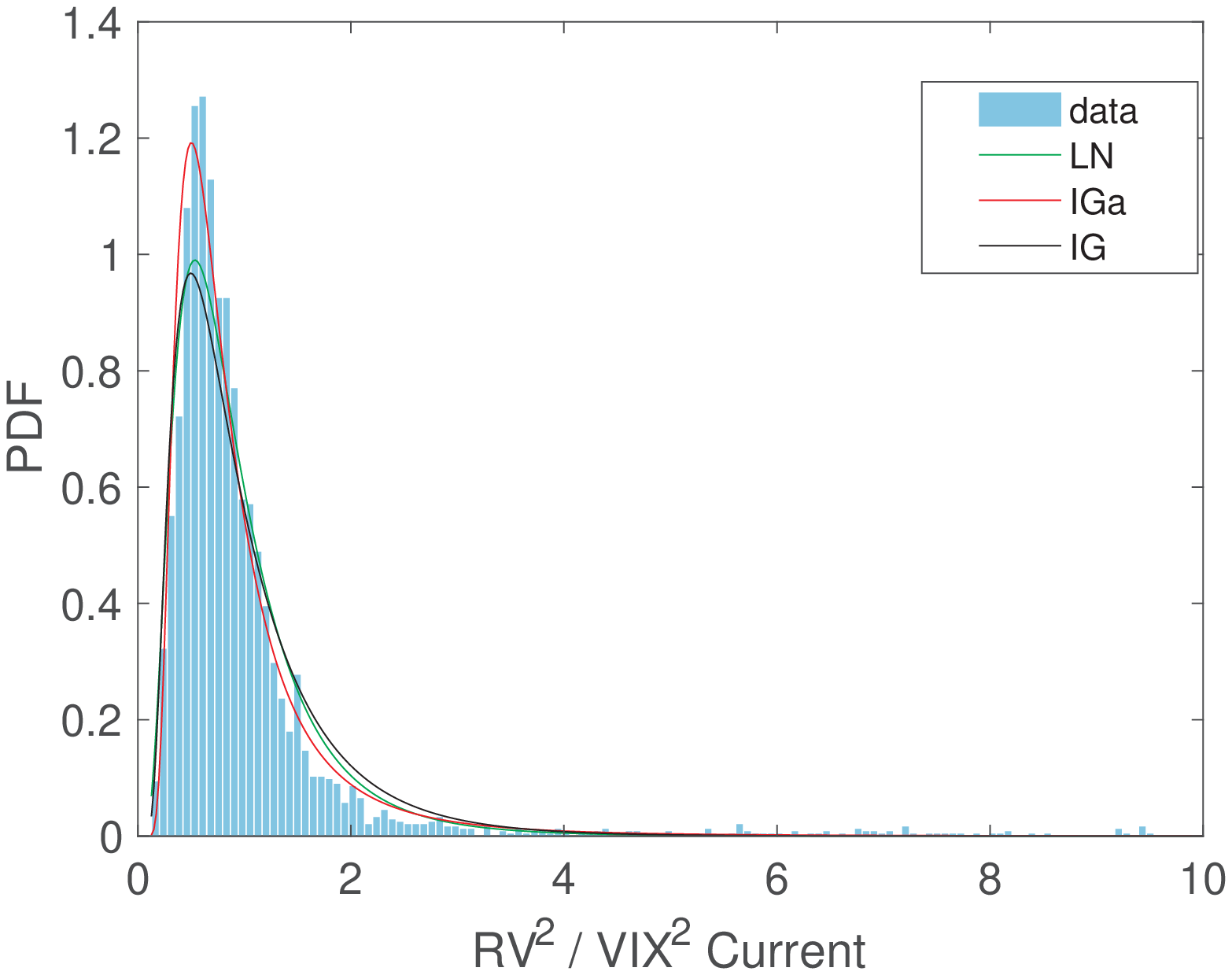}
\end{tabular}
\caption{$\mathrm{RV}^2 / \mathrm{VIX}^2$, from Sep 22nd, 2003 to Dec 30th, 2016.}
\label{RVOverVIXListSRV2OverVIX22017}
\end{figure}

\begin{figure}[!htbp]
\centering
\begin{tabular}{cc}
\includegraphics[width = 0.4 \textwidth]{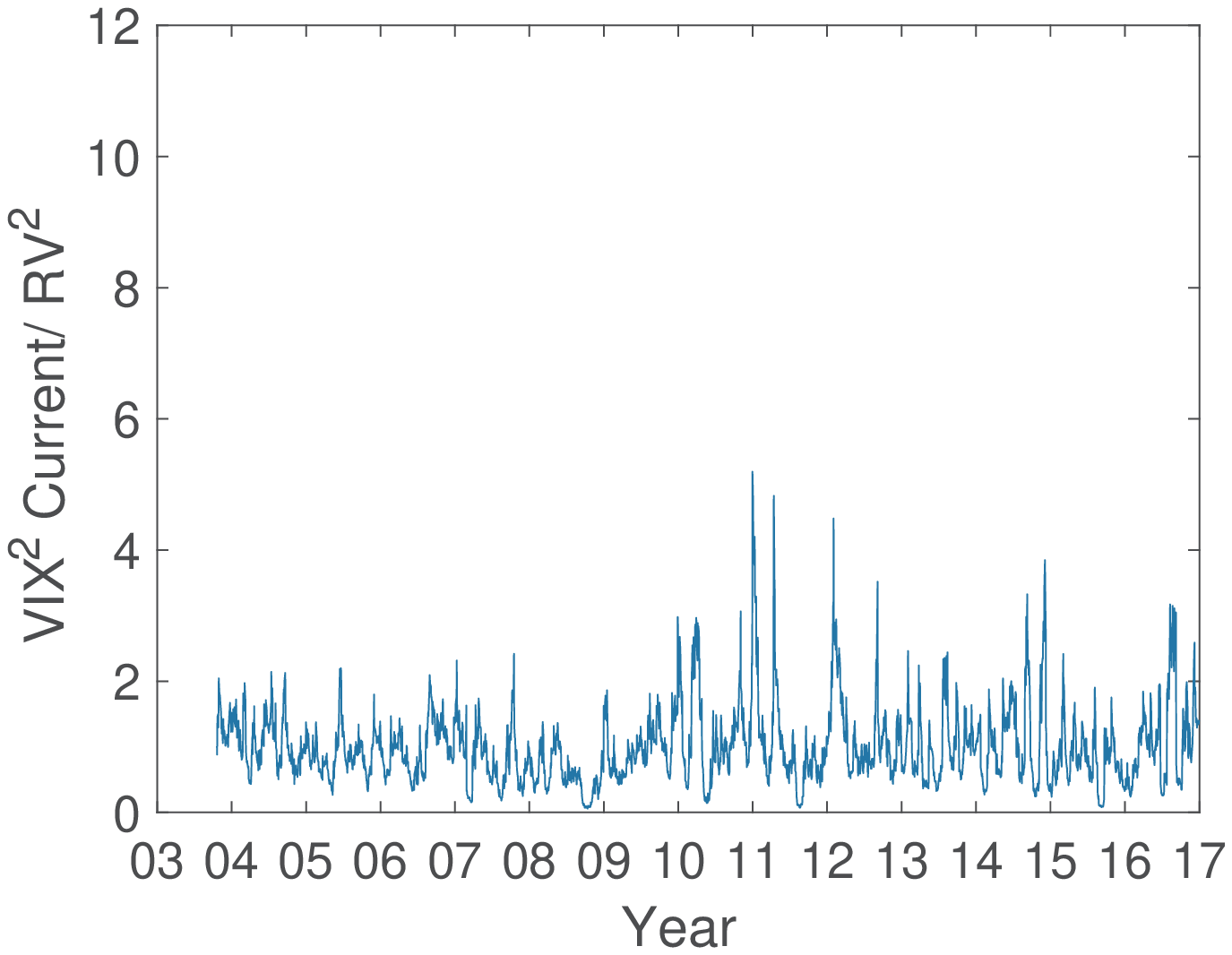} \hspace{0.2cm}
\includegraphics[width = 0.4 \textwidth]{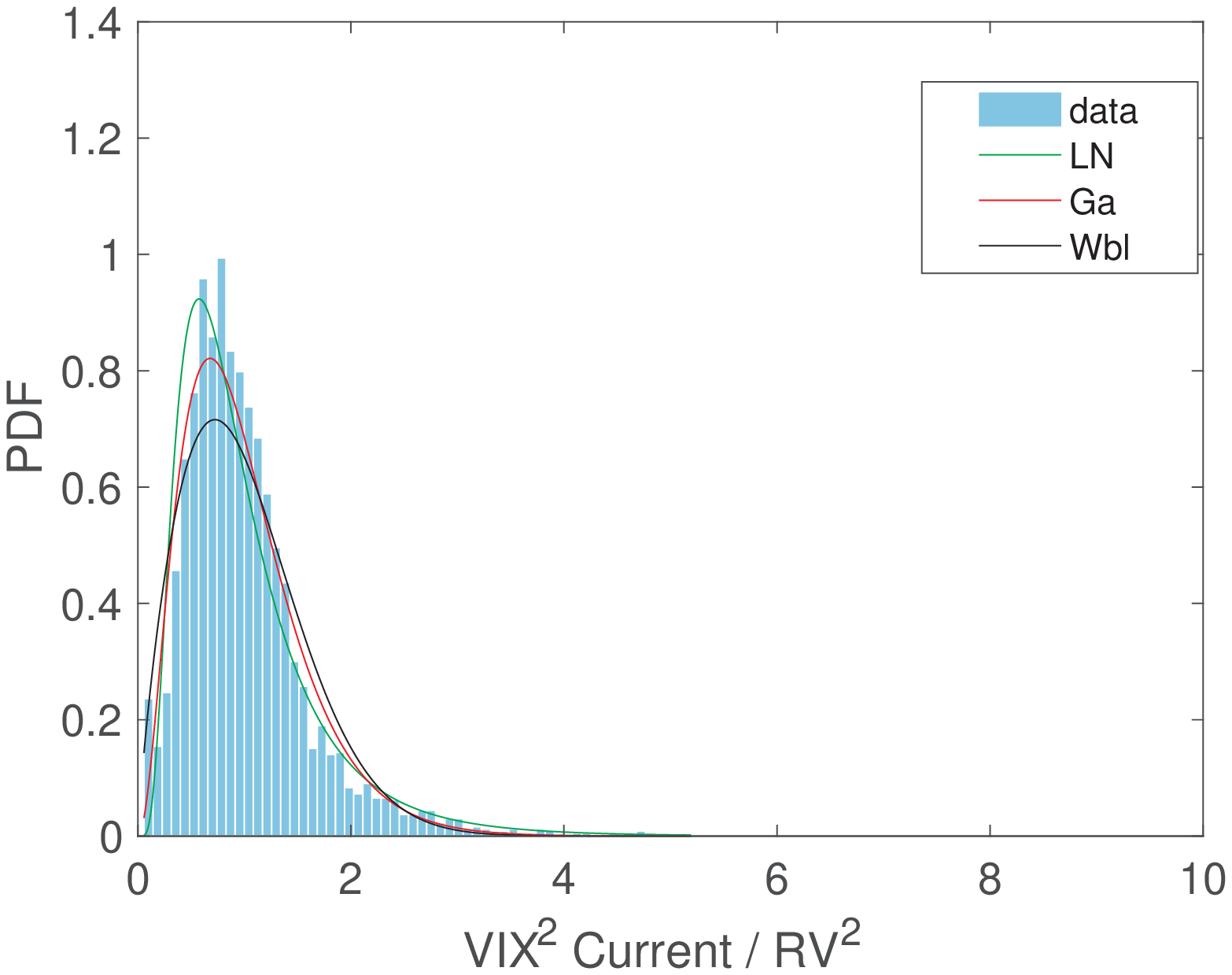}
\end{tabular}
\caption{$ \mathrm{VIX}^2 / \mathrm{RV}^2$, from Sep 22nd, 2003 to Dec 30th, 2016.}
\label{VIXOverRVListSVIX2OverRV22017n}
\end{figure}

\begin{table}[!htb]
\caption{MLE results for ``$\mathrm{RV}^2 / \mathrm{VIX}^2$" and ``$\mathrm{VIX}^2 / \mathrm{RV}^2$"}
\label{MLESRV2OverVIX22017n}
\begin{minipage}{0.5\textwidth}
\begin{center}
\begin{tabular}{ c c c} 
\multicolumn{2}{c}{} \\
\hline
            type &       parameters &          KS Statistic  \\
\hline
Normal & N(          1.0000,           1.0338) &           0.2273 \\
\hline
LogNormal & LN(         -0.2391,           0.6201) &           0.0564 \\
\hline
IGa & IGa(          3.1107,           2.0677) &           0.0360 \\
\hline
Gamma & Gamma(          2.2436,           0.4457) &           0.1178 \\
\hline
Weibull & Weibull(          1.0981,           1.2920) &           0.1344 \\
\hline
IG & IG(          1.0000,           1.9826) &           0.0814 \\
\hline
\end{tabular}
\end{center}
\end{minipage}
\begin{minipage}{.5\textwidth}
\begin{center}
\begin{tabular}{ c c c} 
\multicolumn{2}{c}{} \\
\hline
            type &       parameters &          KS Statistic  \\
\hline
Normal & N(          1.0000,           0.5864) &           0.0959 \\
\hline
LogNormal & LN(         -0.1693,           0.6201) &           0.0564 \\
\hline
IGa & IGa(          2.2436,           1.4914) &           0.1178 \\
\hline
Gamma & Gamma(          3.1107,           0.3215) &           0.0360 \\
\hline
Weibull & Weibull(          1.1282,           1.8115) &           0.0564 \\
\hline
IG & IG(          1.0000,           1.9827) &           0.0948 \\
\hline
\end{tabular}
\end{center}
  \end{minipage}

\end{table}

\newpage

\begin{figure}[!htbp]
\centering
\begin{tabular}{cc}
\includegraphics[width = 0.4 \textwidth]{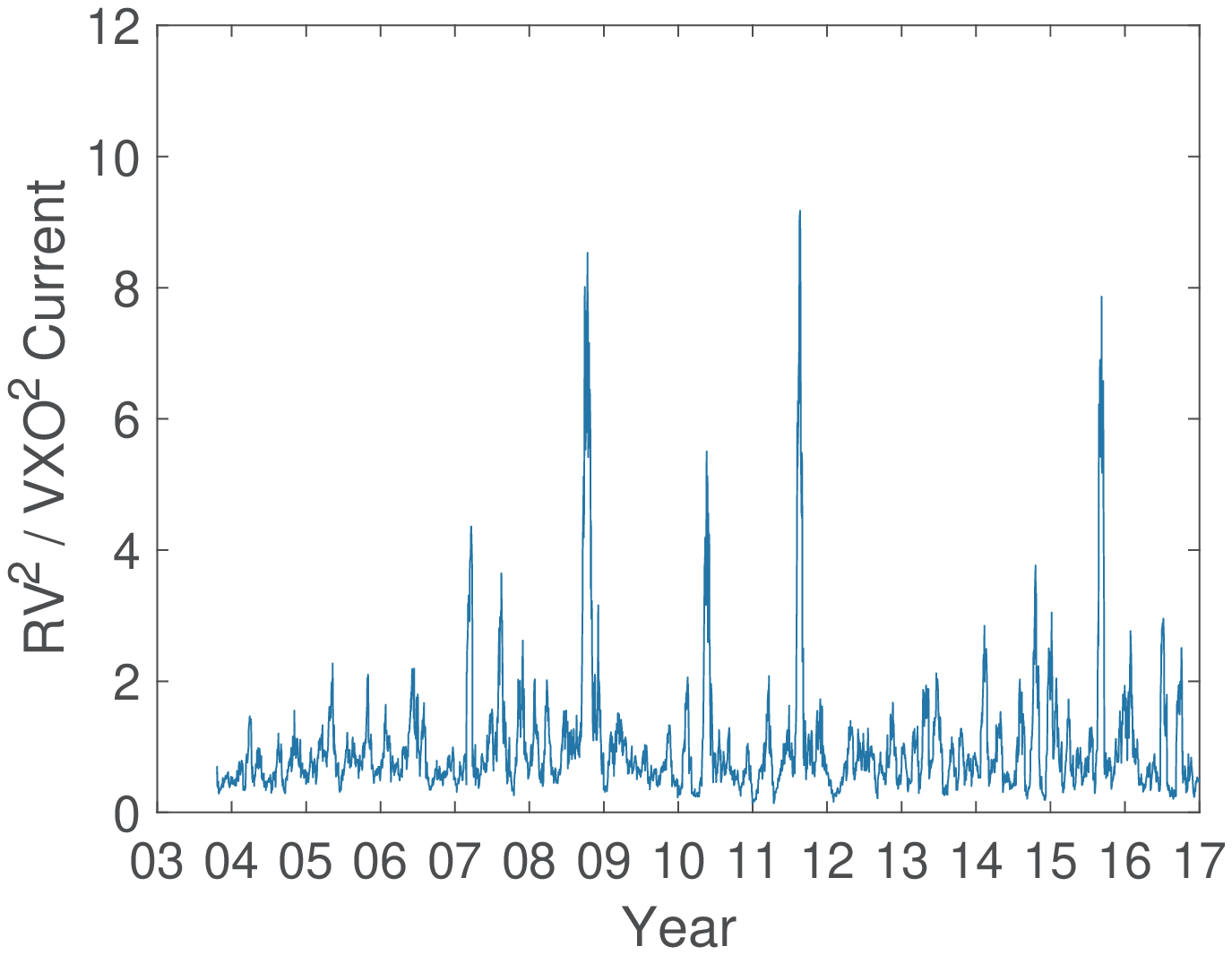} \hspace{0.2cm}
\includegraphics[width = 0.4 \textwidth]{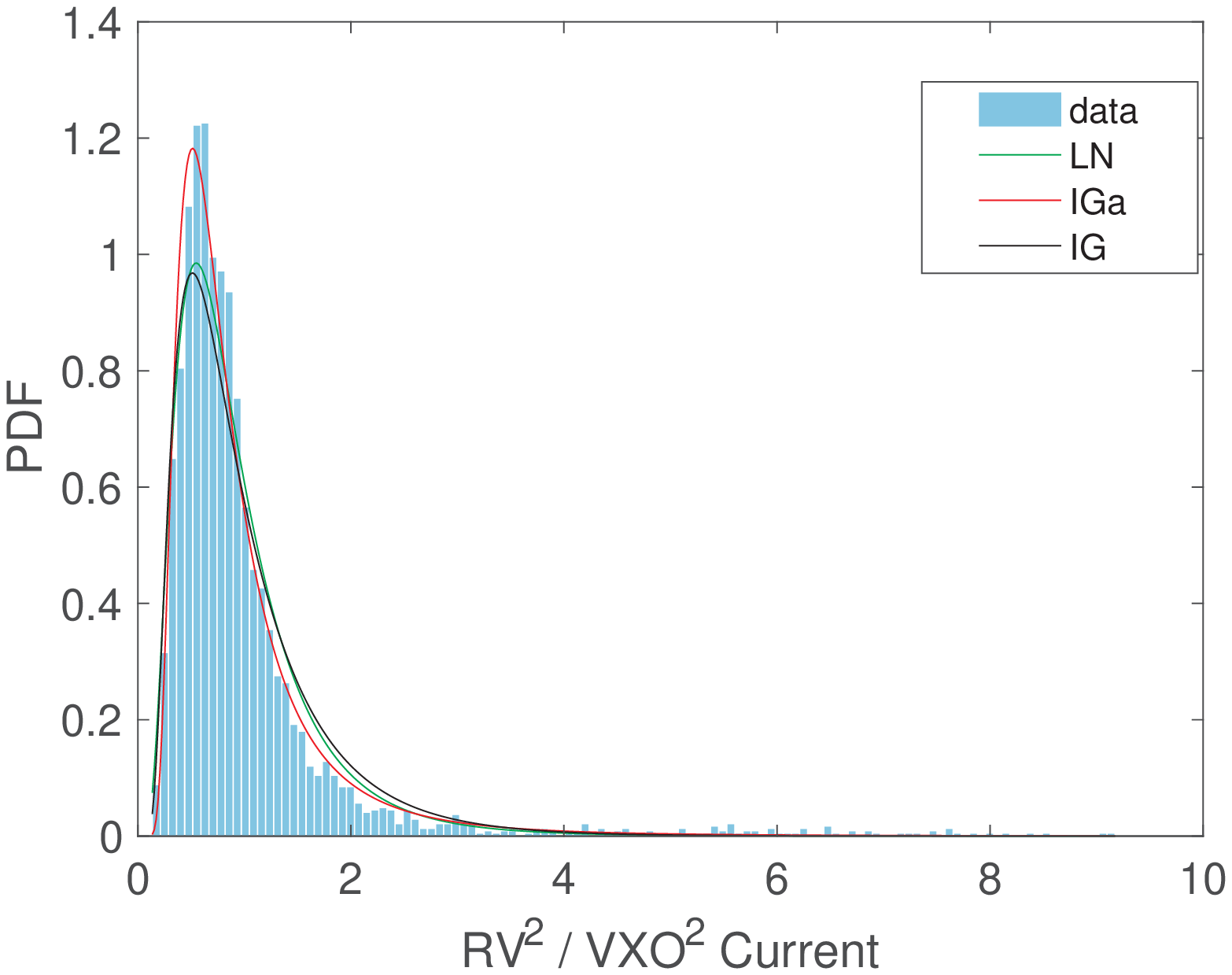}
\end{tabular}
\caption{$\mathrm{RV}^2 / \mathrm{VXO}^2$, from Sep 22nd, 2003 to Dec 30th, 2016.}
\label{RVOverVXOListSRV2OverVXO22017}
\end{figure}

\begin{figure}[!htbp]
\centering
\begin{tabular}{cc}
\includegraphics[width = 0.4 \textwidth]{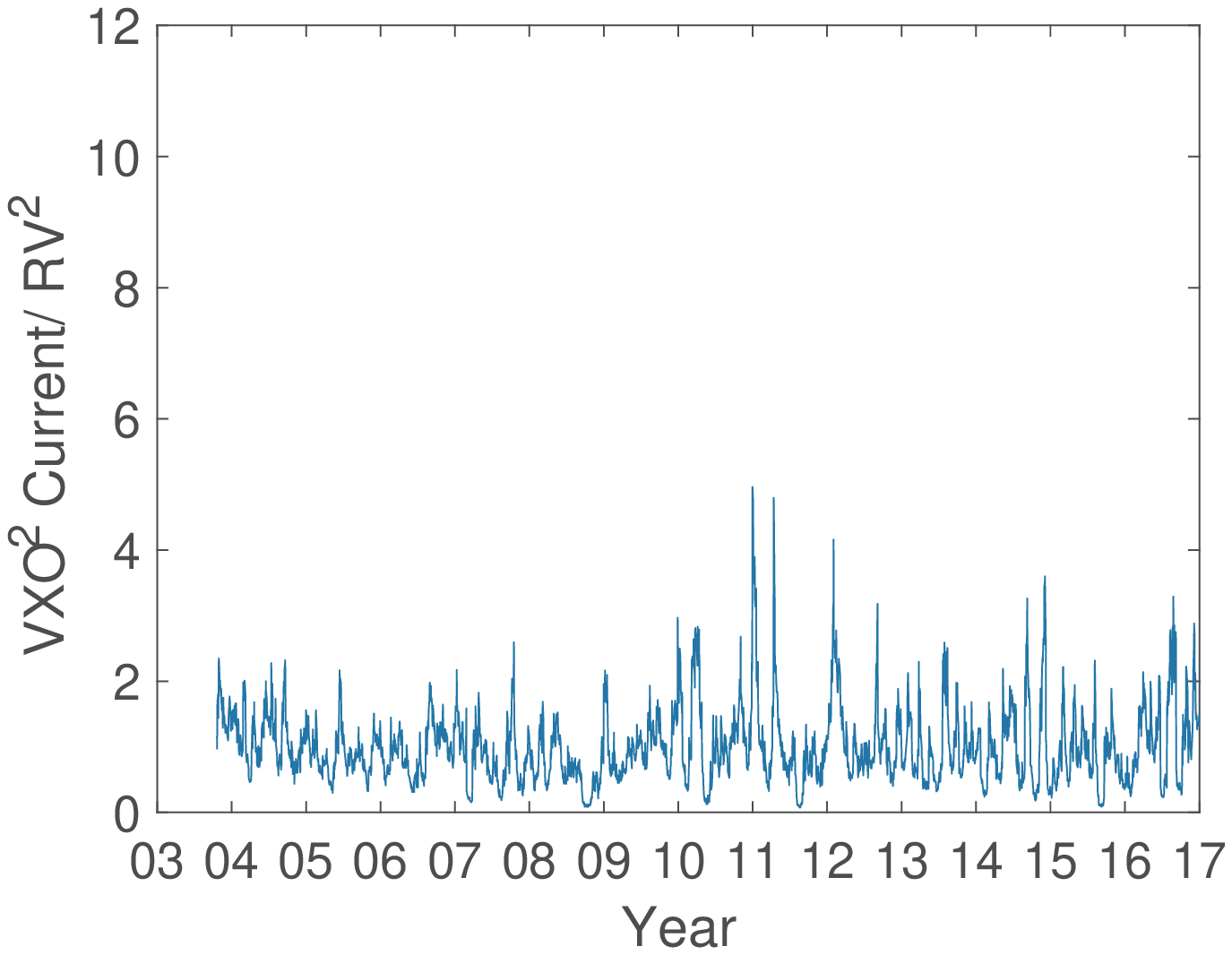} \hspace{0.2cm}
\includegraphics[width = 0.4 \textwidth]{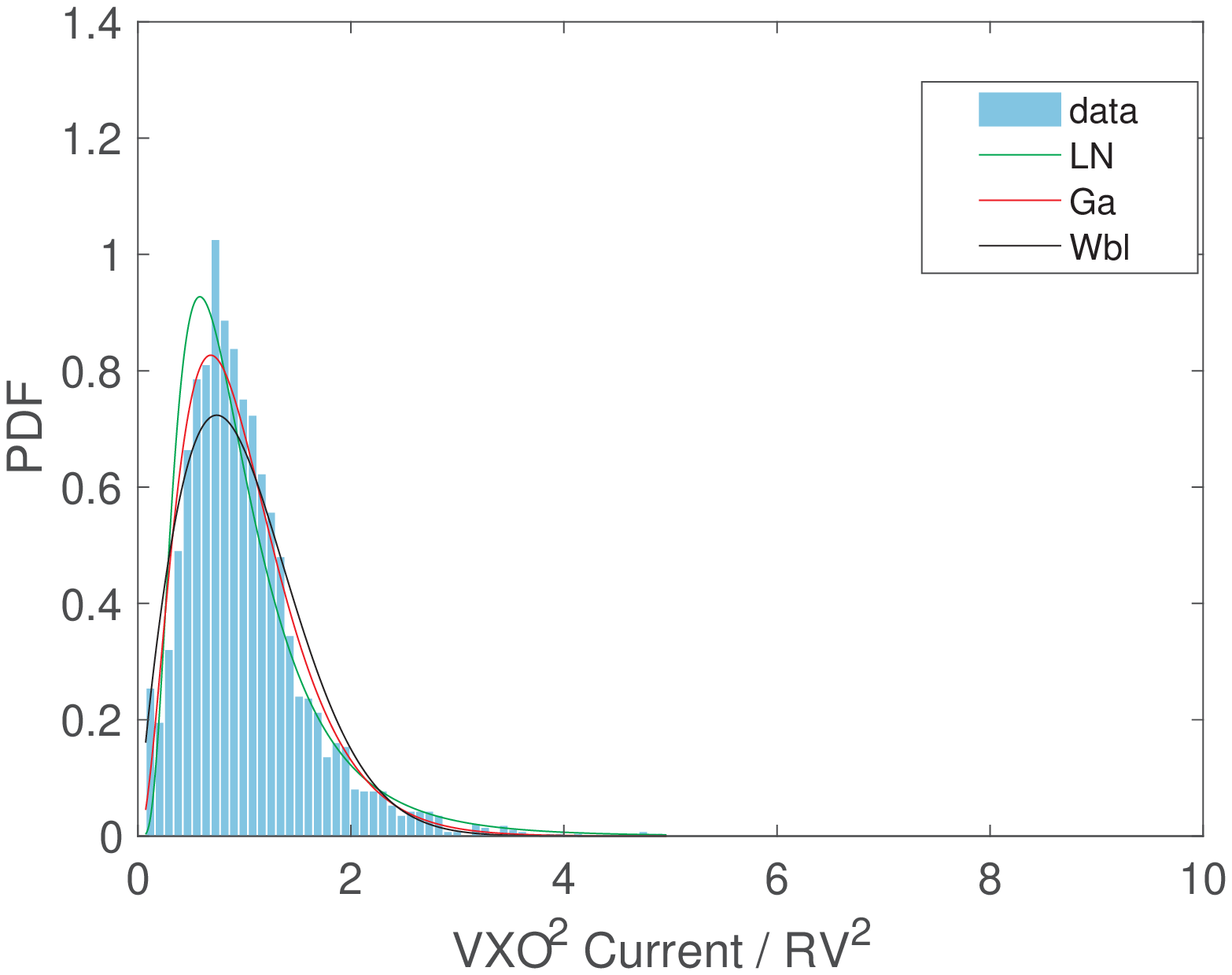}
\end{tabular}
\caption{$ \mathrm{VXO}^2 / \mathrm{RV}^2$, from Sep 22nd, 2003 to Dec 30th, 2016.}
\label{VXOOverRVListSVXO2OverRV22017n}
\end{figure}

\begin{table}[!htb]
\caption{MLE results for ``$\mathrm{RV}^2 / \mathrm{VXO}^2$" and ``$\mathrm{VXO}^2 / \mathrm{RV}^2$"}
\label{MLESRV2OverVXO22017n}
\begin{minipage}{0.5\textwidth}
\begin{center}
\begin{tabular}{ c c c} 
\multicolumn{2}{c}{} \\
\hline
            type &       parameters &          KS Statistic  \\
\hline
Normal & N(          1.0000,           0.9475) &           0.2130 \\
\hline
LogNormal & LN(         -0.2263,           0.6125) &           0.0595 \\
\hline
IGa & IGa(          3.1750,           2.1454) &           0.0282 \\
\hline
Gamma & Gamma(          2.3626,           0.4233) &           0.1169 \\
\hline
Weibull & Weibull(          1.1059,           1.3451) &           0.1270 \\
\hline
IG & IG(          1.0000,           2.0837) &           0.0778 \\
\hline
\end{tabular}
\end{center}
\end{minipage}
\begin{minipage}{.5\textwidth}
\begin{center}
\begin{tabular}{ c c c} 
\multicolumn{2}{c}{} \\
\hline
            type &       parameters &          KS Statistic  \\
\hline
Normal & N(          1.0000,           0.5731) &           0.0911 \\
\hline
LogNormal & LN(         -0.1657,           0.6125) &           0.0595 \\
\hline
IGa & IGa(          2.3626,           1.5965) &           0.1169 \\
\hline
Gamma & Gamma(          3.1750,           0.3150) &           0.0282 \\
\hline
Weibull & Weibull(          1.1289,           1.8480) &           0.0486 \\
\hline
IG & IG(          1.0000,           2.0837) &           0.0930 \\
\hline
\end{tabular}
\end{center}
  \end{minipage}

\end{table}

\clearpage

\subsection{Ratio distribution revisited}\label{RV2/VIX2-r}

Here we repeat the calculation from the Sec. \ref{RV2/VIX2} except with the RV for the month \emph{preceding} the month for which VIX/VXO is calculated. For instance, if on March 31 VIX/VXO predict RV for April, we compare them to RV for March. This is to test the hypothesis that VIX/VXO are pretty much as good a predictor as the RV they are already aware of. Indeed, LN distribution fits best both $RV^2/VIX^2$ and $RV^2/VXO^2$, as well as their inverse, consistent with the fact that for a LN distribution the distribution of the inverse variable is also LN. In Sec. \ref{RV2/VIX2} we hypothesized large spikes of RV as the reason for fat tails. This is congruent with uncertainty -- even with the knowledge of preceding RV -- reflected in heavy LN tails.
\begin{figure}[!htbp]
\centering
\begin{tabular}{cc}
\includegraphics[width = 0.4 \textwidth]{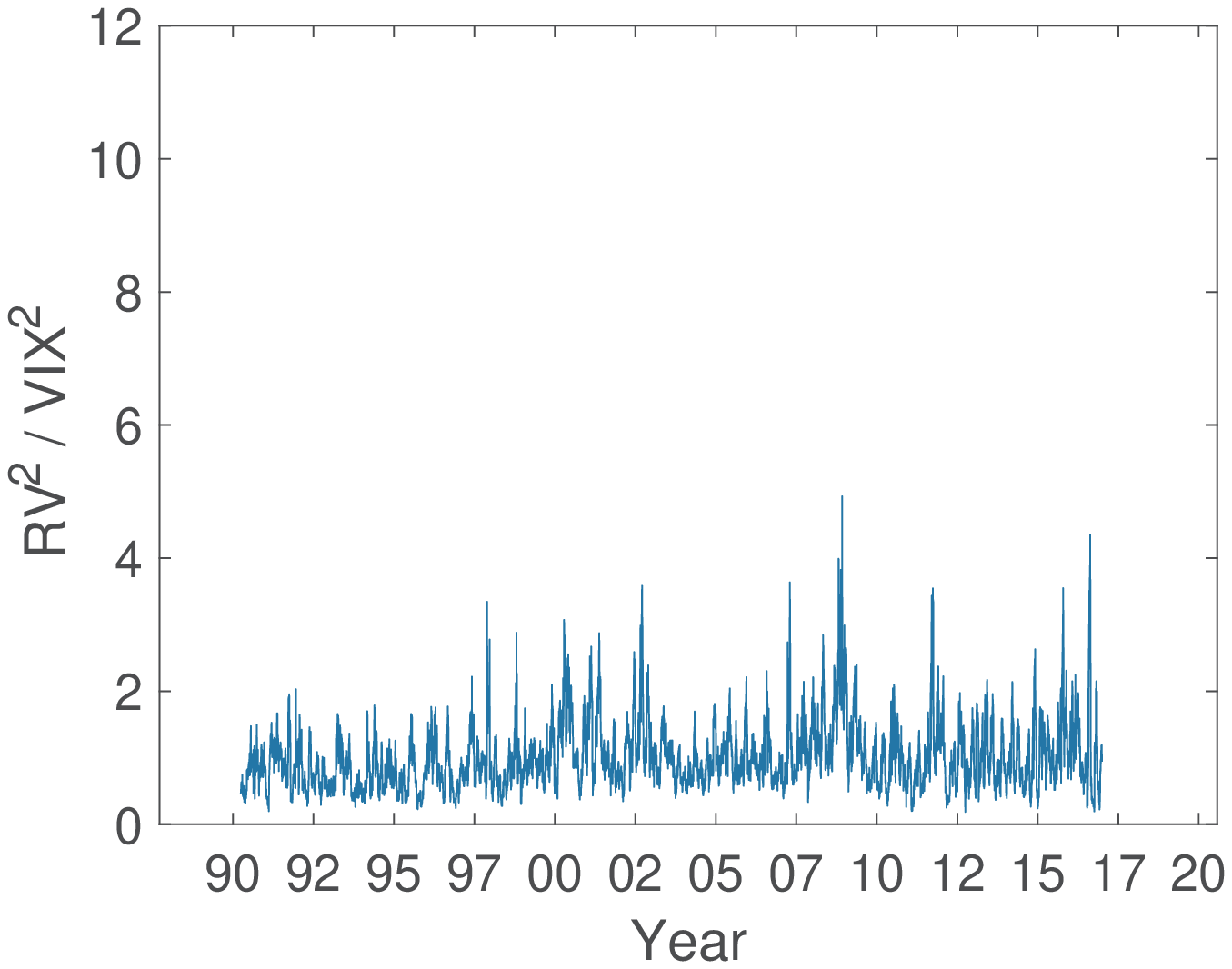} \hspace{0.2cm}
\includegraphics[width = 0.4 \textwidth]{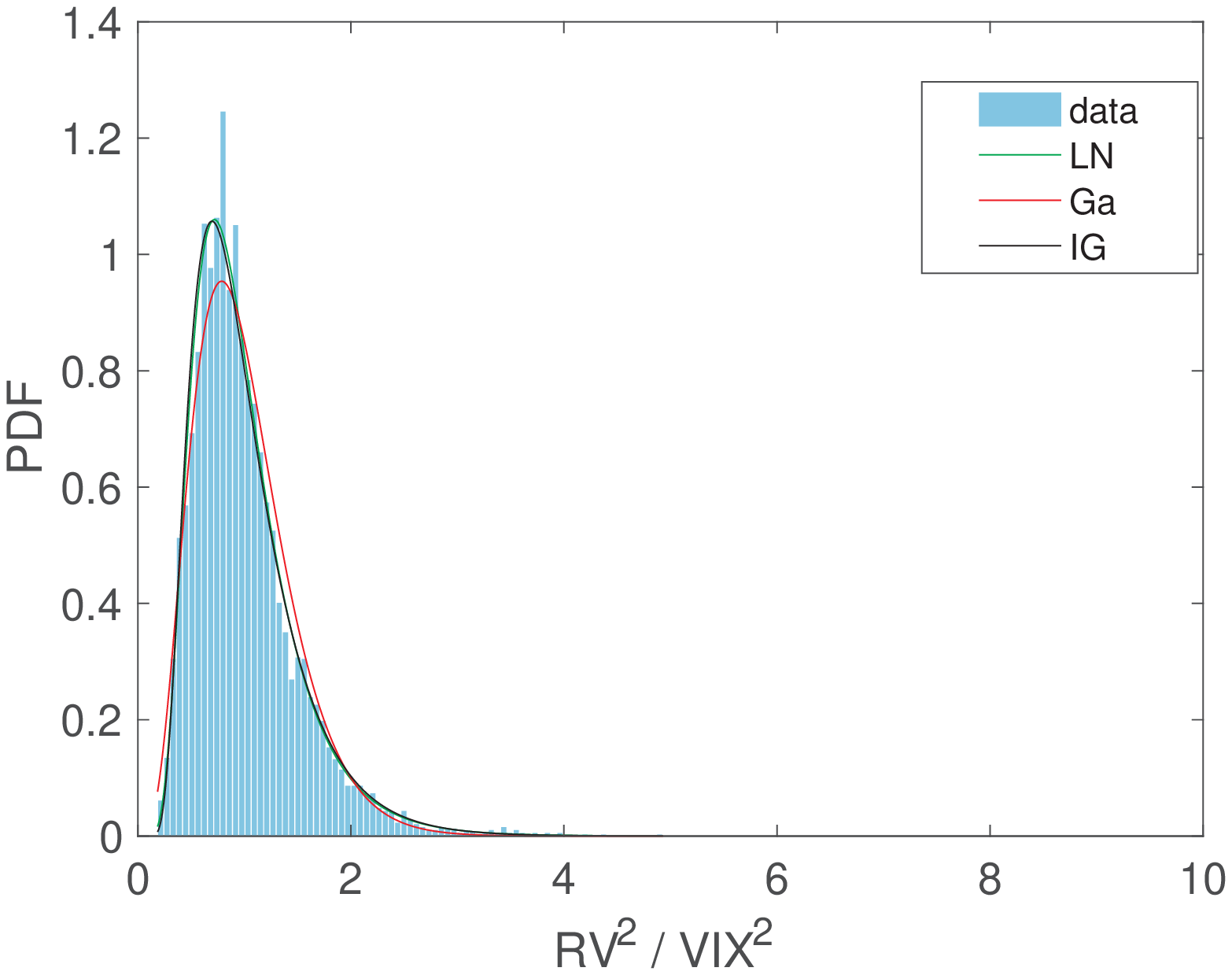}
\end{tabular}
\caption{$\mathrm{RV}^2 / \mathrm{VIX}^2$, from Jan 2nd, 1990 to Dec 30th, 2016.}
\label{RVOverVIXListSRV2OverVIX21990}
\end{figure}

\begin{figure}[!htbp]
\centering
\begin{tabular}{cc}
\includegraphics[width = 0.4 \textwidth]{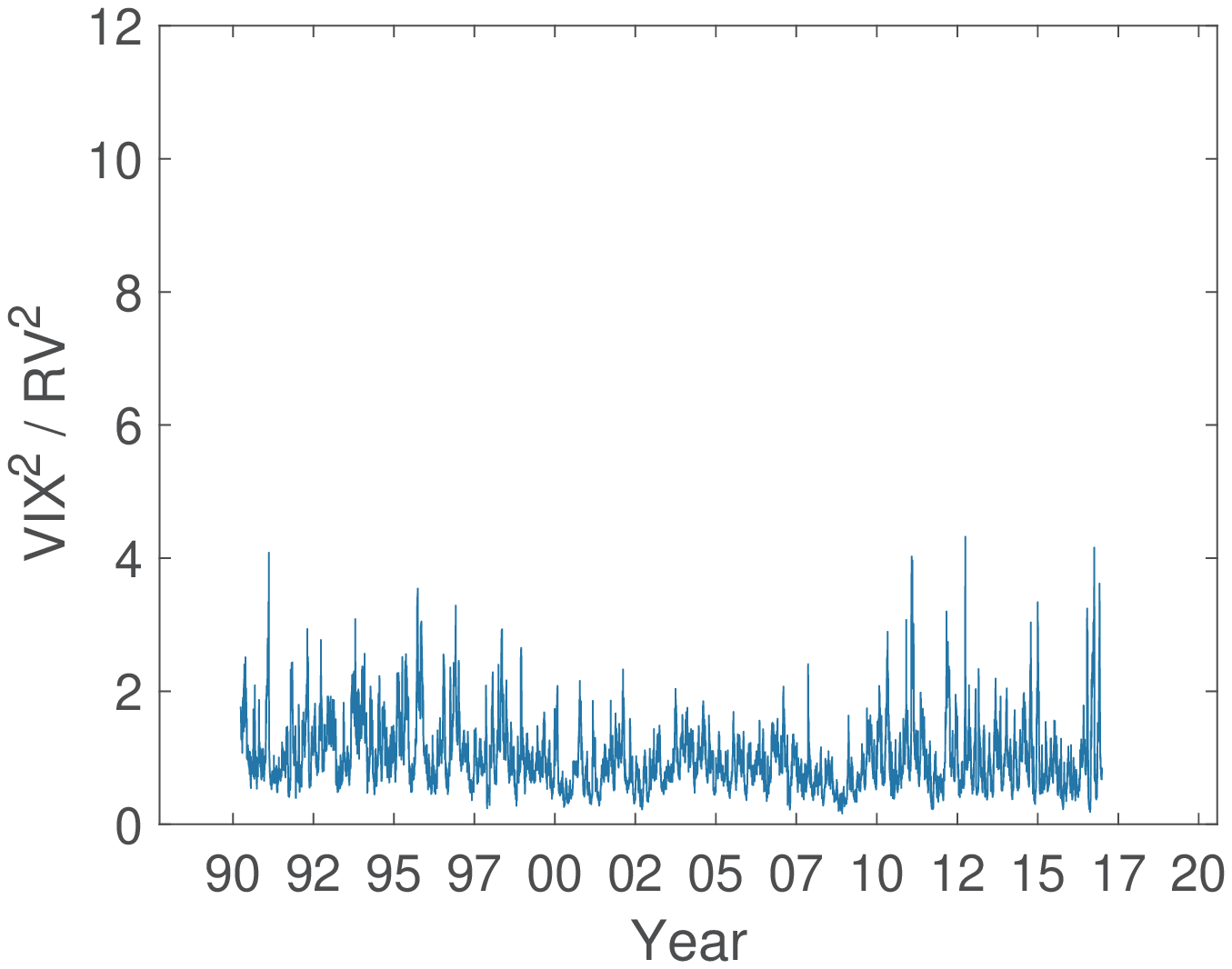} \hspace{0.2cm}
\includegraphics[width = 0.4 \textwidth]{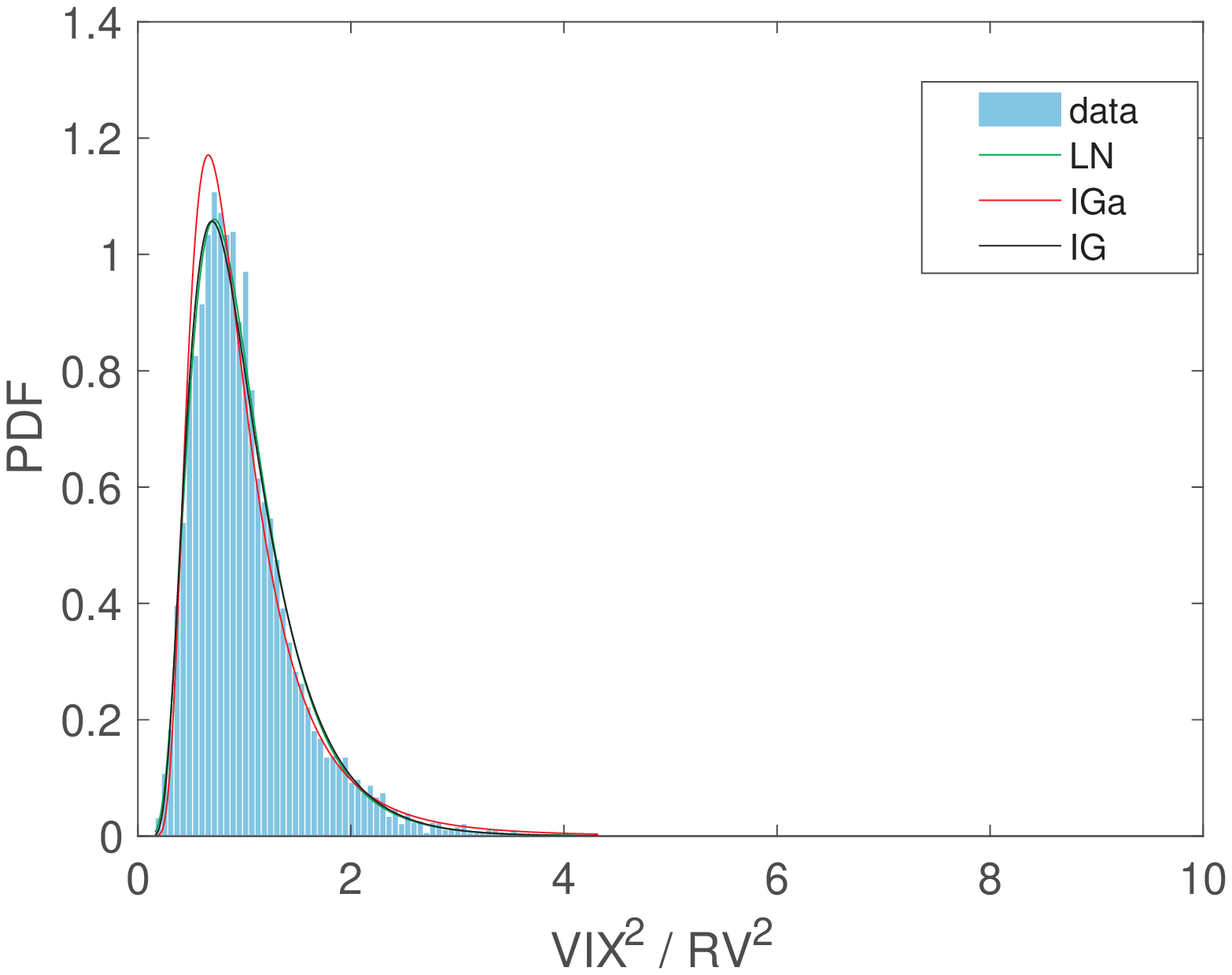}
\end{tabular}
\caption{$ \mathrm{VIX}^2 / \mathrm{RV}^2$, from Jan 2nd, 1990 to Dec 30th, 2016.}
\label{VIXOverRVListSVIX2OverRV21990nn}
\end{figure}

\begin{table}[!htb]
\caption{MLE results for ``$\mathrm{RV}^2 / \mathrm{VIX}^2$" and ``$\mathrm{VIX}^2 / \mathrm{RV}^2$"}
\label{MLESRV2OverVIX21990nn}
\begin{minipage}{0.5\textwidth}
\begin{center}
\begin{tabular}{ c c c} 
\multicolumn{2}{c}{} \\
 \hline
            type &       parameters &          KS Statistic  \\
\hline
Normal & N(          1.0000,           0.4974) &           0.0992 \\
\hline
LogNormal & LN(         -0.1099,           0.4689) &           0.0147 \\
\hline
IGa & IGa(          4.6889,           3.7619) &           0.0431 \\
\hline
Gamma & Gamma(          4.7110,           0.2123) &           0.0381 \\
\hline
Weibull & Weibull(          1.1325,           2.1250) &           0.0672 \\
\hline
IG & IG(          1.0000,           4.0580) &           0.0215 \\
\hline
\end{tabular}
\end{center}
\end{minipage}
\begin{minipage}{.5\textwidth}
\begin{center}
\begin{tabular}{ c c c} 
\multicolumn{2}{c}{} \\
\hline
            type &       parameters &          KS Statistic  \\
\hline
Normal & N(          1.0000,           0.4999) &           0.1059 \\
\hline
LogNormal & LN(         -0.1104,           0.4689) &           0.0147 \\
\hline
IGa & IGa(          4.7110,           3.7796) &           0.0381 \\
\hline
Gamma & Gamma(          4.6889,           0.2133) &           0.0431 \\
\hline
Weibull & Weibull(          1.1329,           2.1186) &           0.0751 \\
\hline
IG & IG(          1.0000,           4.0580) &           0.0163 \\
\hline
\end{tabular}
\end{center}
  \end{minipage}

\end{table}

\newpage

\begin{figure}[!htbp]
\centering
\begin{tabular}{cc}
\includegraphics[width = 0.4 \textwidth]{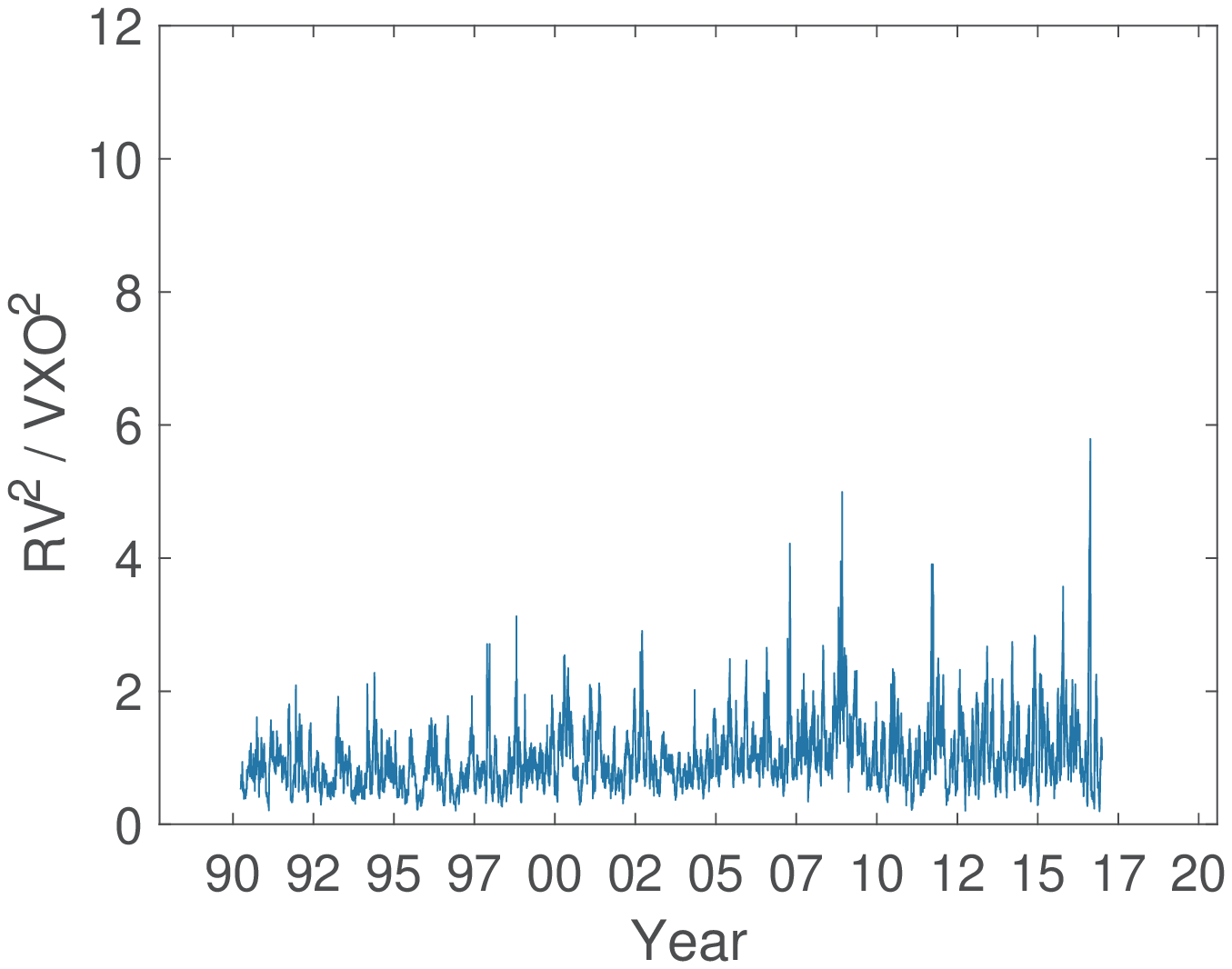} \hspace{0.2cm}
\includegraphics[width = 0.4 \textwidth]{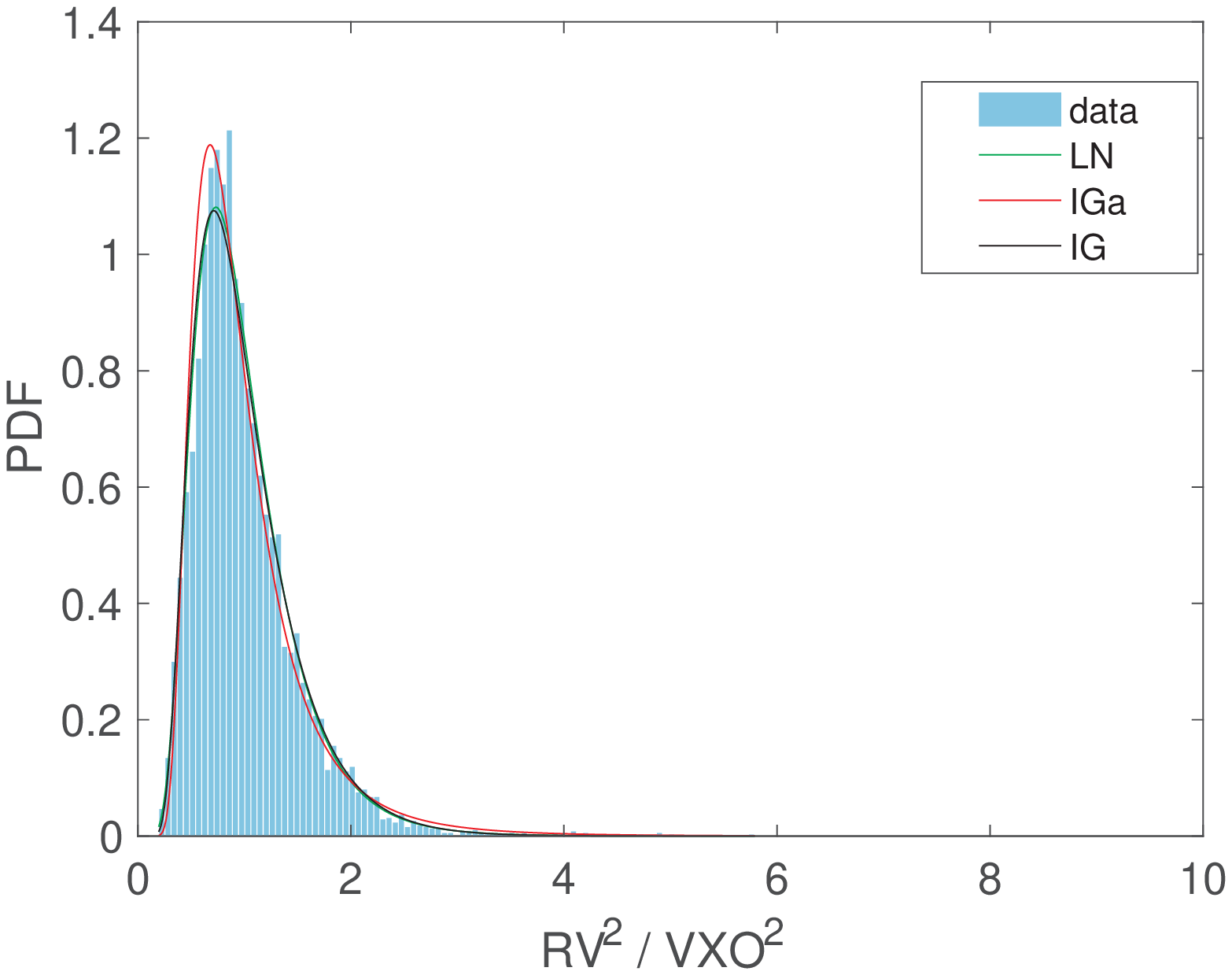}
\end{tabular}
\caption{$\mathrm{RV}^2 / \mathrm{VXO}^2$, Jan 2nd, 1990 to Dec 30th, 2016.}
\label{RVOverVXOListSRV2OverVXO22016}
\end{figure}

\begin{figure}[!htbp]
\centering
\begin{tabular}{cc}
\includegraphics[width = 0.4 \textwidth]{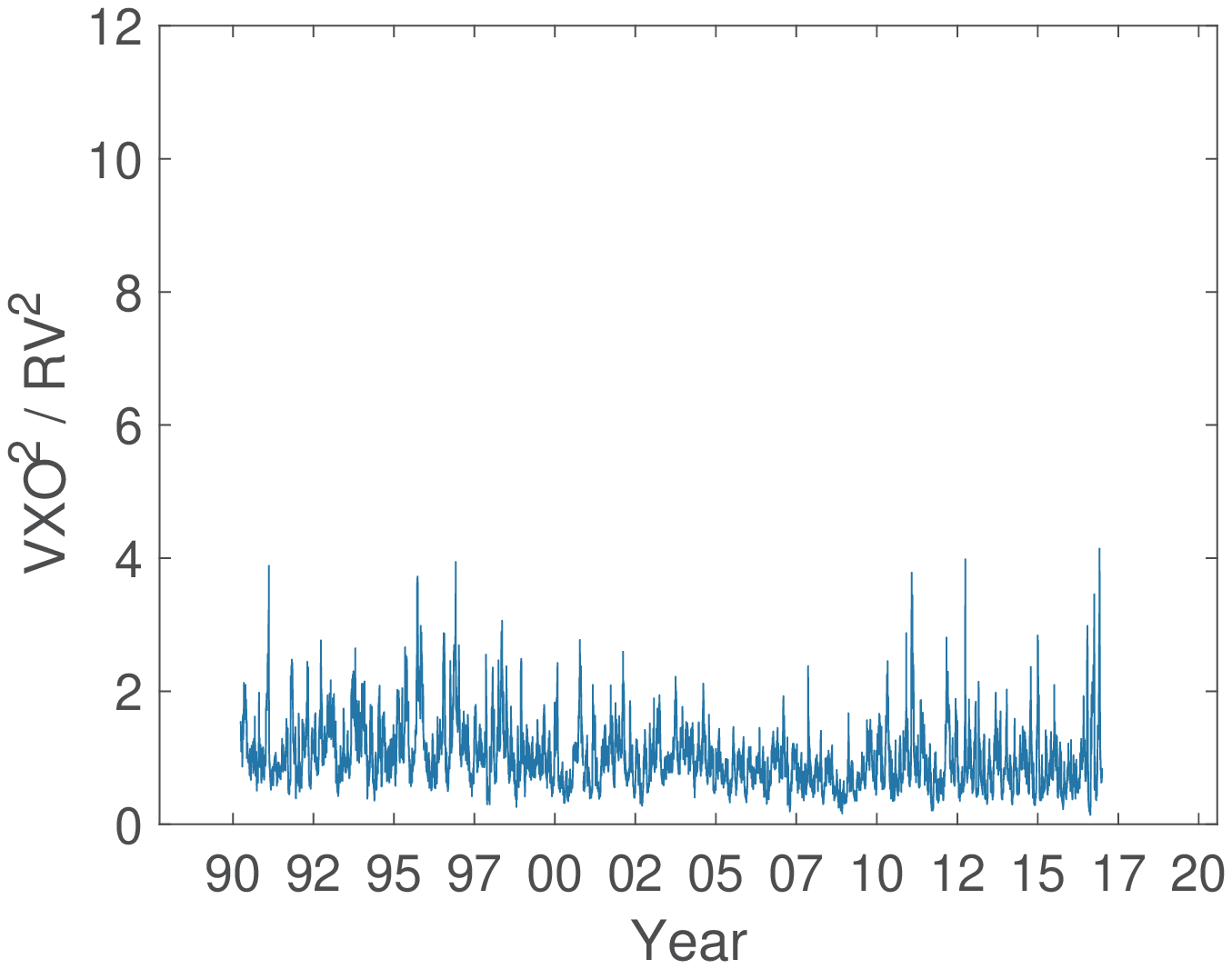} \hspace{0.2cm}
\includegraphics[width = 0.4 \textwidth]{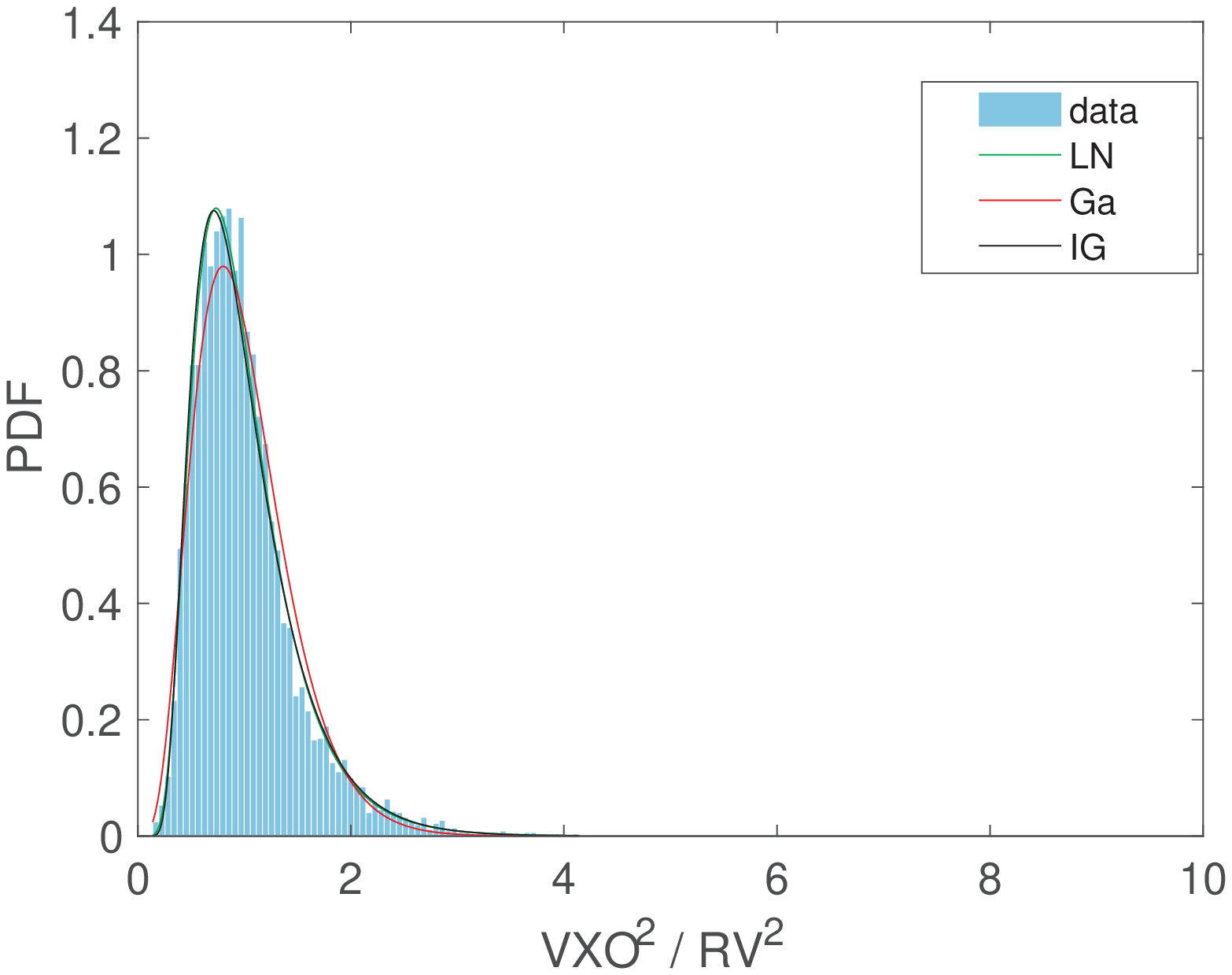}
\end{tabular}
\caption{$ \mathrm{VXO}^2 / \mathrm{RV}^2$, from Jan 2nd, 1990 to Dec 30th, 2016.}
\label{VXOOverRVListSVXO2OverRV22016n}
\end{figure}

\begin{table}[!htb]
\caption{MLE results for ``$\mathrm{RV}^2 / \mathrm{VXO}^2$" and ``$\mathrm{VXO}^2 / \mathrm{RV}^2$"}
\label{MLESRV2OverVXO22016n}
\begin{minipage}{0.5\textwidth}
\begin{center}
\begin{tabular}{ c c c} 
\multicolumn{2}{c}{} \\
\hline
            type &       parameters &          KS Statistic  \\
\hline
Normal & N(          1.0000,           0.4915) &           0.1064 \\
\hline
LogNormal & LN(         -0.1041,           0.4539) &           0.0150 \\
\hline
IGa & IGa(          5.0351,           4.0948) &           0.0331 \\
\hline
Gamma & Gamma(          4.9618,           0.2015) &           0.0454 \\
\hline
Weibull & Weibull(          1.1316,           2.1383) &           0.0730 \\
\hline
IG & IG(          1.0000,           4.3548) &           0.0203 \\
\hline
\end{tabular}
\end{center}
\end{minipage}
\begin{minipage}{.5\textwidth}
\begin{center}
\begin{tabular}{ c c c} 
\multicolumn{2}{c}{} \\
\hline
            type &       parameters &          KS Statistic  \\
\hline
Normal & N(          1.0000,           0.4768) &           0.0933 \\
\hline
LogNormal & LN(         -0.1026,           0.4539) &           0.0150 \\
\hline
IGa & IGa(          4.9618,           4.0352) &           0.0454 \\
\hline
Gamma & Gamma(          5.0351,           0.1986) &           0.0331 \\
\hline
Weibull & Weibull(          1.1319,           2.2099) &           0.0689 \\
\hline
IG & IG(          1.0000,           4.3548) &           0.0212 \\
\hline
\end{tabular}
\end{center}
  \end{minipage}

\end{table}

\clearpage
\section{Variance of realized variance}\label{RV2var}

The goal of this section is to evaluate the expectation value of theoretical variance of realized variance

\begin{equation}
E[(\frac{1}{T}\int_{0}^{T} v_t\mathrm{d}t-E[\frac{1}{T}\int_{0}^{T} v_t\mathrm{d}t])^2]=E[(\frac{1}{T}\int_{0}^{T} v_t\mathrm{d}t-\theta)^2]
\label{VarVar}
\end{equation}
and compare it with the market data using historic squared stock returns. Here $v_t=\sigma_t^2$ is the stochastic variance and $\theta$ is the mean (expectation) value of $v_t$ in the mean-reverting models, 

\begin{equation}
E[v_t]=\theta 
\label{theta}
\end{equation}
Below we discuss two such models -- Heston and multiplicative \cite{liu2017distributions}.

\subsection{Heston Model}

In the Heston model the equation for stochastic variance is given by
\begin{equation}
\mathrm{d}v_t = -\gamma(v_t - \theta)\mathrm{d}t + \kappa \sqrt{v_t}\mathrm{d}W_t
\label{11}
\end{equation}
To evaluate (\ref{VarVar}), we need to know the correlation function \cite{dashti2018correlations}

\begin{equation}
E[v_t v_{t+\tau}]=\theta^2+\frac{\kappa^2\theta}{2\gamma}e^{-\gamma\tau}
\label{meanVolCorHest}
\end{equation}
where

\begin{equation}
\frac{\kappa^2\theta}{2\gamma}=E[v_t^2]-(E[v_t])^2
\label{universalH}
\end{equation}
From (\ref{VarVar}) and (\ref{meanVolCorHest}) we find

\begin{equation}
E[(\frac{1}{T}\int_{0}^{T} v_t\mathrm{d}t-\theta )^2]= \frac{1}{(\gamma T)^2}(\frac{\kappa^2 \theta}{\gamma})(-1+e^{-\gamma T} +\gamma T)
\label{VarVolIntegralHestresult}
\end{equation}
with the following limits 

\begin{equation}
E[(\frac{1}{T}\int_{0}^{T} v_t\mathrm{d}t-\theta)^2] \approx  
\begin{cases}
\frac{\kappa^2 \theta}{2 \gamma} \hspace{0cm} \text{,} \hspace{.1cm} \gamma T \ll 1 \\
\frac{\kappa^2 \theta}{\gamma}(\gamma T)^{-1} \hspace{0cm} \text{,} \hspace{.1cm} \gamma T \gg 1
 \end{cases}
\label{VarVolIntegralHestresultLimits}
\end{equation}

\subsection{Multiplicative Model}

In the multiplicative model the equation for stochastic variance is given by

\begin{equation}
\mathrm{d}v_t = -\gamma(v_t - \theta)\mathrm{d}t + \kappa v_t\mathrm{d}W_t
\label{11}
\end{equation}
In this case \cite{dashti2018correlations}

\begin{equation}
E[v_t v_{t+\tau}]=\theta^2+\frac{\kappa^2\theta^2}{2\gamma-\kappa^2}e^{-\gamma \tau}
\label{meanVolCorMul}
\end{equation}
where

\begin{equation}
\frac{\kappa^2\theta^2}{2\gamma-\kappa^2}=E[v_t^2]-(E[v_t])^2
\label{universalH}
\end{equation}
From (\ref{VarVar}) and (\ref{meanVolCorMul}) we find

\begin{equation}
E[(\frac{1}{T}\int_{0}^{T} v_t\mathrm{d}t-\theta)^2]= \frac{1}{(\gamma T)^2}(\frac{2\kappa^2 \theta^2}{2\gamma-\kappa^2})(-1+e^{-\gamma T} +\gamma T)
\label{VarVolIntegralMultresult}
\end{equation}
with the following limits

\begin{equation}
  E[(\frac{1}{T}\int_{0}^{T} v_t\mathrm{d}t-\theta )^2] \approx
    \begin{cases}
      \frac{\kappa^2 \theta^2}{2\gamma-\kappa^2} \hspace{0cm} \text{,} \hspace{.1cm} & \gamma T \ll 1\\
      \frac{2\kappa^2 \theta^2}{2\gamma-\kappa^2}(\gamma T)^{-1} \hspace{0cm} \text{,} \hspace{.1cm} & \gamma T \gg 1\\
    \end{cases} 
    \label{VarVolIntegralMultresultLimits}      
\end{equation}

\subsection{Numerical simulations}

Clearly, for both models the following holds

\begin{equation}
\frac{E[(\frac{1}{T}\int_{0}^{T} v_t\mathrm{d}t-\theta)^2]}{E[v_t^2]-(E[v_t])^2}= \frac{2}{(\gamma T)^2}(-1+e^{-\gamma T} +\gamma T)
\label{VarVolIntegralrReduced}
\end{equation}
with the following limits

\begin{equation}
\frac{E[(\frac{1}{T}\int_{0}^{T} v_t\mathrm{d}t-\theta)^2]}{E[v_t^2]-(E[v_t])^2}\approx
\begin{cases}
 1  \hspace{0cm} \text{,} \hspace{.1cm} \gamma T \ll 1\\
2 (\gamma T)^{-1} \hspace{0cm} \text{,} \hspace{.1cm} \gamma T \gg 1
\end{cases}
\label{VarVolIntegralrReducedLimits}
\end{equation}
meaning that in the regime of strong correlations in (\ref{meanVolCorHest}) and (\ref{meanVolCorMul}), the value of the variance of realized variance does not depend on $T$ and is consistent with central limit theorem otherwise.

\begin{figure}[!htbp]
\centering
\begin{tabular}{cc}
\includegraphics[width = 0.4 \textwidth]{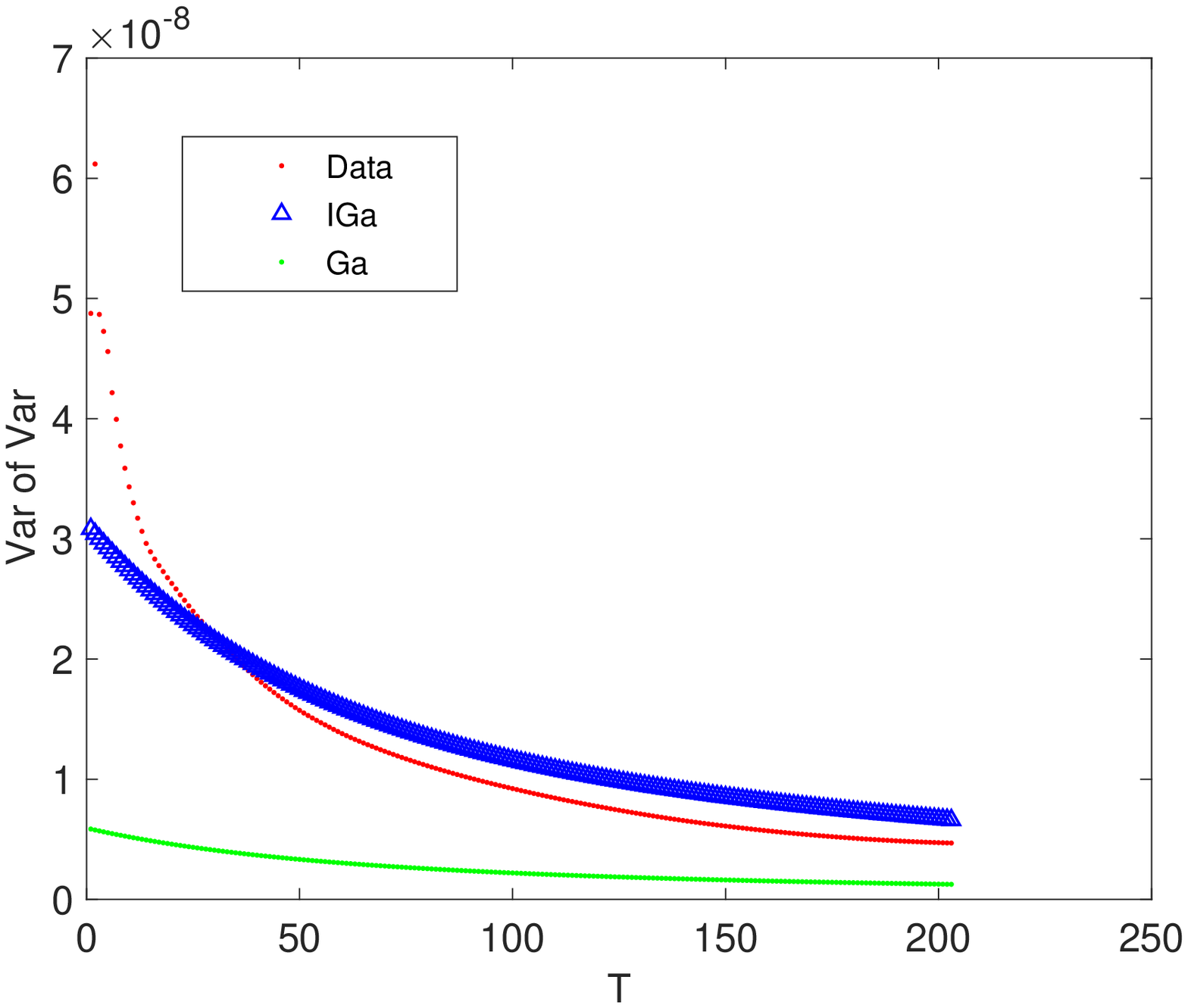}
\includegraphics[width = 0.4 \textwidth]{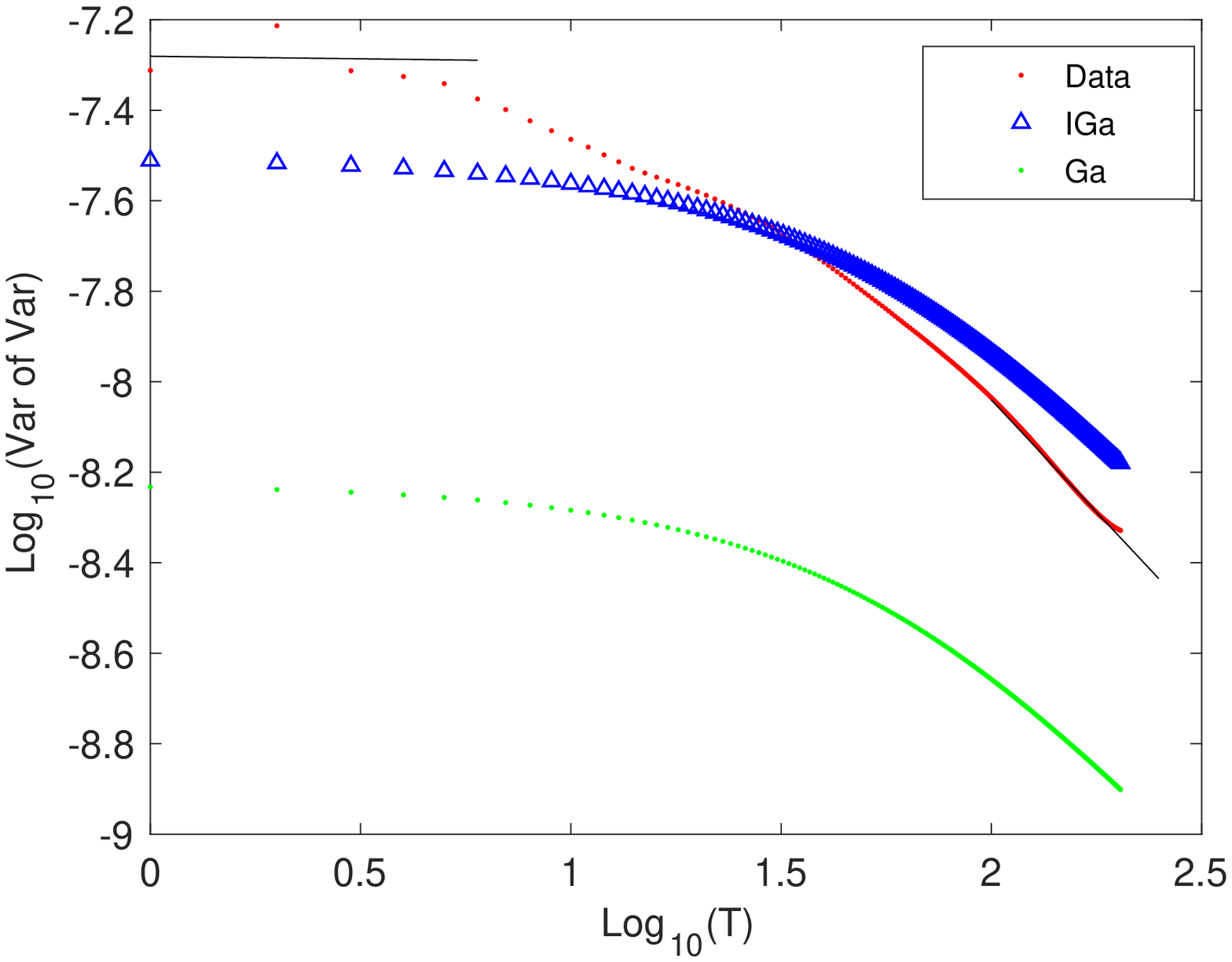}
\end{tabular}
\begin{tabular}{cc}
\includegraphics[width = 0.4 \textwidth]{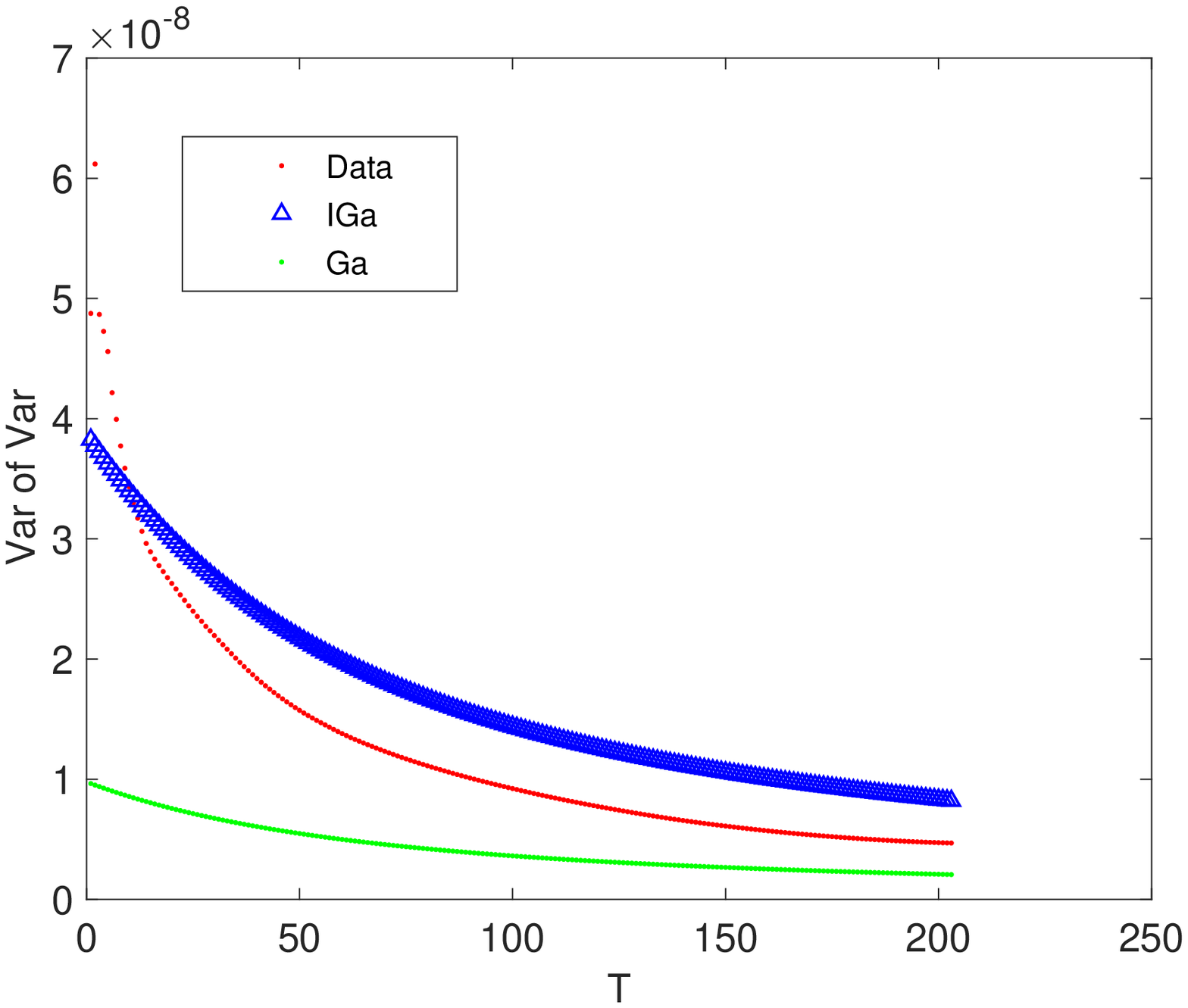}
\includegraphics[width = 0.4 \textwidth]{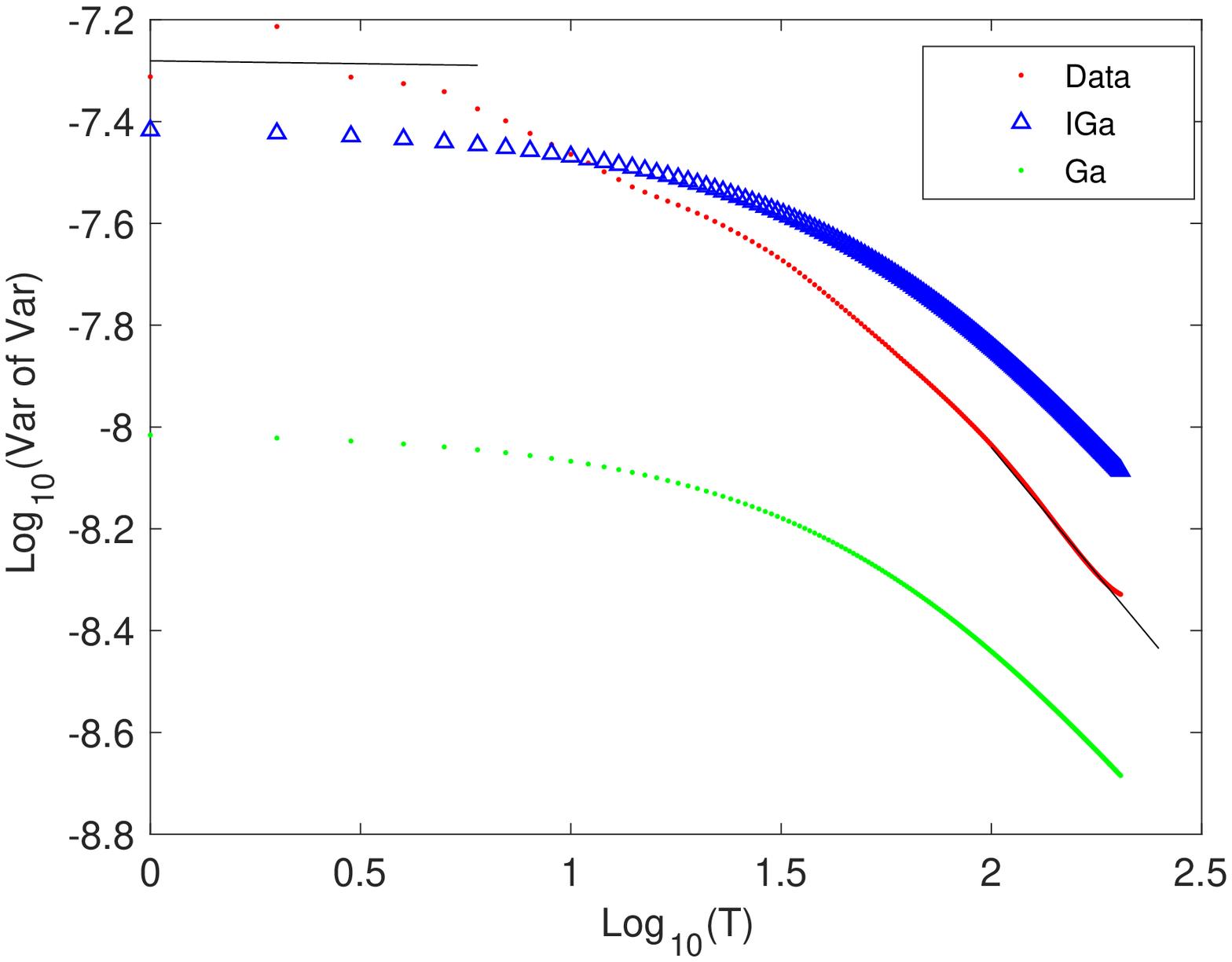}
\end{tabular}
\caption{Historic data vis-a-vis (\ref{VarVolIntegralHestresult}) and (\ref{VarVolIntegralMultresult}). Top row: with parameters from Table\ref{params}. Bottom row: with parameters from Table\ref{paramsone}. Straight lines on the right are best data fits with slopes -0.0113 and -0.992 to compare with the limiting behaviors (\ref{VarVolIntegralHestresult}), (\ref{VarVolIntegralMultresult}) and (\ref{VarVolIntegralrReducedLimits}).}
\label{VarVarTheoryData}
\end{figure}

In Fig. \ref{VarVarTheoryData} we compare the historic market data with theoretical predictions (\ref{VarVolIntegralHestresult}) and (\ref{VarVolIntegralMultresult}), including the limiting behaviors (\ref{VarVolIntegralHestresultLimits}) and (\ref{VarVolIntegralMultresultLimits}). To do so, we need to identify the values of parameters $\theta$, $\kappa$ and $\gamma$. one way to accomplish this is by fitting historic data with (\ref{theta}),  (\ref{meanVolCorHest}) and (\ref{meanVolCorMul}) and the results are summarized in Table \ref{params}. Thus found values of parameters are close to those obtained from leverage \cite{dashti2018correlations}, \cite{perello2004multiple} and by averaging the values of parameters obtained from fitting multi-day stock returns \cite{liu2017distributions}, \cite{dashti2018correlations}. Alternatively, in Table \ref{paramsone} we use parameters obtained from single-day returns  ($\tau=1$ values in \cite{liu2017distributions}, with the same $\gamma$ as in Table \ref{params}). While multiplicative model does better, a continuous model may not be appropriate for single-day returns.

\begin{table}[!htbp]
\centering
\caption{Parameters for Heston and multiplicative models for S\&P 500}
\label{params}
\begin{tabular}{ccc} 
\hline
& &\\

 & Heston model  &  \\
            $\theta$ &      $\gamma$ &         $\kappa$  \\
$9.81 \times 10^{-5}$ & 0.041 &           $2.32\times 10^{-3}$ \\[4ex]
& Multiplicative model & \\
           $\theta$ &      $\gamma$ &         $\kappa$  \\
$9.81 \times 10^{-5}$ & 0.041 &           0.25 \\
\end{tabular}
\end{table}

\begin{table}[!htbp]
\centering
\caption{Parameters for Heston and multiplicative models for S\&P 500 from single-day returns}
\label{paramsone}
\begin{tabular}{ccc} 
\hline
& &\\

 & Heston model  &  \\
            $\theta$ &      $\gamma$ &         $\kappa$  \\
$1.02 \times 10^{-4}$ & 0.041 &           $2.80\times 10^{-3}$ \\[4ex]
& Multiplicative model & \\
           $\theta$ &      $\gamma$ &         $\kappa$  \\
$1.10 \times 10^{-4}$ & 0.041 &           0.25 \\
\end{tabular}
\end{table}

\begin{figure}[!htbp]
\centering
\begin{tabular}{cc}
\includegraphics[width = 0.4 \textwidth]{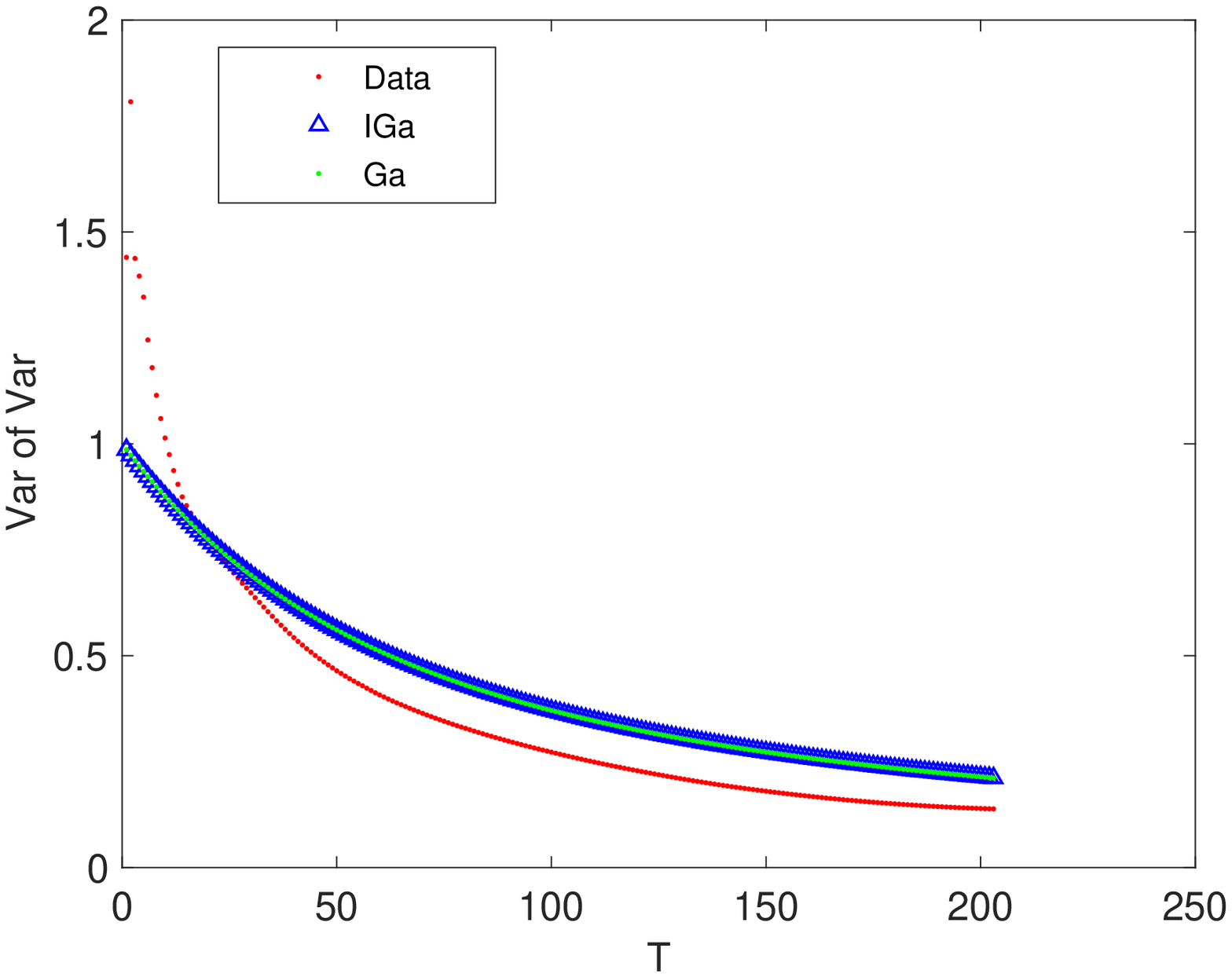}
\includegraphics[width = 0.4 \textwidth]{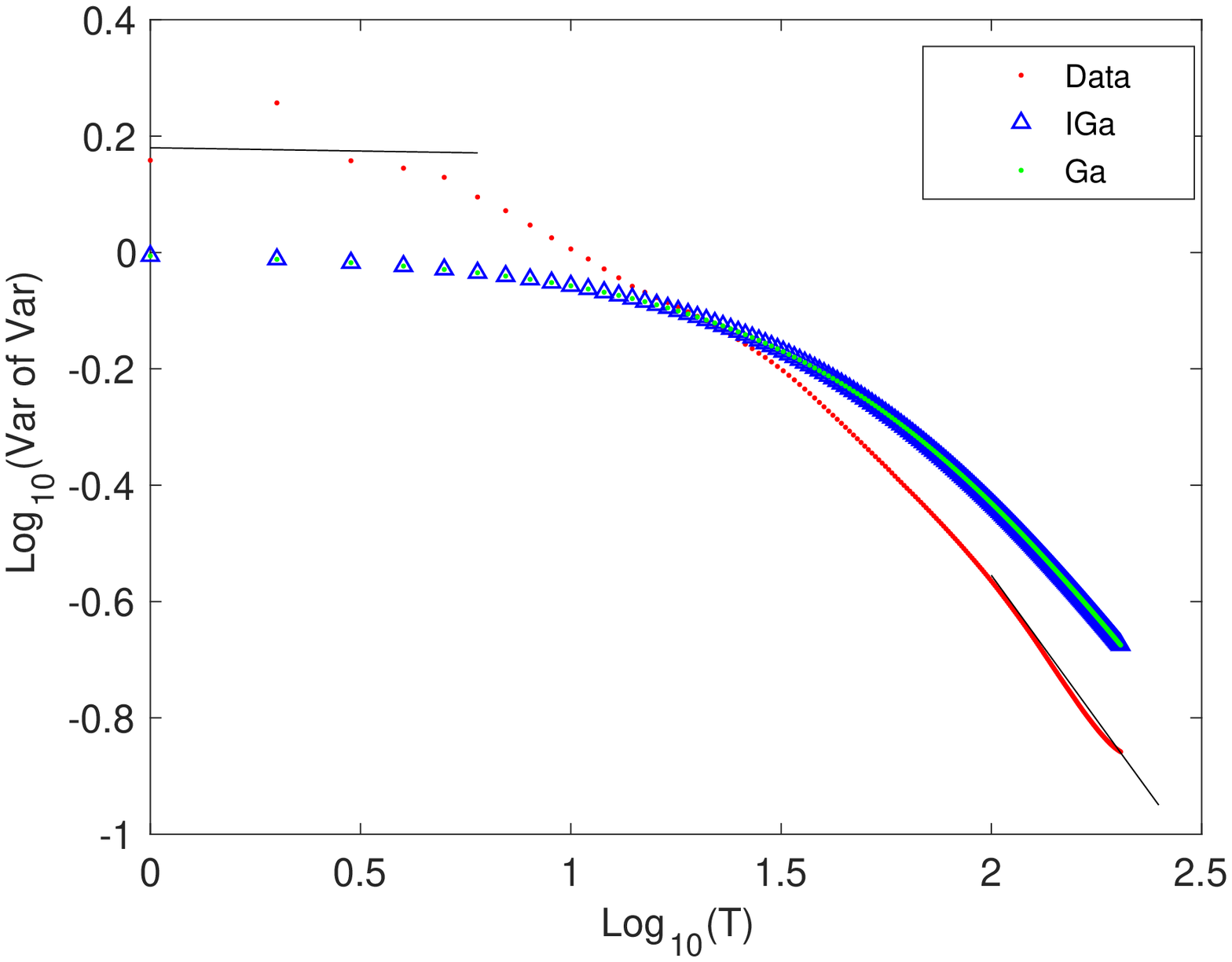}
\end{tabular}
\caption{Historic data vis-a-vis (\ref{VarVolIntegralrReduced}). Straight lines on the right have the same slopes as in Fig. \ref{VarVarTheoryData}.}
\label{VarVarTheoryDataReduced}
\end{figure}

In Fig. \ref{VarVarTheoryDataReduced} the comparison between theory and data is performed for the universal expression (\ref{VarVolIntegralrReduced}), including the limiting behaviors (\ref{VarVolIntegralrReducedLimits}). Clearly, agreement between theory and data in  Fig. \ref{VarVarTheoryDataReduced} is quite good, while it is not as close in Fig. \ref{VarVarTheoryData}. The reason is that former depends only on one parameter $\gamma$, which is well established correlation and relaxation scale for mean reverting models \cite{liu2017distributions}-\cite{liu2017correlation}.

\clearpage

\section{Conclusions}

Previously \cite{liu2017distributions} we showed that the multi-day -- typically longer than several weeks, to account for relaxation processes -- stock returns are better described by the Heston model. Realized volatility, conversely, is calculated from single-day returns, which are presumably on the borderline between intra-day jump processes and continuous processes, such as Heston and multiplicative. Furthermore, realized volatility is square root of the realized variance, which is a sum of roughly 21 realized daily variances representing the trading month. As such, it is clear that the realized variance should be approaching normal distribution. Figs. \ref{sumri2}-\ref{KSsumri2} reflect this conjecture, with the caveat that the distribution maintains a tail similar to that of a single-day variance distribution. 

Regarding the latter, we showed that an exGaussian distribution provides an excellent fit. This, however does not necessarily indicate that the tail is in fact exponential. An exGaussian is a sum of normal and exponential distributions and is an artificial construct, where all parameters are shape parameters and the distribution rescales only when all parameters rescale simultaneously. On the other hand, exGaussian has the right properties for the realized variance distribution and can be described analytically. It is quite possible that a distribution exists, which combines normality with a heavy, perhaps fat, tail that gives comparable Kolmogorov-Smirnov statistics to exGaussian. This is something we intend to study in the future.

With respect to comparison of the realized volatility to predictions of volatility indices, we found no discernible advantages between VIX and VXO. We also found that the ratio of the realized variance to squared VIX and VXO is best fitted by a fat-tailed (power-law) distribution -- in this case Inverse Gamma. This most likely reflects large unexpected spikes of realized volatility not foreseen by volatility indices. The inverse distribution, which is the distribution of the inverse variable, is best fitted with an exponentially decaying distribution -- in this case Gamma. This reflects no unexpected surges in volatility indices relative to realized volatility. When we use realized volatility of the preceding month, however, the ratio distributions are all best fitted with lognormal distribution. This most likely reflects the fact that while we predict future volatility based on what we presently know (past realized volatility and current information \cite{christensen1998relation}), large spikes in volatility still result in heavy lognormal tails of the ratios due to uncertainty associated with such spikes. We will discuss relationship between volatility spikes and tails in a future work.

Finally,  evaluation of the theoretical dependence of variance of realized variance and its limiting behaviors compares well vis-a-vis historic data, with the prediction of the multiplicative model having an edge over the Heston model. The latter may be due that single-day returns, on which realized variance is based, are better described by the Student's distribution \cite{liu2017distributions}, \cite{fuentes2009universal}. Interestingly, 21 days, over which the realized variance is calculated, is also roughly the relaxation time of stochastic volatility, $\gamma^{-1}$. Additionally, the mean value of the volatility in the mean-reverting models may be stochastic itself \cite{perello2004multiple} -- with similar time scales -- which may as well affect the comparison. Ideally, one would also want to study the higher moments of the realized variance distribution. The difficulty, however, is that decoupling the higher-order correlation functions of stochastic variance into pair-wise correlation functions leads trivially to a normal distribution, which is not the case per above. This is a challenging problem that also needs to be addressed. 

\clearpage

\appendix

\section{VIX Current and VXO Current}
Here we split 2003-2016 data in two roughly equal time periods, before and after the financial crisis.

\subsection{Visual Comparison}
\begin{figure}[!htbp]
\centering
\begin{tabular}{cc}
\includegraphics[width = 0.49 \textwidth]{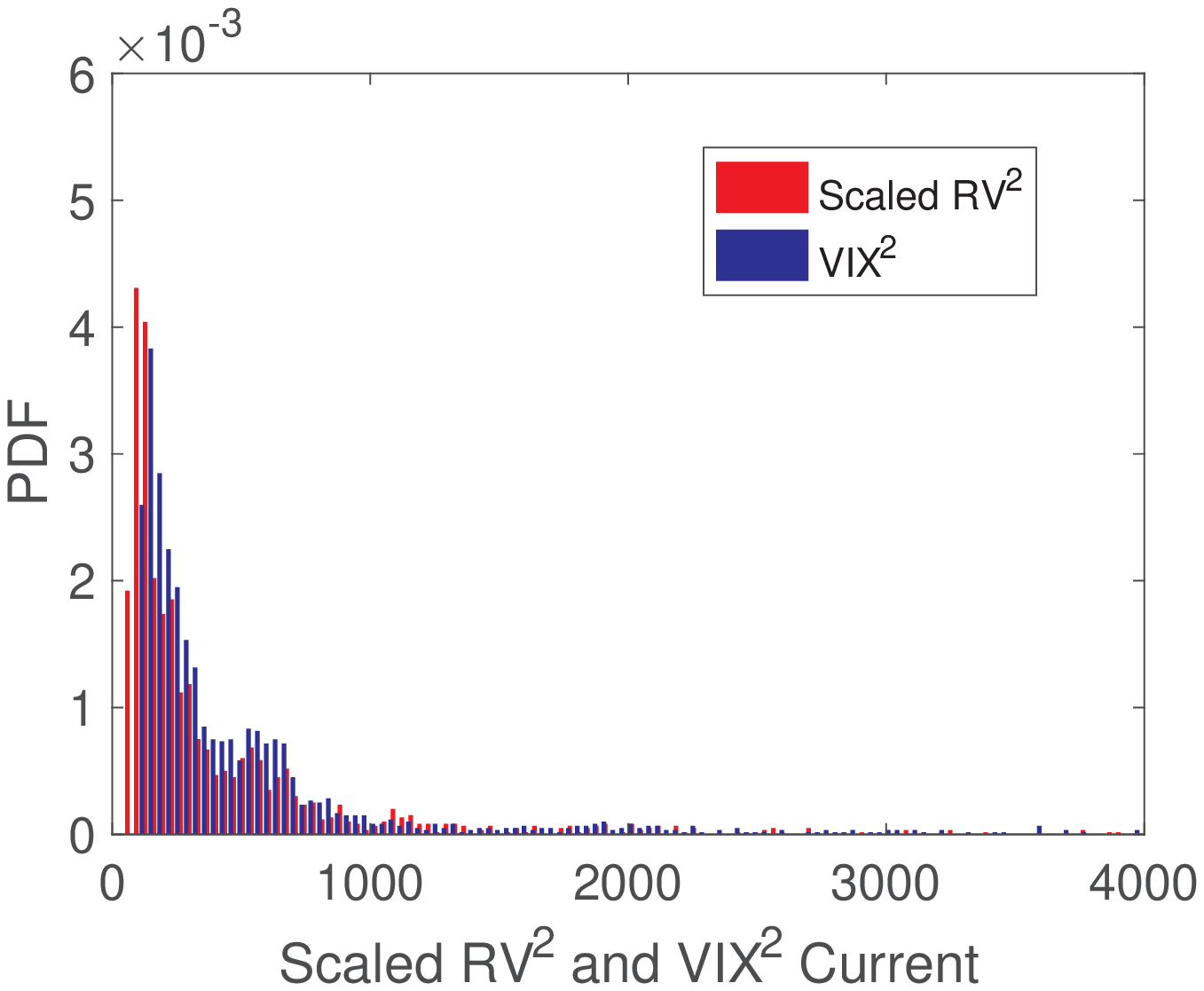} \hspace{.2cm}
\includegraphics[width = 0.49 \textwidth]{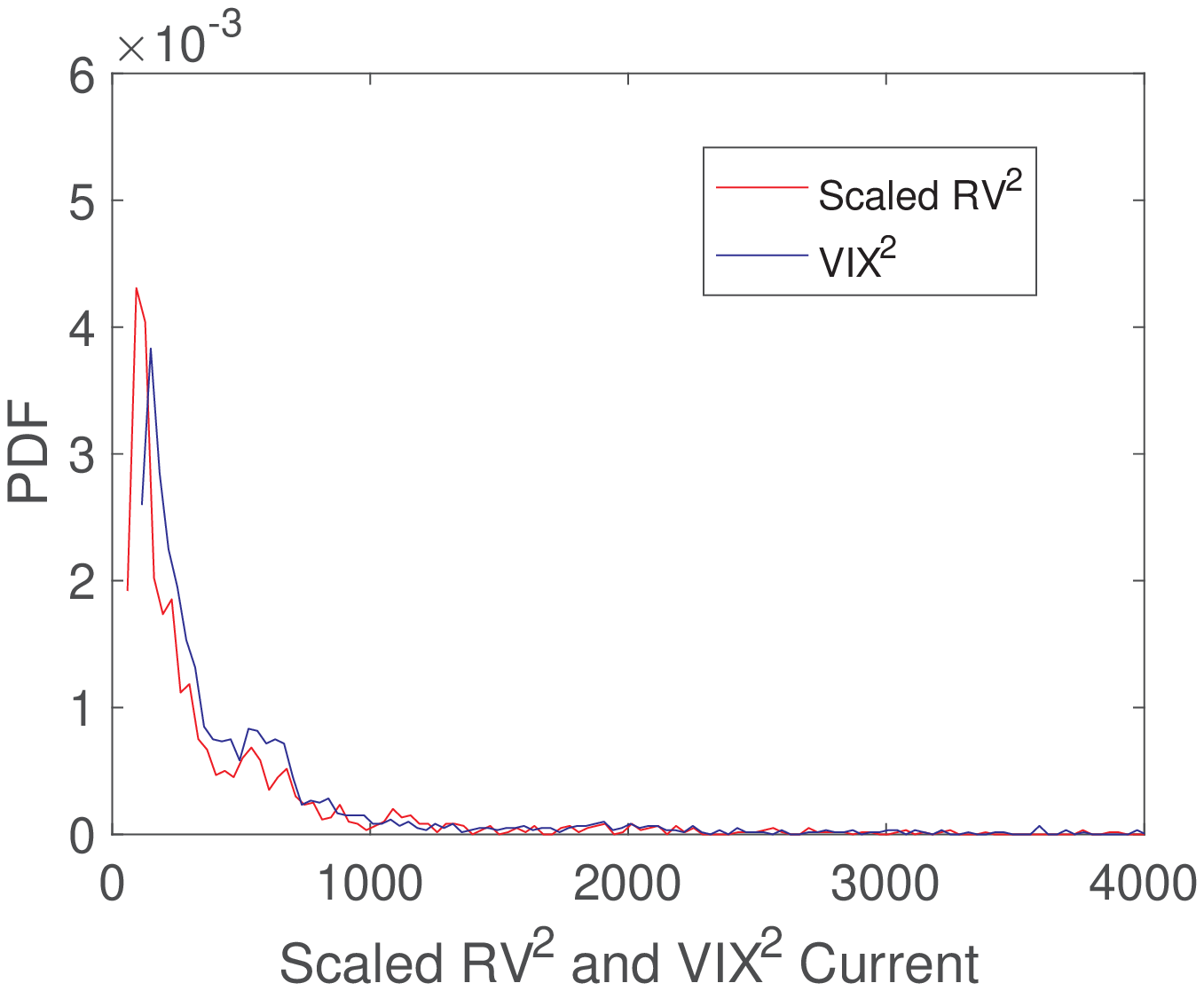}
\end{tabular}
\caption{PDFs of scaled $RV^2$ and  $VIX^2$ from Sep 22nd, 2003 to Aug 30th, 2010.}
\label{histogramScaleRVVIX20032010ScaleRVVIXHistogram}
\end{figure}

\begin{figure}[!htbp]
\centering
\begin{tabular}{cc}
\includegraphics[width = 0.49 \textwidth]{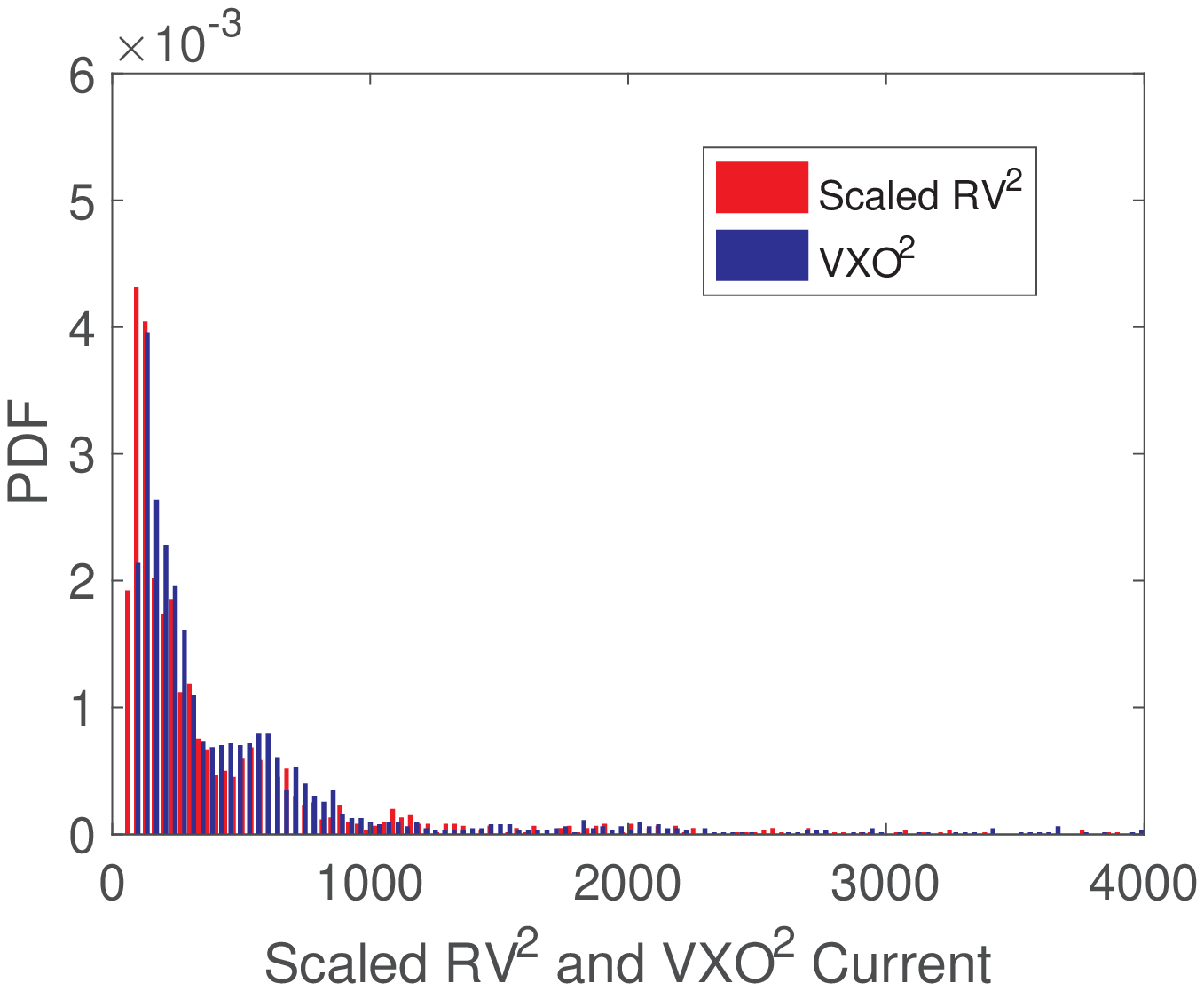} \hspace{.2cm}
\includegraphics[width = 0.49 \textwidth]{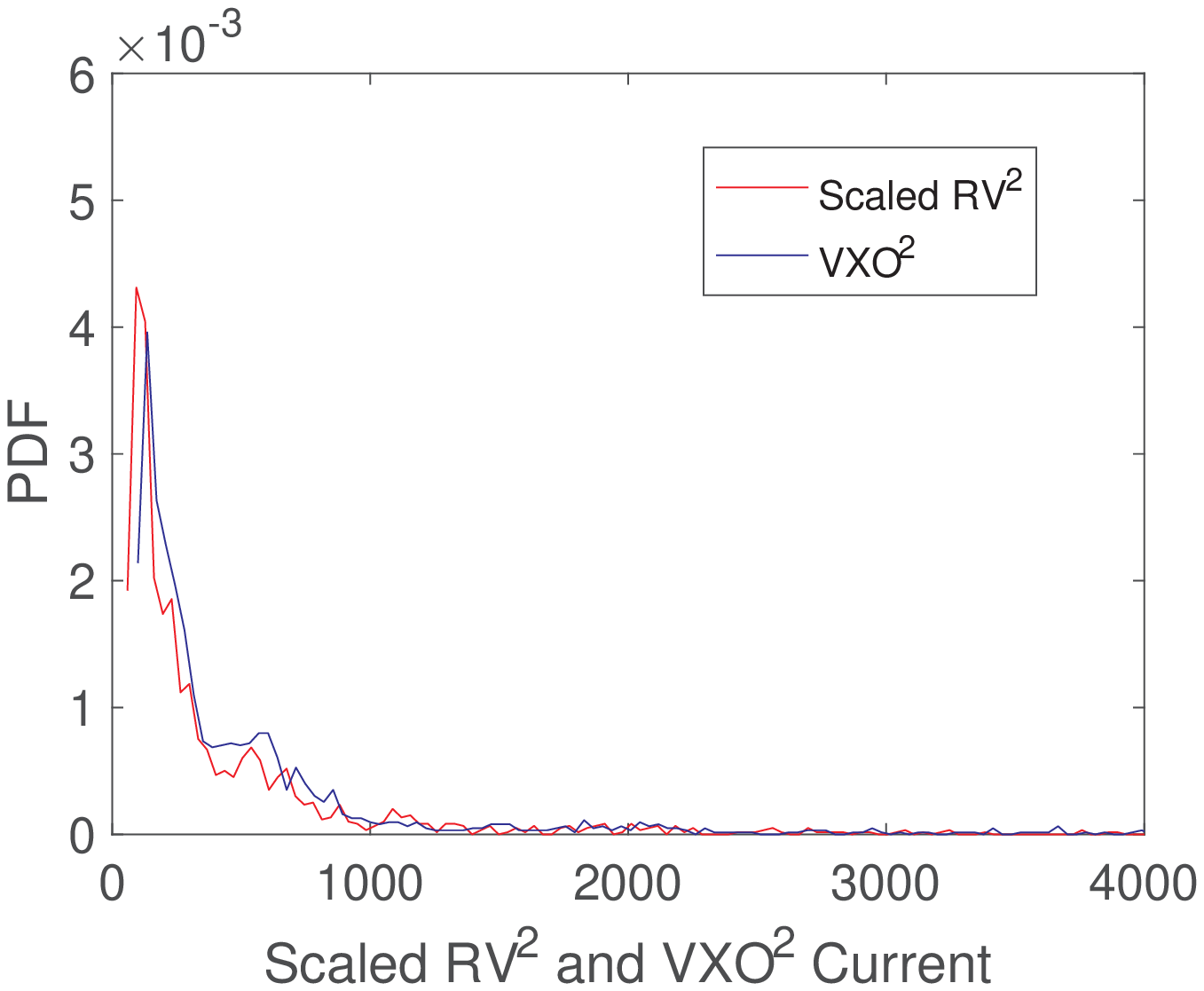}
\end{tabular}
\caption{PDFs of scaled $RV^2$ and  $VXO^2$ from Sep 22nd, 2003 to Aug 30th, 2010.}
\label{histogramScaleRVVXO20032010ScaleRVVXOHistogram}
\end{figure}

\begin{figure}[!htbp]
\centering
\begin{tabular}{cc}
\includegraphics[width = 0.49 \textwidth]{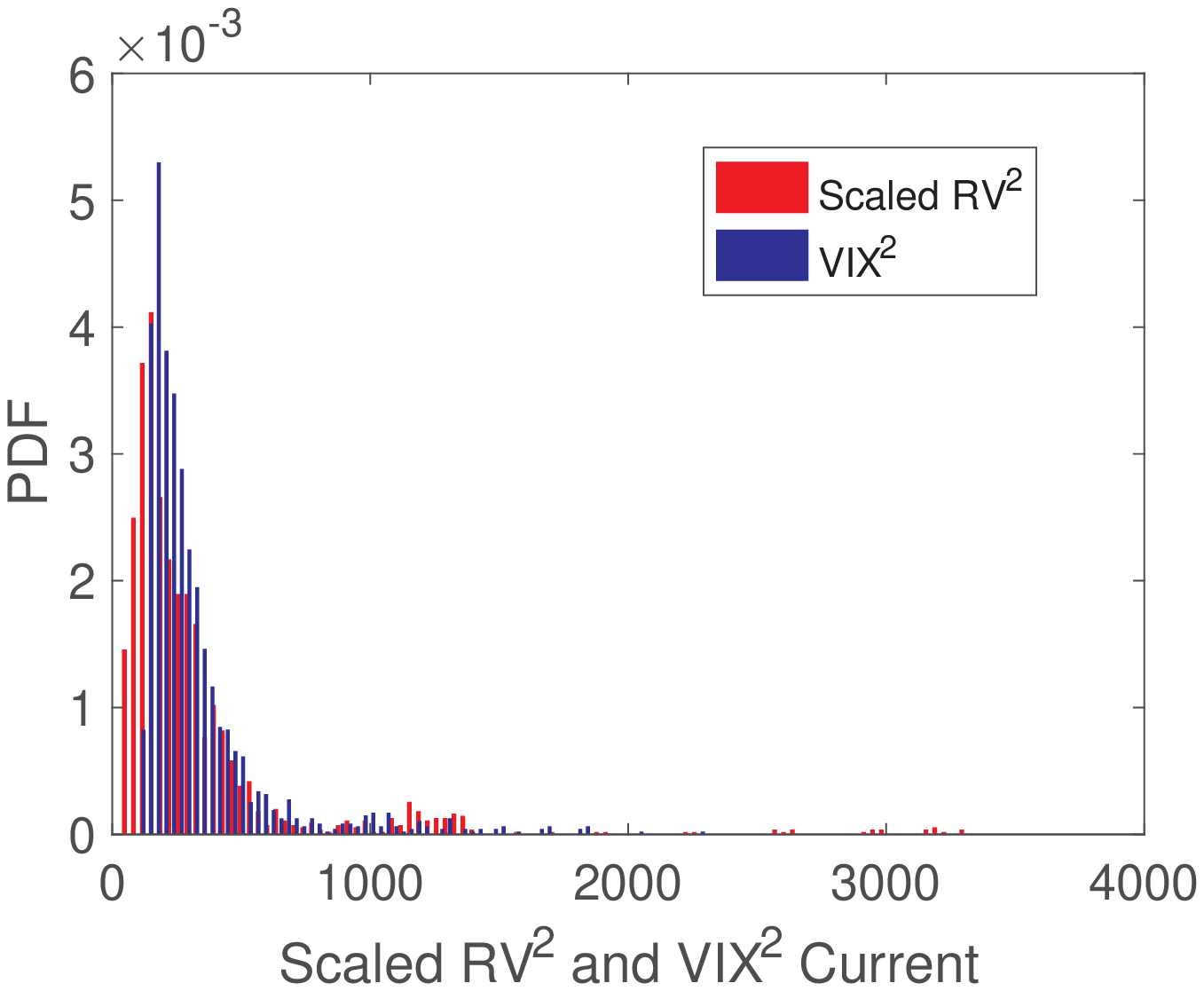} \hspace{.2cm}
\includegraphics[width = 0.49 \textwidth]{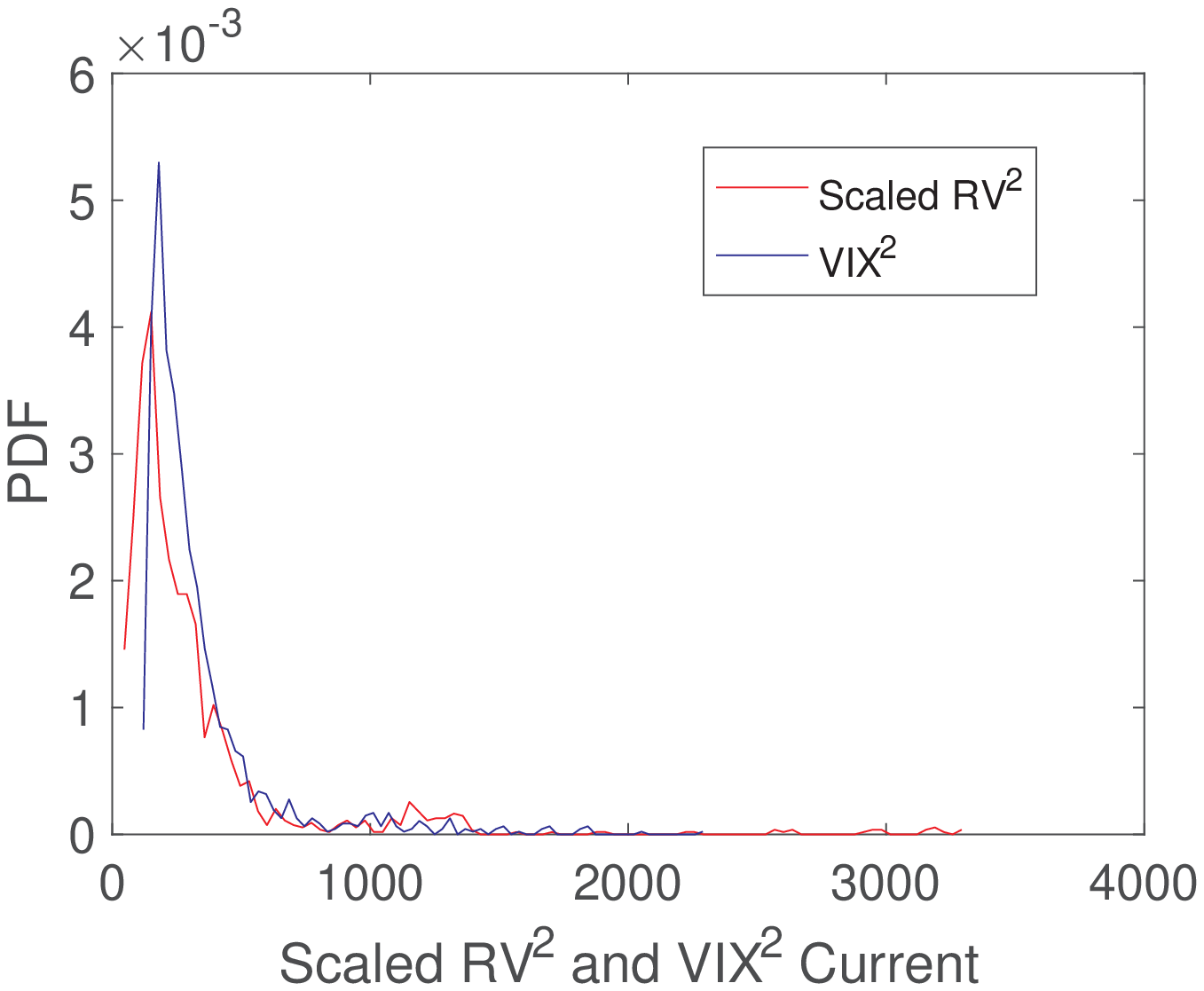}
\end{tabular}
\caption{PDFs of scaled $RV^2$ and  $VIX^2$ from Aug 31st, 2010 to Dec 30th, 2016.}
\label{histogramScaleRVVIX20102016ScaleRVVIXHistogram}
\end{figure}

\begin{figure}[!htbp]
\centering
\begin{tabular}{cc}
\includegraphics[width = 0.49 \textwidth]{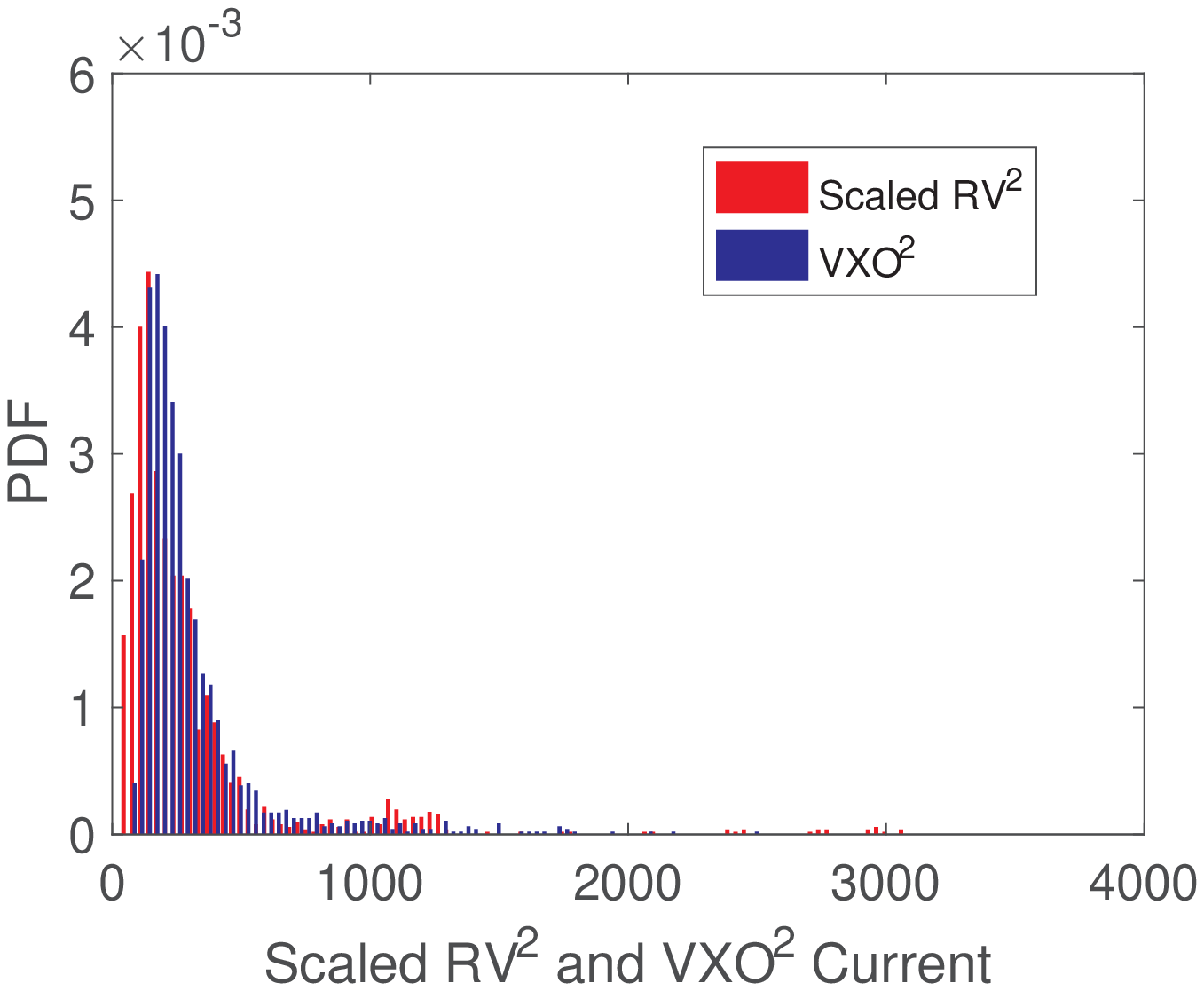} \hspace{.2cm}
\includegraphics[width = 0.49 \textwidth]{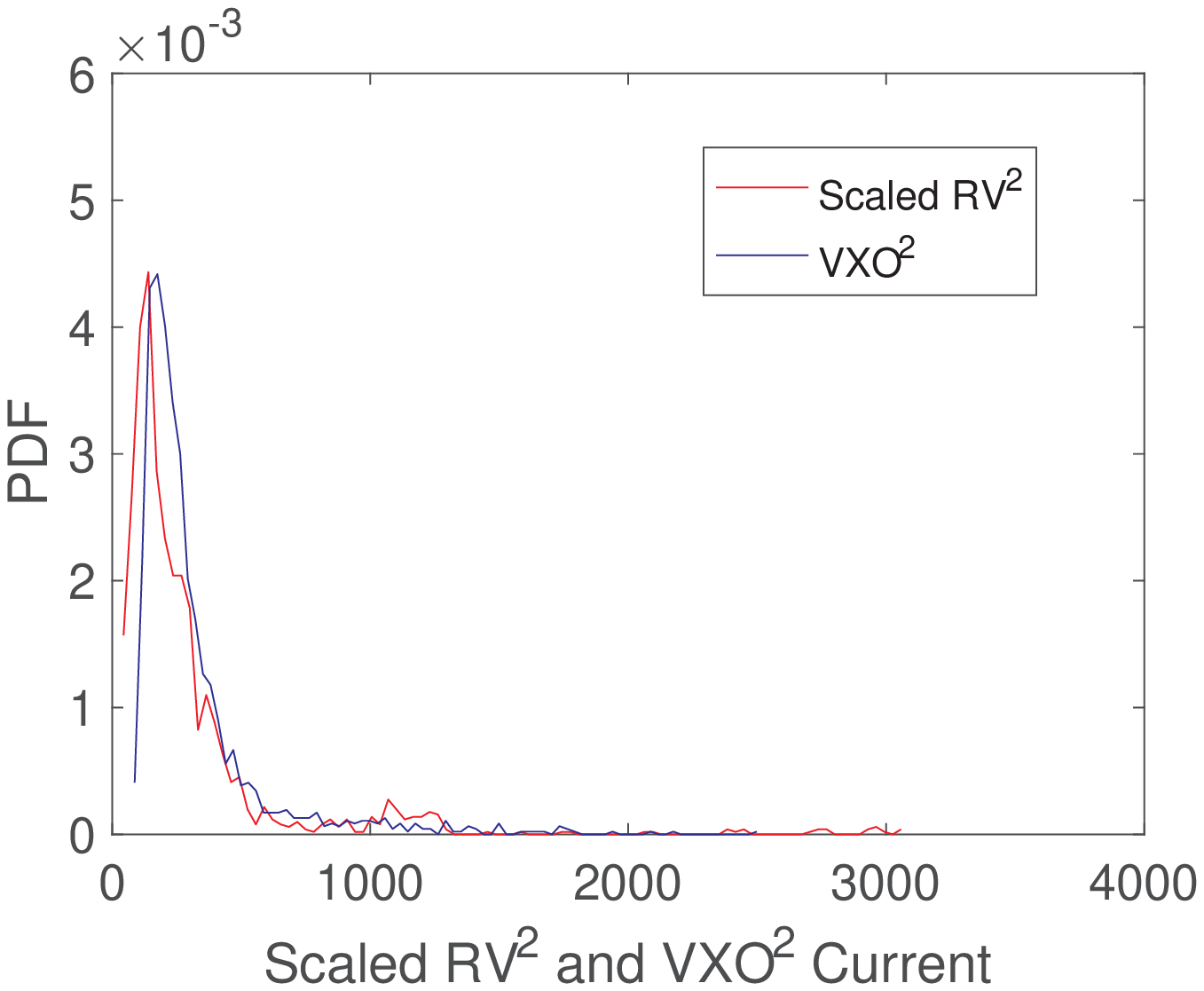}
\end{tabular}
\caption{PDFs of scaled $RV^2$ and  $VXO^2$ from Aug 31st, 2010 to Dec 30th, 2016.}
\label{histogramScaleRVVXO20102016ScaleRVVXOHistogram}
\end{figure}

\clearpage

\subsection{Ratio Distribution}
\begin{figure}[!htbp]
\centering
\begin{tabular}{cc}
\includegraphics[width = 0.4 \textwidth]{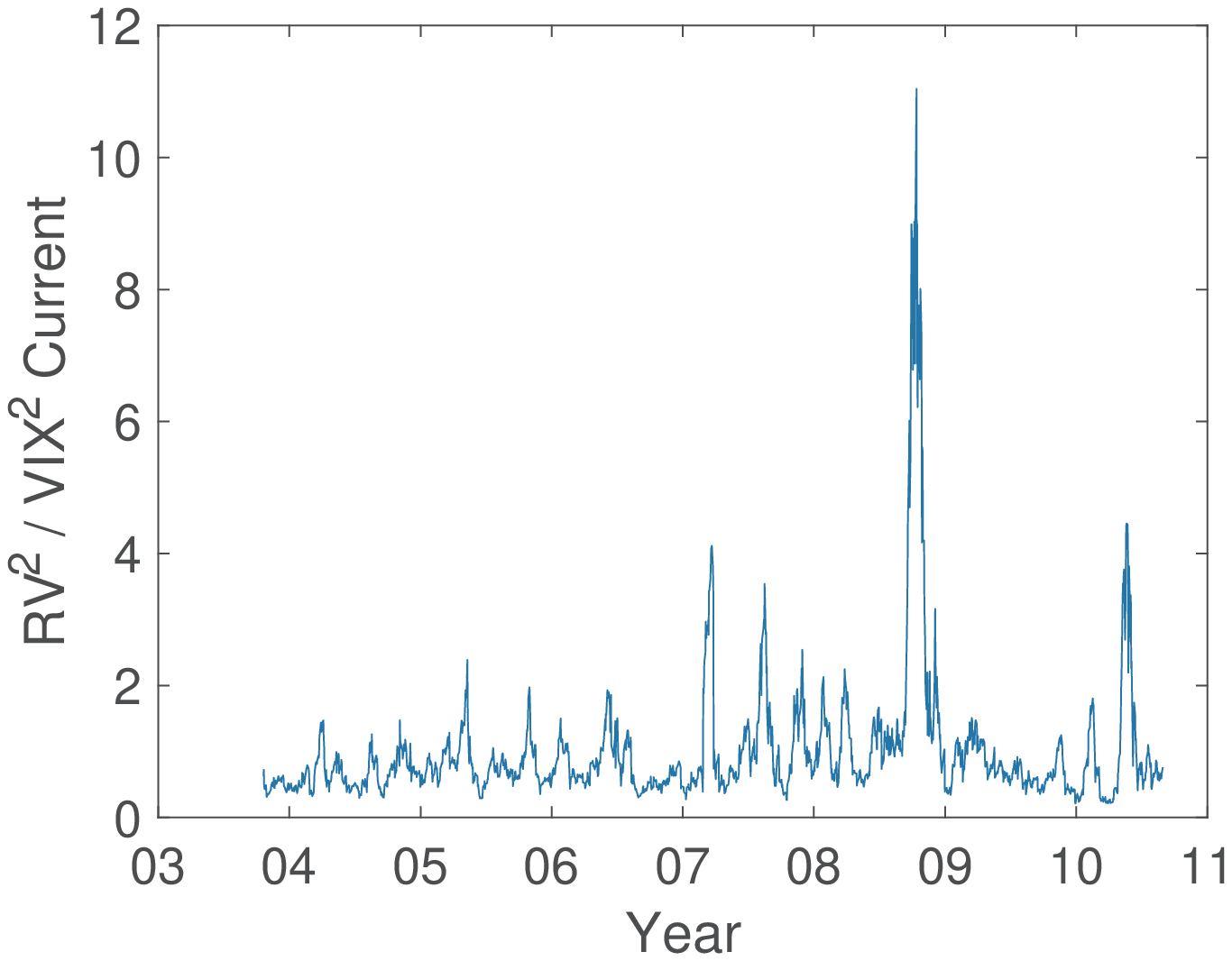} \hspace{.2cm}
\includegraphics[width = 0.4 \textwidth]{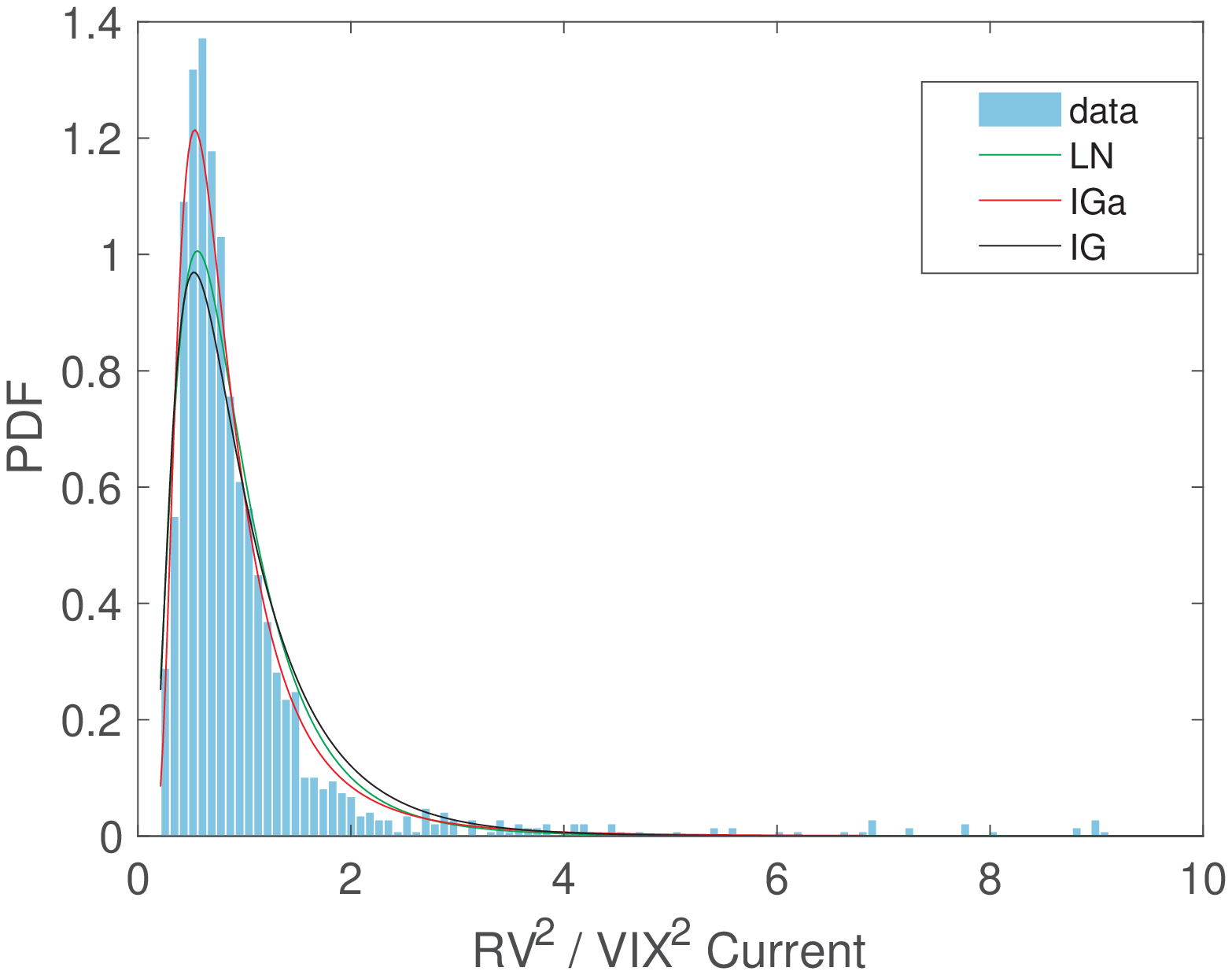}
\end{tabular}
\caption{$\mathrm{RV}^2 / \mathrm{VIX}^2$, from Sep 22nd, 2003 to Aug 30th, 2010.}
\label{RVOverVIXListSRV2OverVIX22003}
\end{figure}

\begin{figure}[!htbp]
\centering
\begin{tabular}{cc}
\includegraphics[width = 0.4 \textwidth]{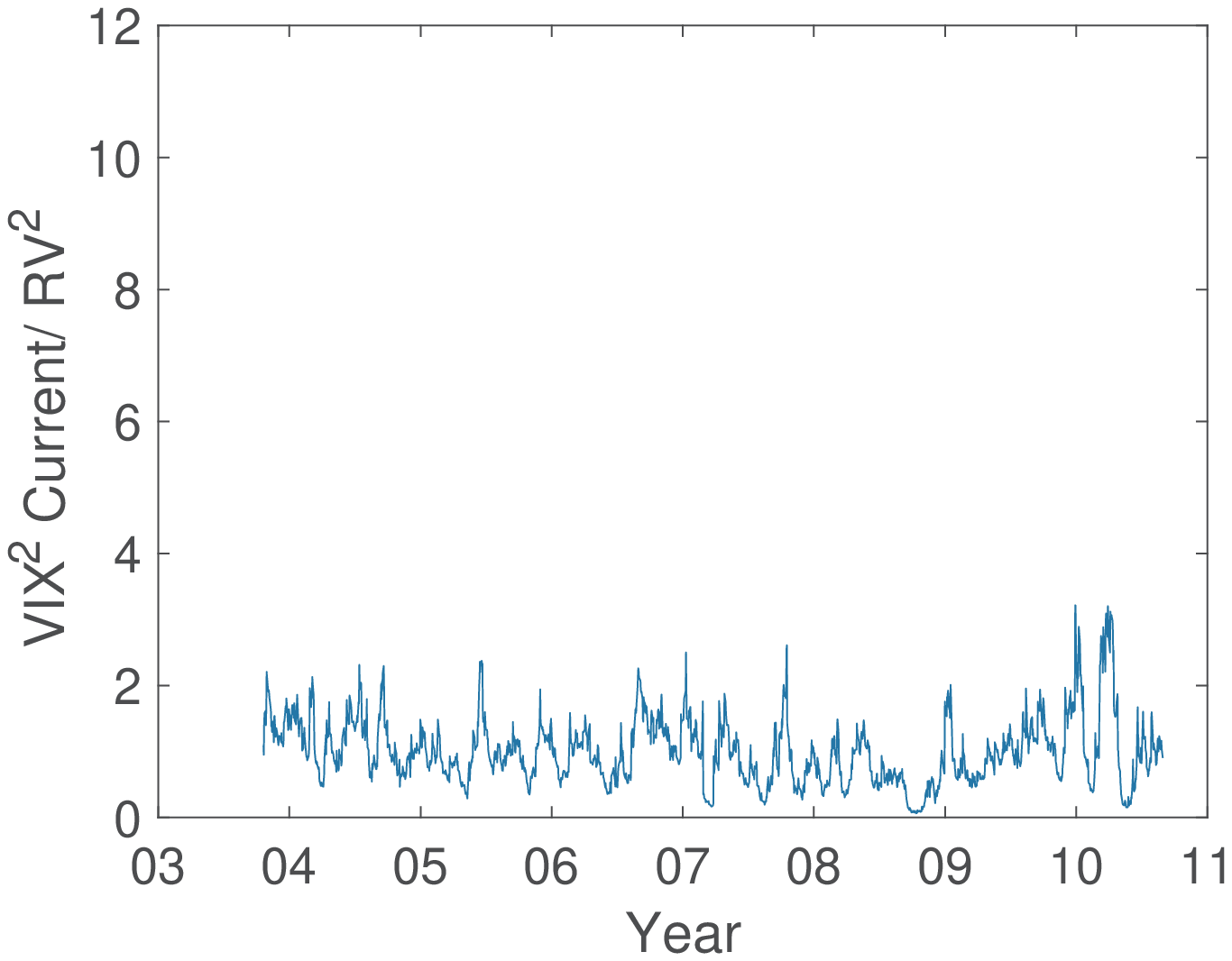} \hspace{.2cm}
\includegraphics[width = 0.4 \textwidth]{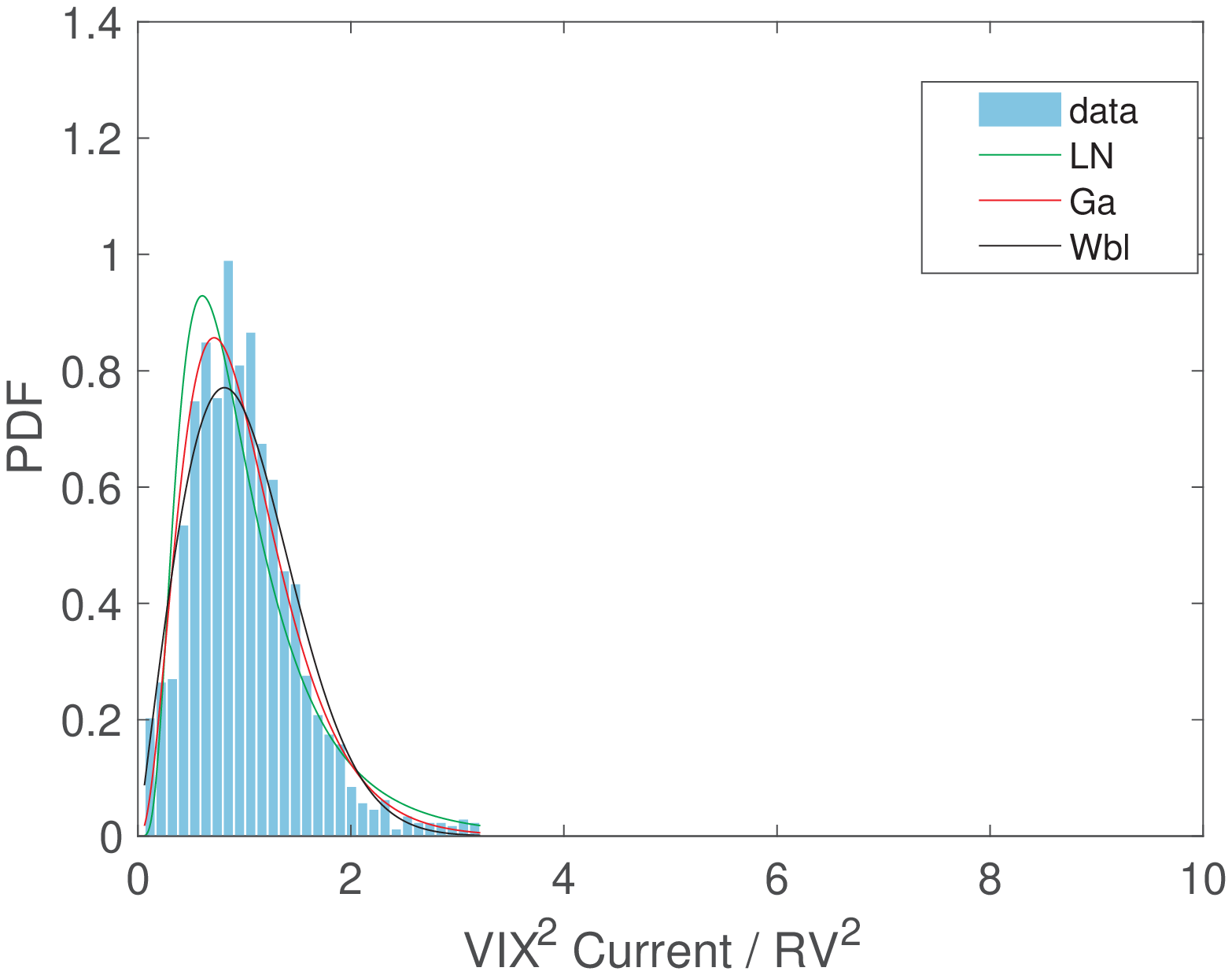}
\end{tabular}
\caption{$ \mathrm{VIX}^2 / \mathrm{RV}^2$, from Sep 22nd, 2003 to Aug 30th, 2010.}
\label{VIXOverRVListSVIX2OverRV22003n}
\end{figure}

\begin{table}[!htbp]
\caption{MLE results for ``$\mathrm{RV}^2 / \mathrm{VIX}^2$" and ``$\mathrm{VIX}^2 / \mathrm{RV}^2$"}
\label{MLESRV2OverVIX22003n}
\begin{minipage}{0.5\textwidth}
\begin{center}
\begin{tabular}{ c c c} 
\multicolumn{2}{c}{} \\
\hline
            type &       parameters &          KS Statistic  \\
\hline
Normal & N(          1.0000,           1.0244) &           0.2338 \\
\hline
LogNormal & LN(         -0.2280,           0.5943) &           0.0763 \\
\hline
IGa & IGa(          3.5292,           2.4224) &           0.0383 \\
\hline
Gamma & Gamma(          2.3463,           0.4262) &           0.1297 \\
\hline
Weibull & Weibull(          1.1013,           1.3060) &           0.1588 \\
\hline
IG & IG(          1.0000,           2.1887) &           0.0969 \\
\hline
\end{tabular}
\end{center}
\end{minipage}
\begin{minipage}{.5\textwidth}
\begin{center}
\begin{tabular}{ c c c} 
\multicolumn{2}{c}{} \\
\hline
            type &       parameters &          KS Statistic  \\
\hline
Normal & N(          1.0000,           0.5124) &           0.0613 \\
\hline
LogNormal & LN(         -0.1483,           0.5943) &           0.0763 \\
\hline
IGa & IGa(          2.3463,           1.6105) &           0.1297 \\
\hline
Gamma & Gamma(          3.5292,           0.2833) &           0.0383 \\
\hline
Weibull & Weibull(          1.1290,           2.0447) &           0.0392 \\
\hline
IG & IG(          1.0000,           2.1887) &           0.1128 \\
\hline
\end{tabular}
\end{center}
  \end{minipage}

\end{table}

\newpage

\begin{figure}[!htbp]
\centering
\begin{tabular}{cc}
\includegraphics[width = 0.4 \textwidth]{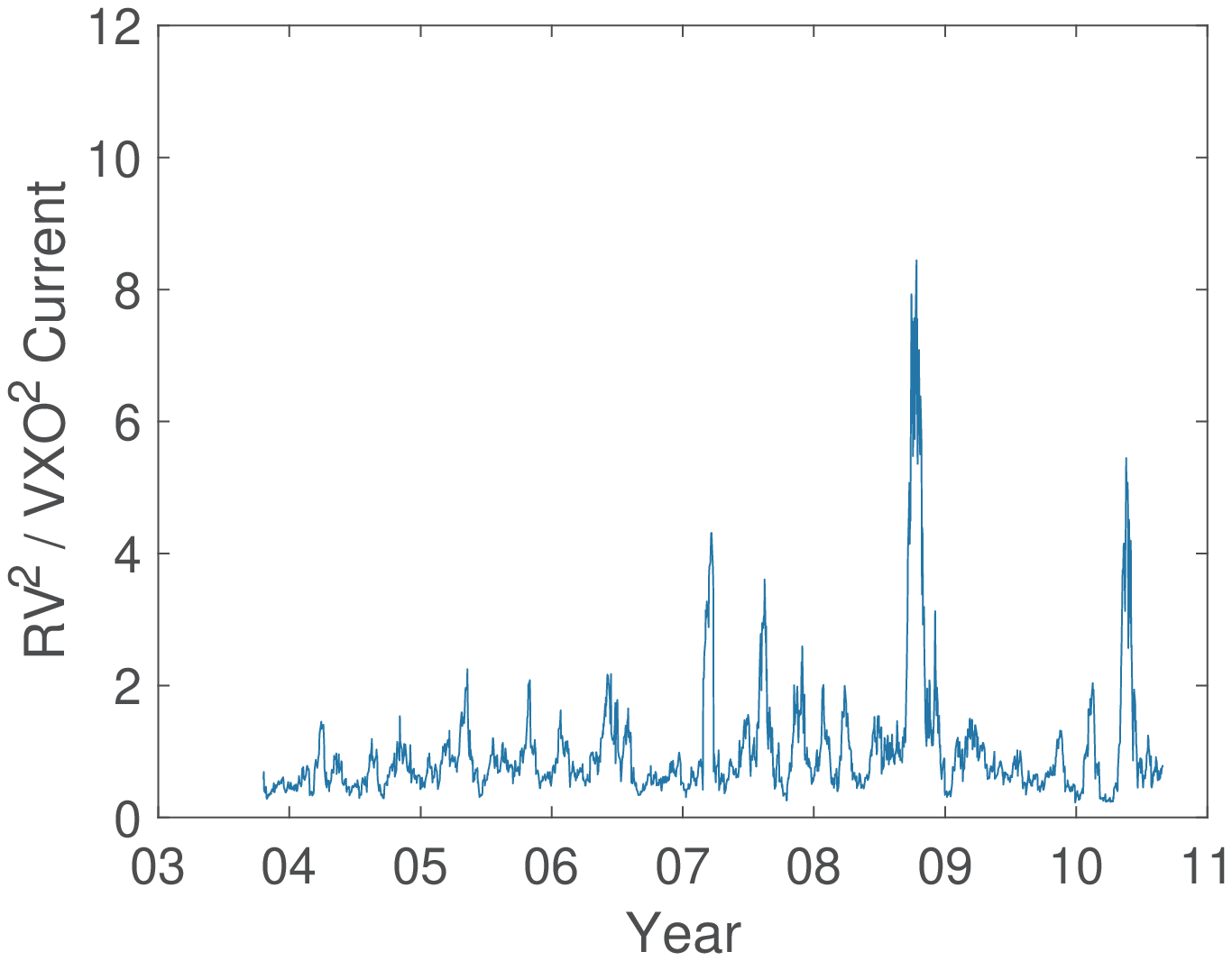} \hspace{.2cm}
\includegraphics[width = 0.4 \textwidth]{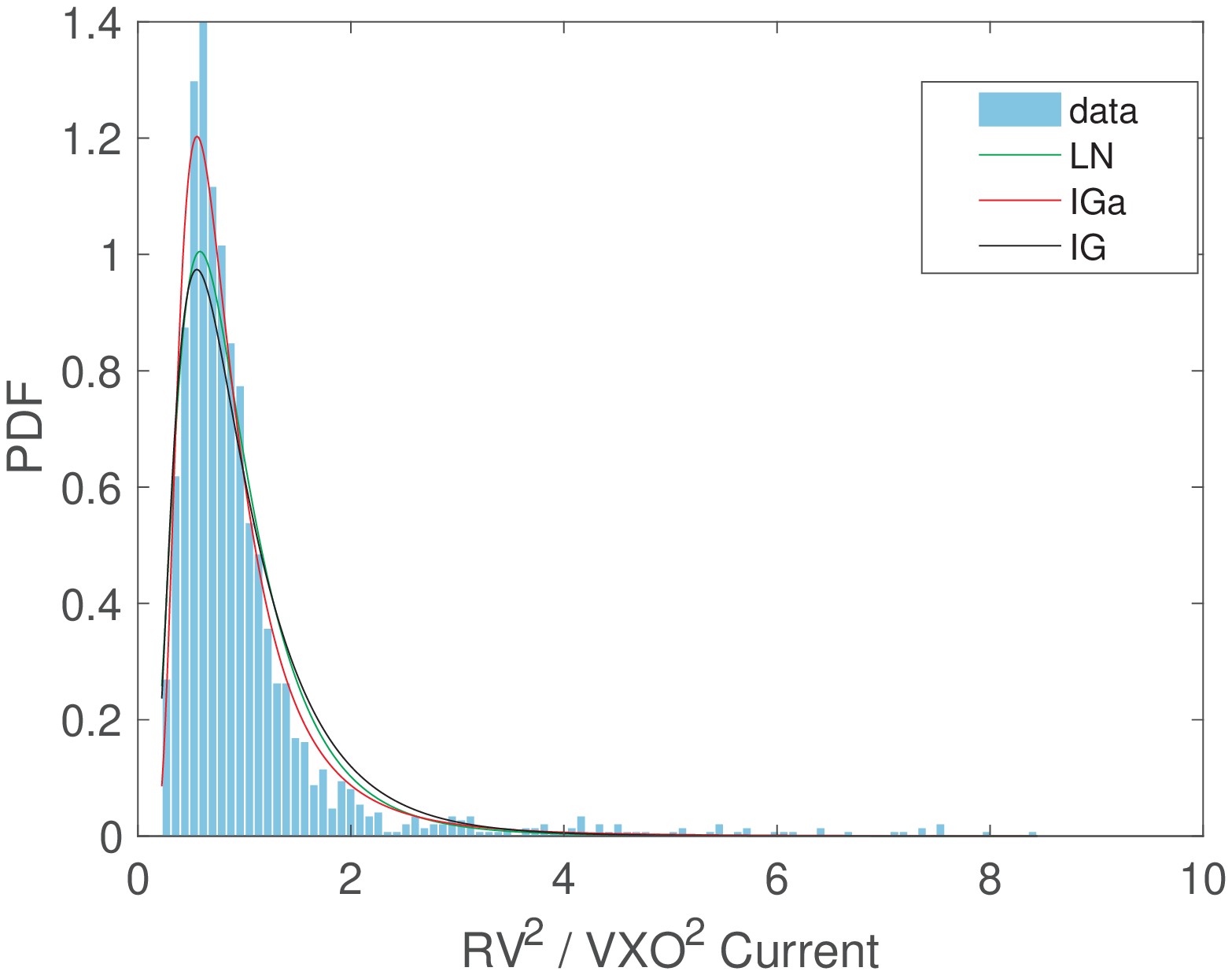}
\end{tabular}
\caption{$\mathrm{RV}^2 / \mathrm{VXO}^2$, from Sep 22nd, 2003 to Aug 30th, 2010.}
\label{RVOverVXOListSRV2OverVXO22003}
\end{figure}

\begin{figure}[!htbp]
\centering
\begin{tabular}{cc}
\includegraphics[width = 0.4 \textwidth]{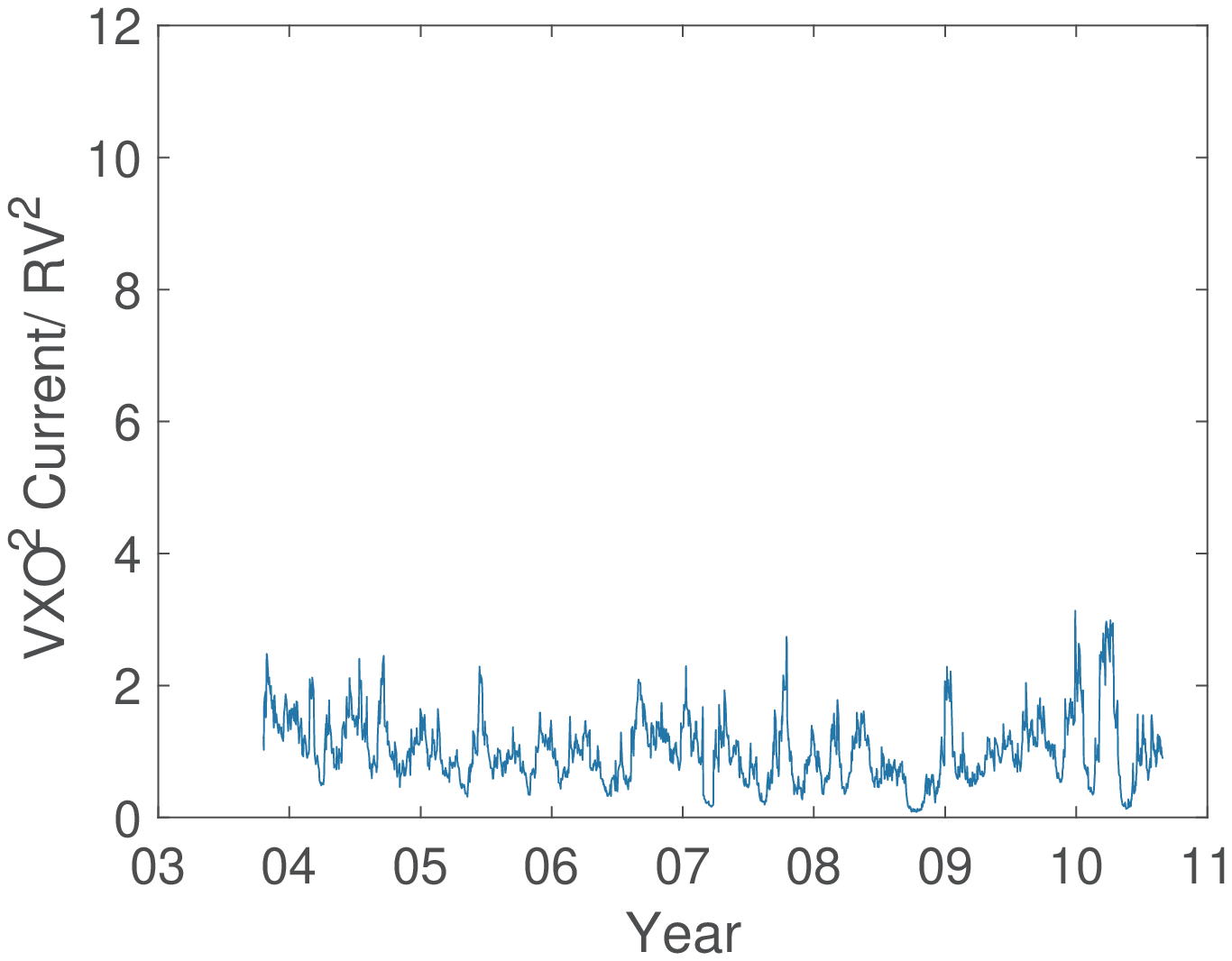} \hspace{.2cm}
\includegraphics[width = 0.4 \textwidth]{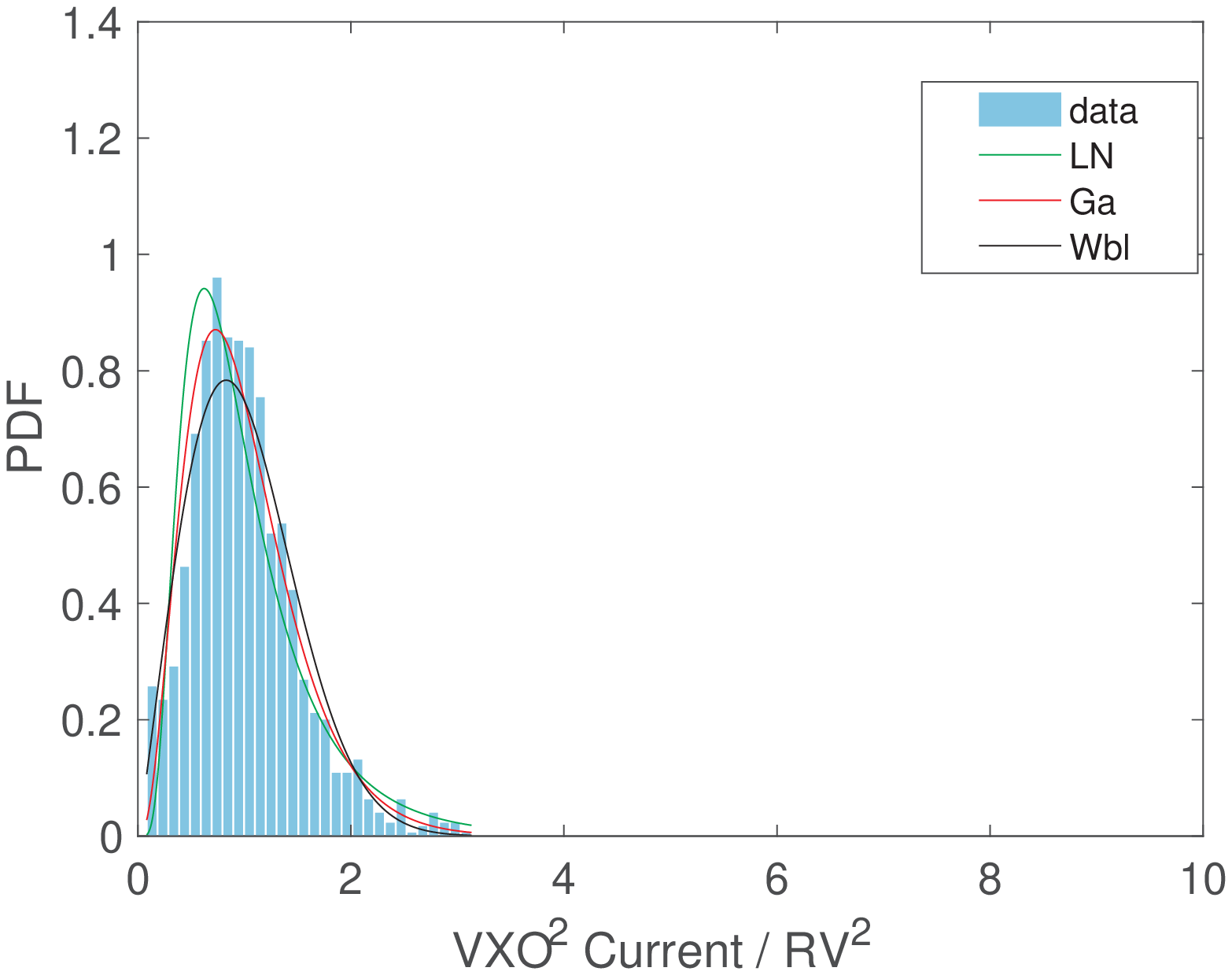}
\end{tabular}
\caption{$ \mathrm{VXO}^2 / \mathrm{RV}^2$, from Sep 22nd, 2003 to Aug 30th, 2010.}
\label{VXOOverRVListSVXO2OverRV22003n}
\end{figure}

\begin{table}[!htbp]
\caption{MLE results for ``$\mathrm{RV}^2 / \mathrm{VXO}^2$" and ``$\mathrm{VXO}^2 / \mathrm{RV}^2$"}
\label{MLESRV2OverVXO22003n}
\begin{minipage}{0.5\textwidth}
\begin{center}
\begin{tabular}{ c c c} 
\multicolumn{2}{c}{} \\
\hline
            type &       parameters &          KS Statistic  \\
\hline
Normal & N(          1.0000,           0.9130) &           0.2237 \\
\hline
LogNormal & LN(         -0.2069,           0.5763) &           0.0715 \\
\hline
IGa & IGa(          3.6954,           2.6086) &           0.0358 \\
\hline
Gamma & Gamma(          2.5708,           0.3890) &           0.1274 \\
\hline
Weibull & Weibull(          1.1119,           1.3839) &           0.1435 \\
\hline
IG & IG(          1.0000,           2.4002) &           0.0922 \\
\hline
\end{tabular}
\end{center}
\end{minipage}
\begin{minipage}{.5\textwidth}
\begin{center}
\begin{tabular}{ c c c} 
\multicolumn{2}{c}{} \\
\hline
            type &       parameters &          KS Statistic  \\
\hline
Normal & N(          1.0000,           0.5009) &           0.0635 \\
\hline
LogNormal & LN(         -0.1414,           0.5763) &           0.0715 \\
\hline
IGa & IGa(          2.5708,           1.8147) &           0.1274 \\
\hline
Gamma & Gamma(          3.6954,           0.2706) &           0.0358 \\
\hline
Weibull & Weibull(          1.1296,           2.0966) &           0.0375 \\
\hline
IG & IG(          1.0000,           2.4002) &           0.1055 \\
\hline
\end{tabular}
\end{center}
  \end{minipage}

\end{table}

\newpage

\begin{figure}[!htbp]
\centering
\begin{tabular}{cc}
\includegraphics[width = 0.4 \textwidth]{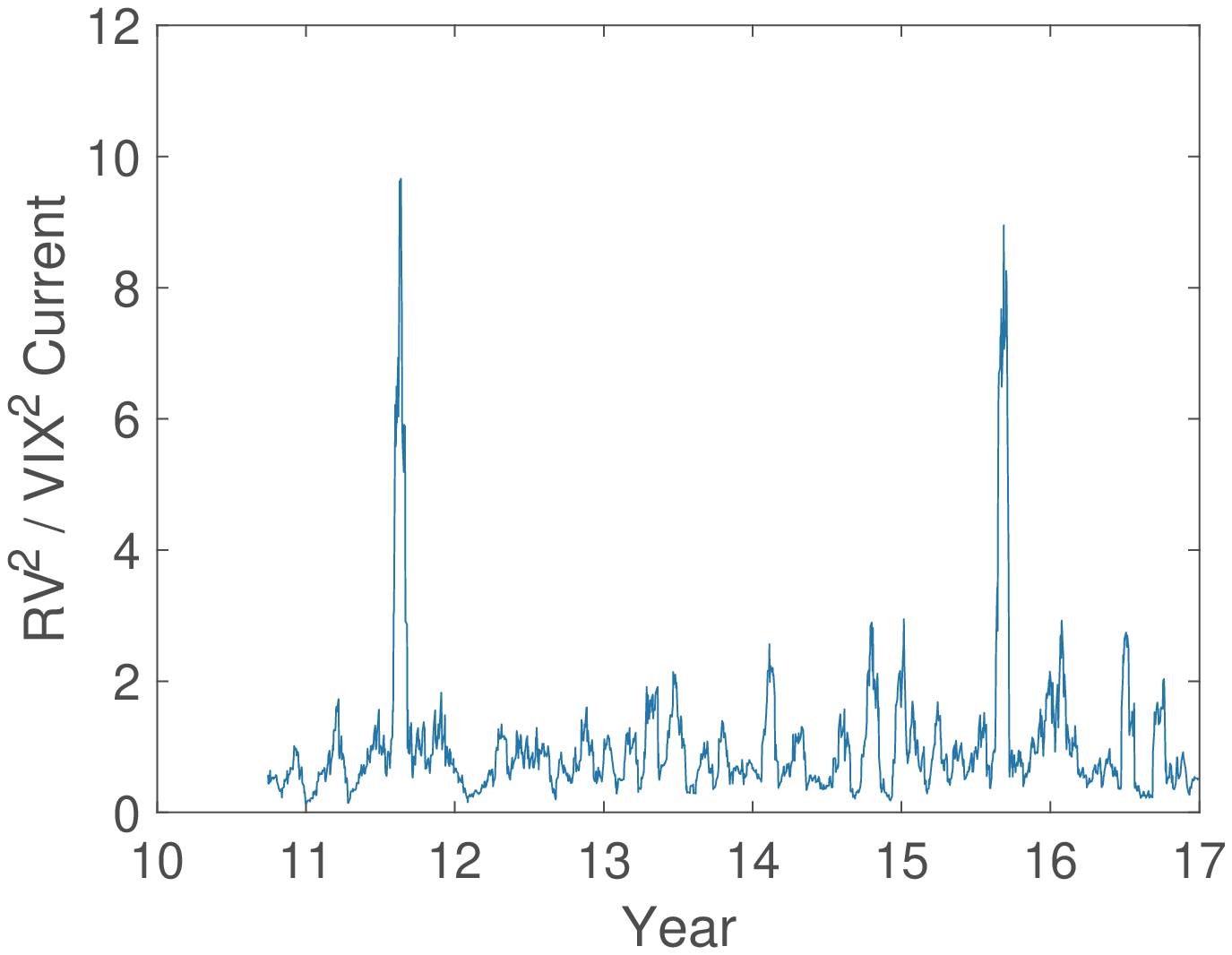} \hspace{.2cm}
\includegraphics[width = 0.4 \textwidth]{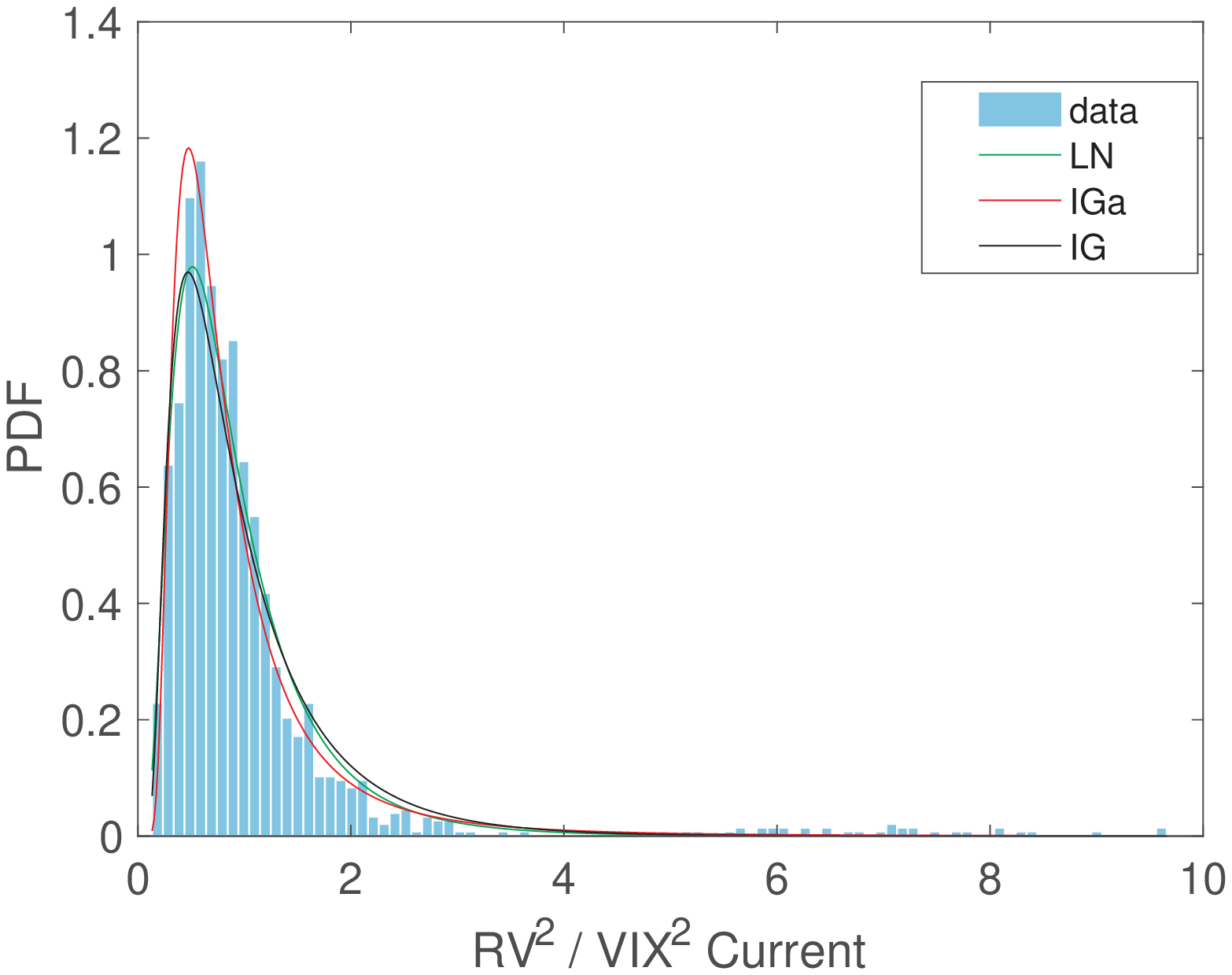}
\end{tabular}
\caption{$\mathrm{RV}^2 / \mathrm{VIX}^2$, from Aug 31st, 2010 to Dec 30th, 2016.}
\label{RVOverVIXListSRV2OverVIX22010}
\end{figure}

\begin{figure}[!htbp]
\centering
\begin{tabular}{cc}
\includegraphics[width = 0.4 \textwidth]{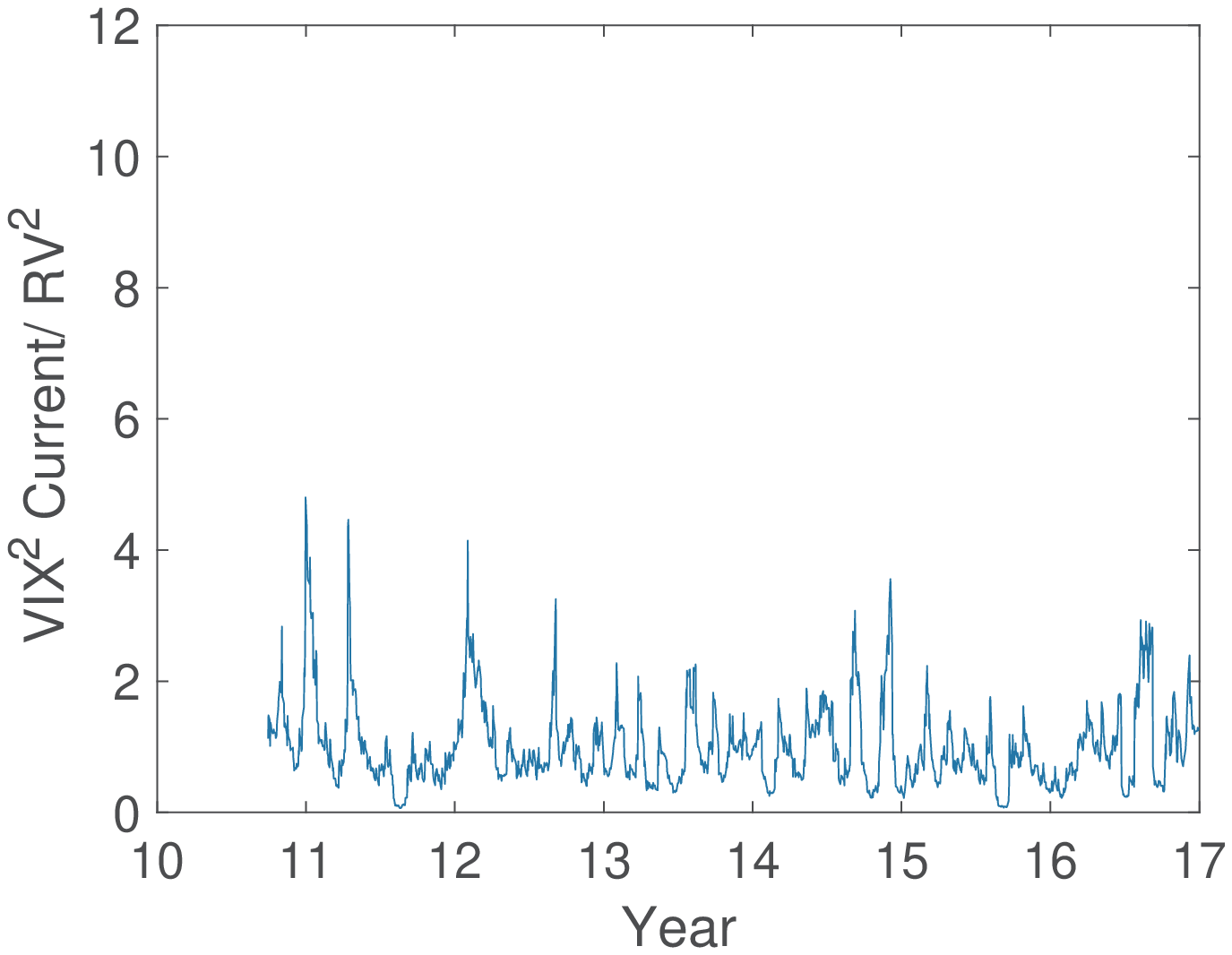} \hspace{.2cm}
\includegraphics[width = 0.4 \textwidth]{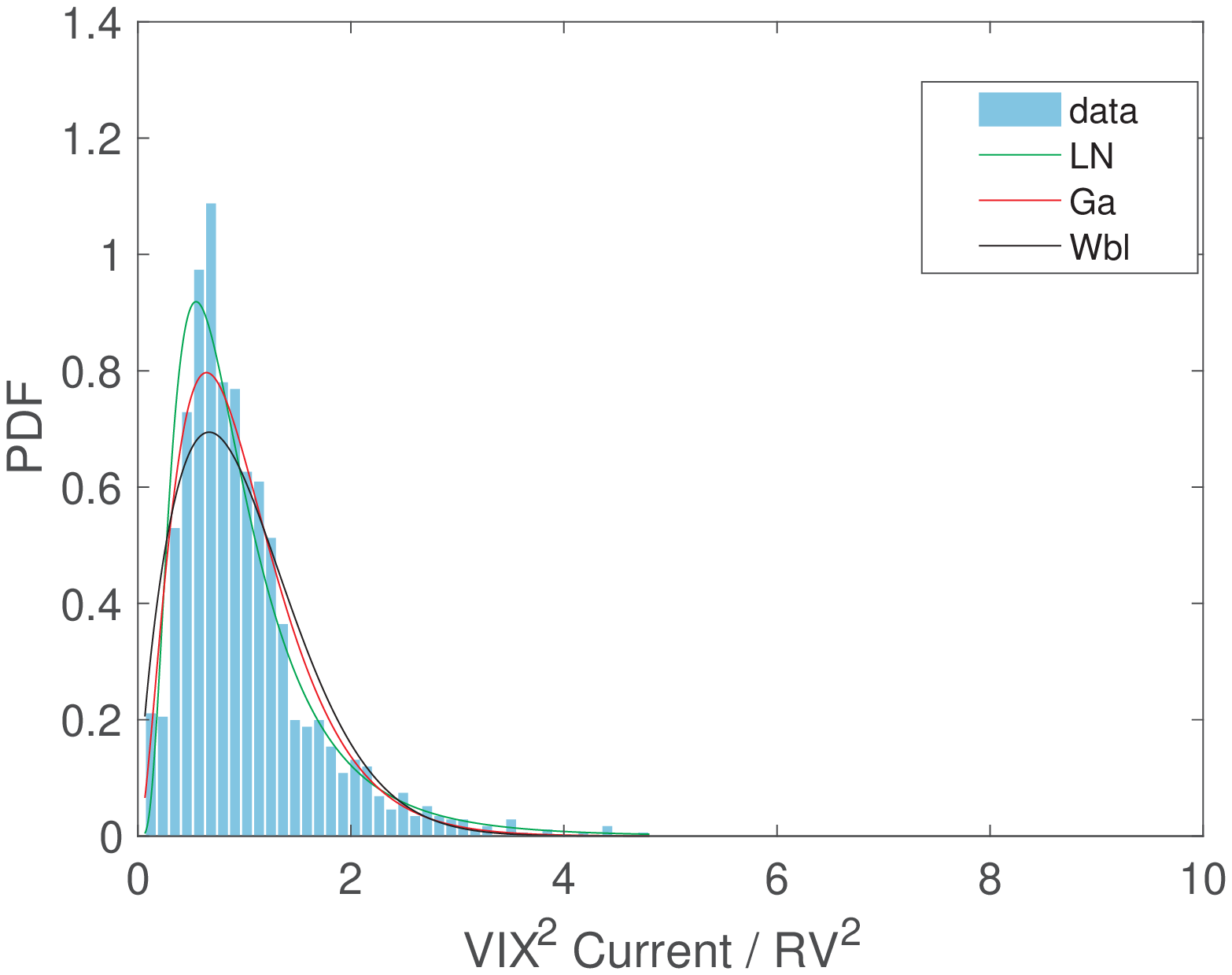}
\end{tabular}
\caption{$ \mathrm{VIX}^2 / \mathrm{RV}^2$, from Aug 31st, 2010 to Dec 30th, 2016.}
\label{VIXOverRVListSVIX2OverRV22010n}
\end{figure}

\begin{table}[!htbp]
\caption{MLE results for ``$\mathrm{RV}^2 / \mathrm{VIX}^2$" and ``$\mathrm{VIX}^2 / \mathrm{RV}^2$"}
\label{MLESRV2OverVIX22010n}
\begin{minipage}{0.5\textwidth}
\begin{center}
\begin{tabular}{ c c c} 
\multicolumn{2}{c}{} \\
\hline
            type &       parameters &          KS Statistic  \\
\hline
Normal & N(          1.0000,           1.0445) &           0.2245 \\
\hline
LogNormal & LN(         -0.2513,           0.6450) &           0.0510 \\
\hline
IGa & IGa(          2.8188,           1.8171) &           0.0420 \\
\hline
Gamma & Gamma(          2.1414,           0.4670) &           0.1082 \\
\hline
Weibull & Weibull(          1.0947,           1.2762) &           0.1251 \\
\hline
IG & IG(          1.0000,           1.8141) &           0.0745 \\
\hline
\end{tabular}
\end{center}
\end{minipage}
\begin{minipage}{.5\textwidth}
\begin{center}
\begin{tabular}{ c c c} 
\multicolumn{2}{c}{} \\
\hline
            type &       parameters &          KS Statistic  \\
\hline
Normal & N(          1.0000,           0.6306) &           0.1114 \\
\hline
LogNormal & LN(         -0.1877,           0.6450) &           0.0510 \\
\hline
IGa & IGa(          2.1414,           1.3804) &           0.1082 \\
\hline
Gamma & Gamma(          2.8188,           0.3548) &           0.0420 \\
\hline
Weibull & Weibull(          1.1263,           1.7028) &           0.0654 \\
\hline
IG & IG(          1.0000,           1.8141) &           0.0868 \\
\hline
\end{tabular}
\end{center}
  \end{minipage}
\end{table}

\newpage

\begin{figure}[!htbp]
\centering
\begin{tabular}{cc}
\includegraphics[width = 0.4 \textwidth]{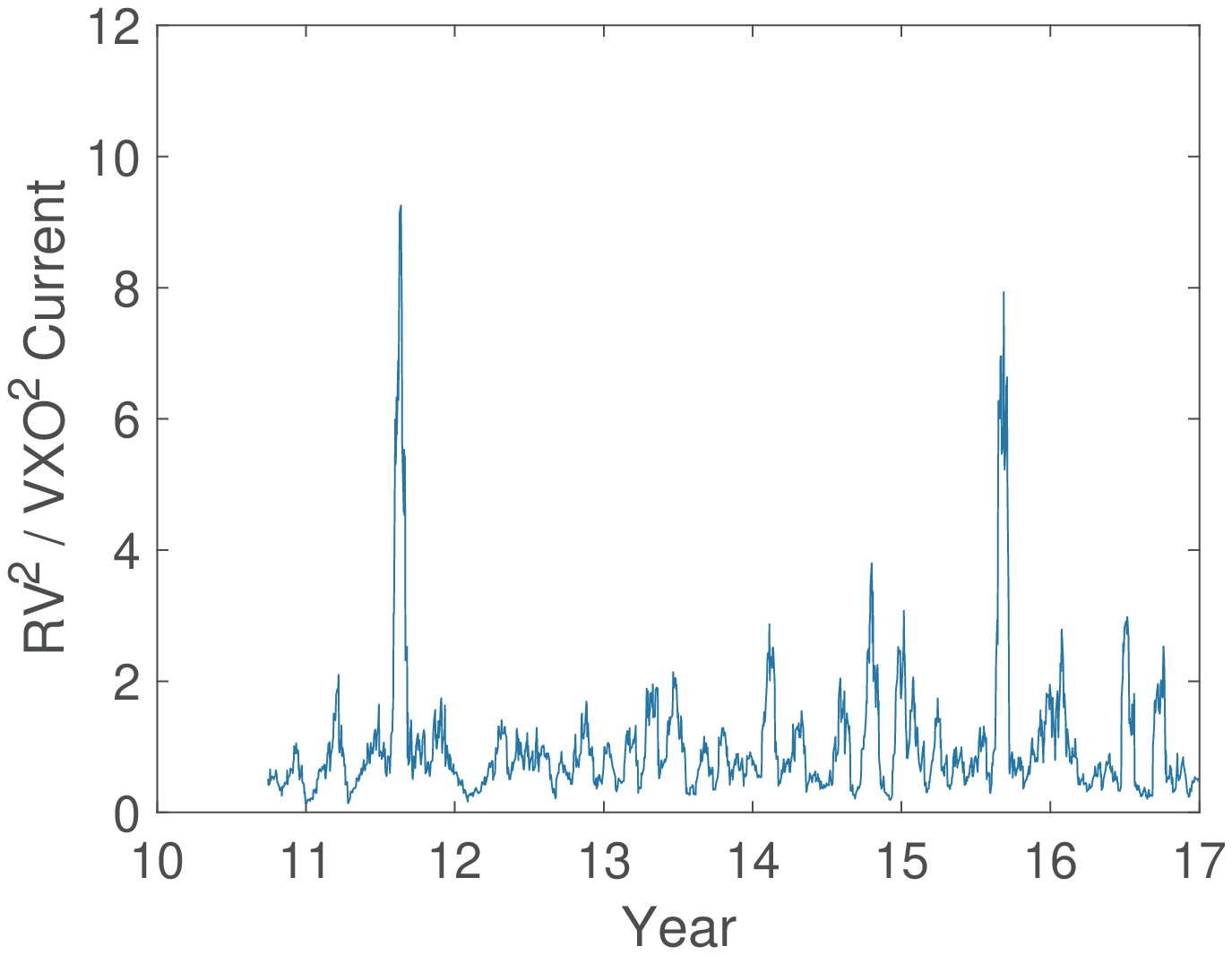} \hspace{.2cm}
\includegraphics[width = 0.4 \textwidth]{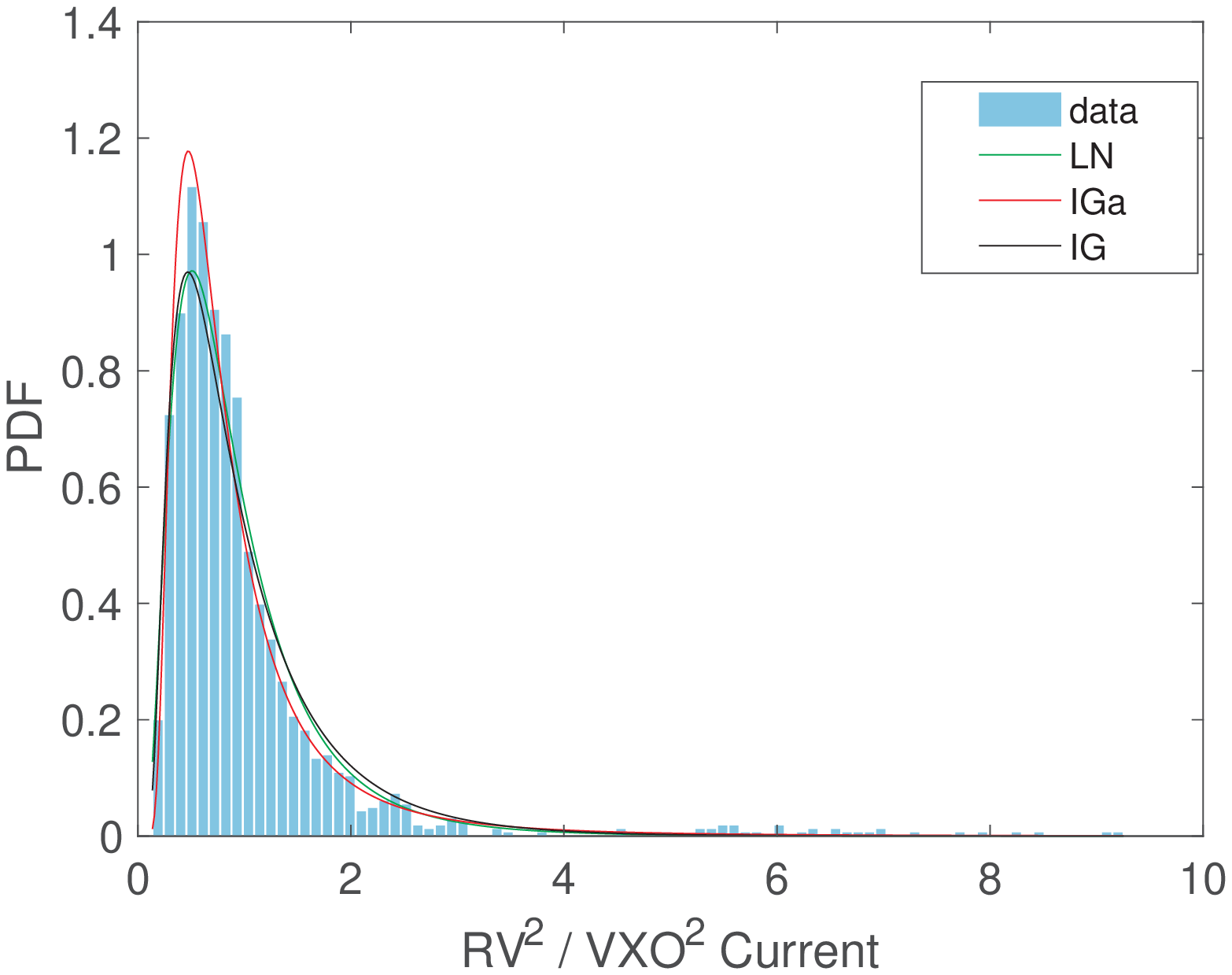}
\end{tabular}
\caption{$\mathrm{RV}^2 / \mathrm{VXO}^2$, from Aug 31st, 2010 to Dec 30th, 2016.}
\label{RVOverVXOListSRV2OverVXO22010}
\end{figure}

\begin{figure}[!htbp]
\centering
\begin{tabular}{cc}
\includegraphics[width = 0.4 \textwidth]{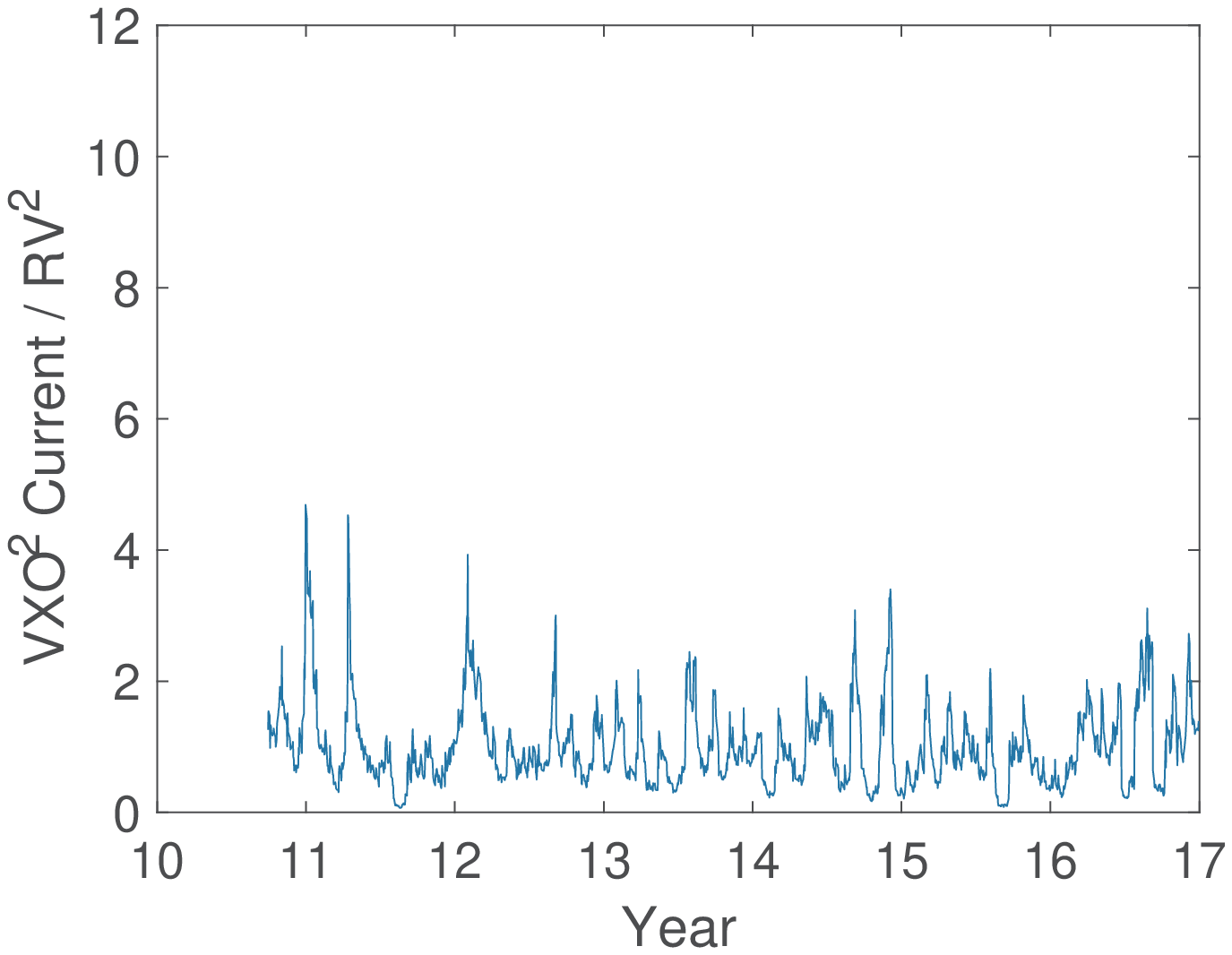} \hspace{.2cm}
\includegraphics[width = 0.4 \textwidth]{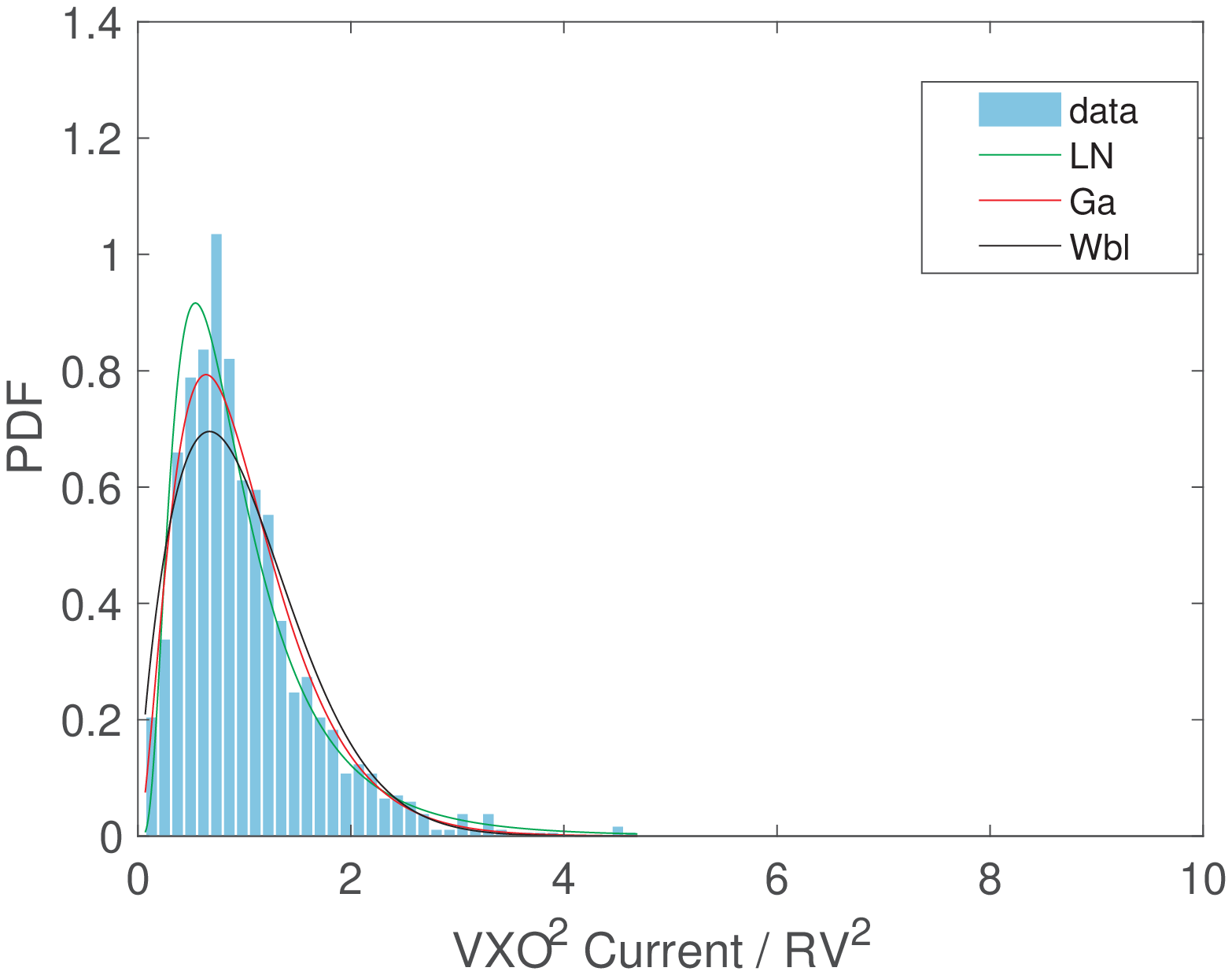}
\end{tabular}
\caption{$ \mathrm{VXO}^2 / \mathrm{RV}^2$, from Aug 31st, 2010 to Dec 30th, 2016.}
\label{VXOOverRVListSVXO2OverRV22010n}
\end{figure}

\begin{table}[!htbp]
\caption{MLE results for ``$\mathrm{RV}^2 / \mathrm{VXO}^2$" and ``$\mathrm{VXO}^2 / \mathrm{RV}^2$"}
\label{MLESRV2OverVXO22010n}
\begin{minipage}{0.5\textwidth}
\begin{center}
\begin{tabular}{ c c c} 
\multicolumn{2}{c}{} \\
\hline
            type &       parameters &          KS Statistic  \\
\hline
Normal & N(          1.0000,           0.9878) &           0.2060 \\
\hline
LogNormal & LN(         -0.2494,           0.6511) &           0.0504 \\
\hline
IGa & IGa(          2.7773,           1.7884) &           0.0301 \\
\hline
Gamma & Gamma(          2.1563,           0.4638) &           0.1090 \\
\hline
Weibull & Weibull(          1.0984,           1.3018) &           0.1153 \\
\hline
IG & IG(          1.0000,           1.8085) &           0.0670 \\
\hline
\end{tabular}
\end{center}
\end{minipage}
\begin{minipage}{.5\textwidth}
\begin{center}
\begin{tabular}{ c c c} 
\multicolumn{2}{c}{} \\
\hline
            type &       parameters &          KS Statistic  \\
\hline
Normal & N(          1.0000,           0.6258) &           0.1065 \\
\hline
LogNormal & LN(         -0.1907,           0.6511) &           0.0504 \\
\hline
IGa & IGa(          2.1563,           1.3885) &           0.1090 \\
\hline
Gamma & Gamma(          2.7773,           0.3601) &           0.0301 \\
\hline
Weibull & Weibull(          1.1262,           1.7084) &           0.0516 \\
\hline
IG & IG(          1.0000,           1.8085) &           0.0843 \\
\hline
\end{tabular}
\end{center}
  \end{minipage}
\end{table}

\clearpage

\bibliography{mybib}

\end{document}